\documentclass{aa}  

\usepackage{graphicx}
\usepackage{txfonts}
\begin{document}

   \title{Optical polarization of high-energy BL Lac objects}

   \author{T. Hovatta\inst{1,}\inst{2}
          \and
          E. Lindfors\inst{3}
          \and
          D. Blinov\inst{4,}\inst{5,}\inst{6}
          \and
          V. Pavlidou\inst{4,}\inst{5}
          \and
           K. Nilsson\inst{7}
           \and
          S. Kiehlmann\inst{1,}\inst{2}
         \and
          E. Angelakis\inst{8}
          \and
          V. Fallah Ramazani\inst{3}
          \and
          I. Liodakis\inst{4,}\inst{5}
          \and
          I. Myserlis\inst{8}
          \and
          G.~V. Panopoulou\inst{4,}\inst{5}
          \and
          T. Pursimo\inst{9}
          }

   \institute{Aalto University Mets\"ahovi Radio Observatory,
     Mets\"ahovintie 114, 02540 Kylm\"al\"a, Finland\\
              \email{talvikki.hovatta@aalto.fi}
\and
Aalto University Department of Radio Science and Engineering,P.O. BOX 13000, FI-00076 AALTO, Finland
         \and
             Tuorla Observatory, Department of Physics and Astronomy, University of Turku Finland
             \and 
             Department of Physics and Institute for Plasma Physics,
             University of Crete, 71003, Heraklion, Greece
             \and
             Foundation for Research and Technology - Hellas, IESL,
             Voutes, 71110 Heraklion, Greece
             \and
             Astronomical Institute, St. Petersburg State
             University,Universitetsky pr. 28, Petrodvoretz, 198504
             St. Petersburg, Russia
             \and
             Finnish Centre for Astronomy with ESO (FINCA), University
             of Turku, Finland
\and
             Max-Planck-Institut f\"ur Radioastronomie, Auf dem
             H\"ugel 69, 53121 Bonn, Germany
\and
             Nordic Optical Telescope, Apartado 474, E-38700 Santa
             Cruz de La Palma, Santa Cruz de Tenerife, Spain 
             }

   \date{Received XX; accepted XX}

% \abstract{}{}{}{}{} 
% 5 {} token are mandatory
 
  \abstract
  % context heading (optional)
  % {} leave it empty if necessary  
   {We investigate the optical polarization properties of high-energy
     BL~Lac objects using data from the RoboPol blazar monitoring
     program and the Nordic Optical Telescope.}
  % aims heading (mandatory)
   {We wish to understand if there are differences in the BL~Lac
     objects that are detected with the current-generation TeV
     instruments compared to those that have not yet been detected.}
  % methods heading (mandatory)
   {We use a maximum likelihood method to investigate the optical
     polarization fraction and its variability in these sources. In
     order to study the polarization position angle variability, we
     calculate the time derivative of the electric vector position
     angle (EVPA) change. We also study
     the spread in the Stokes $Q/I - U/I$--plane and rotations in the
     polarization plane.}
  % results heading (mandatory)
   { The mean polarization fraction of the TeV-detected BL~Lacs is 5\%
     while the non-TeV sources show a higher mean
     polarization fraction of 7\%. This difference in polarization fraction disappears when the
     dilution by the unpolarized light of the host galaxy is accounted
   for. The TeV sources show somewhat lower fractional
   polarization variability amplitudes than the non-TeV sources. Also the fraction of sources with a
   smaller spread in the $Q/I - U/I$--plane and a clumped
   distribution of points away from the origin, possibly indicating a
   preferred polarization angle, is larger in the TeV than in the non-TeV
   sources. These differences between TeV and non-TeV samples seems to arise from differences
   between intermediate and high spectral peaking sources instead of
   the TeV detection.  When the EVPA variations are studied, the rate of EVPA change
   is similar in both samples.  We detect significant EVPA rotations in both TeV
 and non-TeV sources, showing that rotations can occur in high
 spectral peaking BL Lac objects when the monitoring cadence is dense
 enough. Our simulations show that we cannot exclude a random walk
 origin for these rotations.}
  % conclusions heading (optional), leave it empty if necessary 
   { These results indicate that there
   are no intrinsic differences in the polarization properties of the TeV-detected and
   non-TeV-detected high-energy BL~Lac objects.  This suggests that
   the polarization properties are not directly related to the
   TeV-detection, but instead the
   TeV loudness is connected to the general flaring activity,  redshift, and the
   synchrotron peak location.}

   \keywords{Polarization --
               BL Lacertae objects: general  --
               Galaxies: jets
               }

   \maketitle
%
%________________________________________________________________

\section{Introduction}
BL~Lac objects are a type of active galactic nuclei characterized by
weak or absent emission lines \citep{stocke91}. They are typically bright and highly variable at
all wavelengths from radio to very high-energy (VHE) gamma
rays. Their spectral energy distribution (SED) consists of two humps,
the first due to synchrotron radiation, peaking at optical to X-ray
wavelengths, and the second due to inverse Compton or some hadronic
process, peaking at gamma-ray energies \citep[e.g.,][]{bottcher13}.

Traditionally, BL~Lac objects were classified as radio or X-ray
selected based on the wavelength where they were first discovered
\citep[e.g.,][]{stickel91,stocke91}. \cite{padovani95} refined the classification based on the
location of their synchrotron peak to low and
high-peaking BL~Lac objects. In this paper we use the classification from
the 3rd Fermi Gamma-Ray Space Telescope (hereafter {\it Fermi}) AGN
catalogue (3LAC) \citep{3LAC} where the sources with synchrotron peak frequency
$\nu_p < 10^{14}$\,Hz are called low synchrotron peaked (LSP),
sources with $10^{14} < \nu_p < 10^{15}$\,Hz are intermediate
synchrotron peaked (ISP), and sources with $\nu_p > 10^{15}$\,Hz are
high synchrotron peaked (HSP).

Another characteristic of BL~Lacs is their high and variable optical
polarization \citep[e.g.,][]{angel80,stocke85}. The typical
polarization fraction of BL~Lacs is $5-10$\%
\citep[e.g.,][]{angel80,smith07,heidt11} with a duty cycle of high
polarization ($p \ge 4\%$) varying from $\sim 40$ to $70$\% depending on the study. 

BL~Lac objects are also the most numerous extragalactic source class
detected in the VHE ($> 100$GeV) energies according to the TeVCat
catalogue\footnote{http://tevcat.uchicago.edu/} of sources detected by
TeV instruments. This is especially true for the ISP and HSP type
BL~Lacs, which can be explained by their high synchrotron peak
frequencies that also shifts their high-energy SED peak to higher
energies making them more easily detectable at TeV energies. They are
typically located also at relatively low redshifts so that their high-energy
emission is not greatly attenuated by the extragalactic background
light (EBL).  However, many of the sources show featureless
  spectra making it very difficult to determine their redshifts and
  the amount of EBL attenuation.

Apart from the location of their SED peaks, it is still unclear what makes an object TeV loud. It seems to be
connected to optical \citep{reinthal12,aleksic15_0806,ahnen16_1722}
and GeV \citep{aleksic14_J2001} flaring activity, indicating that
all ISP and HSP sources could be detected at very high energies if
observed during a high flux state. In this paper we study the role of optical
polarization variability by comparing a complete sample of
TeV-detected BL~Lacs with a sample of BL~Lacs not
detected in VHE bands. We concentrate on the ISP and HSP sources as
the true nature of the LSP BL~Lacs and their classification as BL~Lacs
is uncertain \citep[e.g.,][]{giommi12}. We use data from the RoboPol
blazar monitoring programme \citep{pavlidou14}, where a total number of 88
BL~Lac objects have been observed. Additionally, we use data of the
high-energy BL~Lac objects obtained at the Nordic Optical Telescope
as a part of a BL~Lac monitoring program (PI E. Lindfors).

Our paper is constructed as follows: in
Sect.~\ref{sample} we describe our sample selection and the
observations used in this paper. The results from our analysis of the
optical polarization fraction and position angle variability are given
in Sect.~\ref{results}. Our discussion and conclusions are described
in Sects.~\ref{discussion} and \ref{conclusions}. In all the
statistical tests we use a limit $p=0.05$ as the acceptance limit.

%__________________________________________________________________

\section{Sample and observations}\label{sample}
We selected our TeV-detected sample from the TeVCat
catalogue in January 2014, and it
includes 32 ISP and HSP BL~Lacs North of declination
$0^\circ$, detected by the TeV instruments before 2014.
Our TeV sample sources for which we have reliable polarization
measurements (29 objects) are
tabulated in Table~\ref{tevtable}.

In addition to the TeV sample, we constructed a sample of
ISP and HSP BL~Lac objects that have not been detected by the
TeV instruments. The observing strategies of the current-generation TeV instruments
result in a biased set of TeV-detected objects, as pointed
observations are typically done only when the sources are flaring at
some other wavelength \citep[e.g.,][]{reinthal12}. In order to verify
that our  non-TeV sources are indeed faint in the TeV
bands, we took advantage of the second {\it Fermi} high-energy catalog 2FHL
\citep{2FHL2015}, which due to its all-sky nature does not suffer from similar selection effects.
We used the highest energy band of 171-585\,GeV in 2FHL and selected all
ISP and HSP objects from the RoboPol main programme sample
\citep{pavlidou14} that are not detected in this energy bin.
This way our non-TeV sample selection is not affected by the pointing
strategies of the TeV instruments, and they have a similar observing
cadence as our TeV-detected sources. Our  non-TeV sample includes 19 objects tabulated in
Table~\ref{controltable}. Three of them have
been targeted by the VERITAS telescope with short exposure times, but
none of them showed signal higher than $0.3\sigma$ \citep{veritas16_UL}.

\subsection{Redshift distribution}
\begin{figure}
   \centering
   \includegraphics[width=9cm]{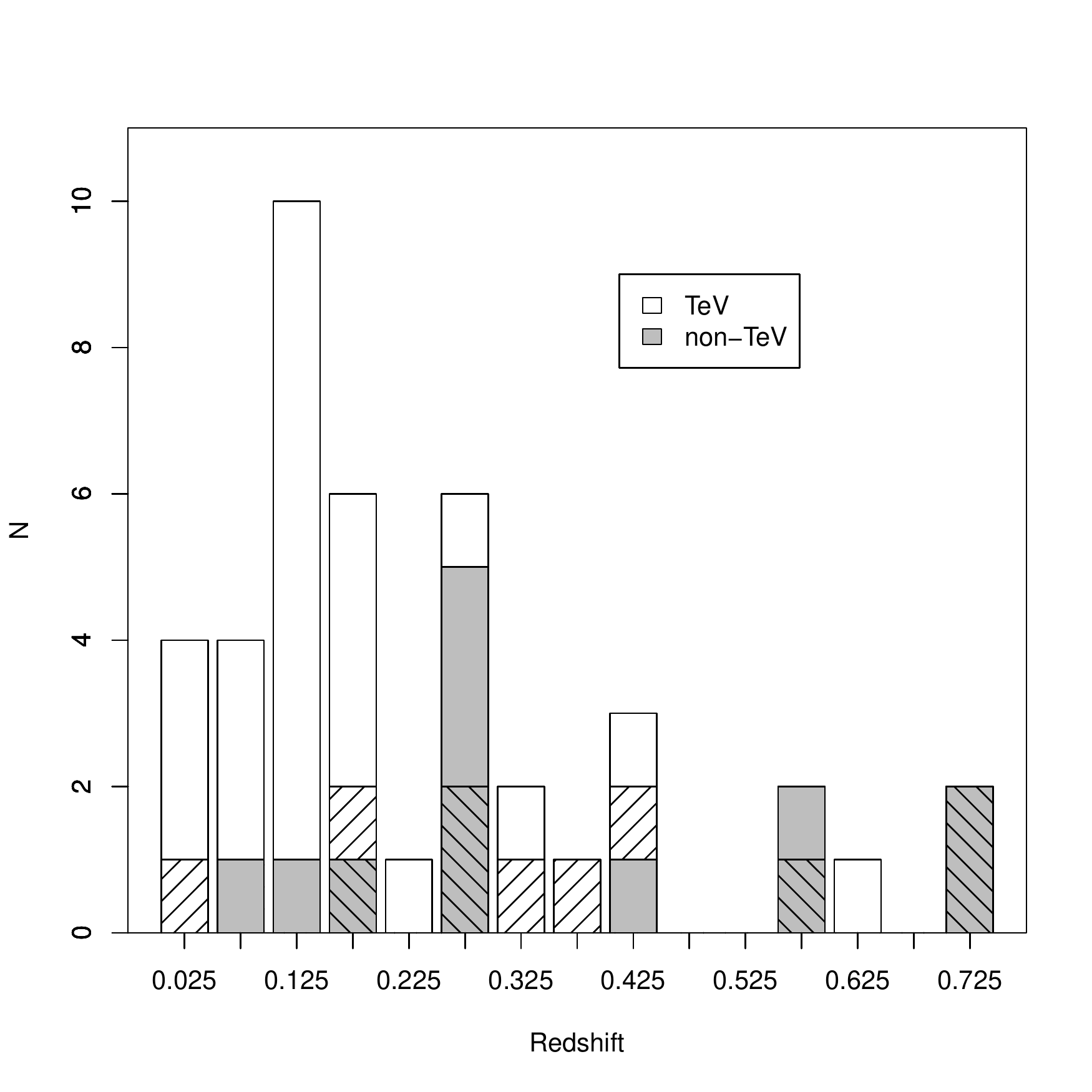}
   \caption{Stacked histogram of the redshift in the TeV (white) and non-TeV
     (gray) samples. The hatched bars show the fraction of sources
     with lower limit redshifts.}
             \label{redshift}%
\end{figure}

The redshift distributions of the TeV and  non-TeV sample sources are
shown in Fig.~\ref{redshift}. The mean redshift for the TeV sources
($0.222 \pm 0.035$) is smaller than for the  non-TeV sources ($0.446 \pm
0.070$). Because of the lower limits, the means were estimated through
a Kaplan-Meier estimator implemented in the
ASURV package \citep{lavalley92}. Similarly, we use Gehan's
generalized Wilcoxon test from the ASURV package to estimate the probability that the distributions come from the same
population. The test gives $p=0.01$, indicating that the  non-TeV sources are at higher redshifts,
assuming that the lower limits are accurate.  This redshift
  difference affects the TeV-detection of the sources as the EBL
  attenuation factor for redshift of 0.45 at 200\,GeV is about three times higher
  than for redshift of 0.2, albeit still less than one
  \citep{franceschini08, dominguez11}.
However, considering the recent detections of TeV 
emission from objects at $z > 0.9$ by the MAGIC telescopes
\citep{sitarek15, ahnen15_1441}, it is likely that this is not the
only reason why the sources are not detected by the TeV instruments.

\subsection{SED classification}\label{SEDs}
 There are several ways to model the SEDs of blazars, for example,
  by using a parabolic fit \citep[e.g.,][]{nieppola06}, a third
  degree polynomial fit \citep[e.g.,][]{3LAC}, or an empirical relation
  between the radio-optical and optical-X-ray spectral indices \citep[e.g.,][]{ackermann11} . In BL~Lac
  objects the host galaxy contribution in the optical band may also
  shift the peak frequency to a lower value, depending on how the fit
  is done. For example, VER~J0521+211 in our TeV sample is
classified as HSP in TeVCat, while it is listed as an ISP in the 3LAC
catalogue. We take all our SED classifications from 3LAC \citep{3LAC},
where they
have been uniformly estimated for both of our samples. Our TeV sample includes only three ISP sources while
these form the majority (13/19) of the non-TeV sample.  Sometimes the sources are also seen to change their SED
peak frequency during flaring \citep[e.g.,][]{pian98, giommi00, ahnen16_1011}, which further
complicates the classification. Thus, it is clear that our TeV and
non-TeV samples differ on their SED properties, which along with the
redshift difference may explain why the non-TeV objects have not been
detected at TeV energies.
Therefore, whenever possible, we check how  the SED-classification difference affects our results.

\subsection{RoboPol observations}\label{obs}
Optical R-band polarization observations were obtained for the TeV and
and  non-TeV sources
with the RoboPol instrument, mounted on the 1.3m telescope at Skinakas
Observatory\footnote{\url{http://skinakas.physics.uoc.gr}}, in Crete. 
The polarimeter contains a fixed set of two Wollaston prisms and
half-wave plates, allowing simultaneous measurements 
of the Stokes $I,Q/I,U/I$ parameters for all point sources within the
$13\arcmin \times 13\arcmin$ field. The R-band magnitudes were calculated
using calibrated field stars either from the
literature\footnote{\url{https://www.lsw.uni-heidelberg.de/projects/extragalactic/charts/}}
or from the Palomar Transient Factory $R$-band catalogue
\citep{ofek12b} or USNO-B1.0 catalogue \citep{Monet2003}, depending
on their availability. 
The exposure time was adjusted based the brightness of the target and
sky conditions, and varied between 100\,s and 1800\,s.

The observations were reduced using the pipeline described in
\cite{king14}, which uses aperture photometry. The TeV sources were
measured with a fixed $3\arcsec$ aperture diameter. In the  non-TeV sources
where the host galaxy contribution is smaller (see Sect.~\ref{host}),
in order to optimize the SNR, 
the aperture size was defined as $2.5 \times \text{FWHM}$, where FWHM
is an average full width at half maximum of stellar images. The average FWHM for RoboPol images is equal to $2.07\arcsec$.
Stability of the instrumental polarization was controlled by nightly observations of polarized and zero-polarized standard stars.

Four sources, 1ES~0033+595, 1ES 1440+122, Mrk~421 and 1ES
1741+196, had confusing field sources entering the aperture, preventing us from
obtaining good-quality observations with RoboPol.  In addition, three sources, 1ES~0229+200, Mrk~501, and 1ES~2344+514, had too
bright a host galaxy for reliable polarization measurements with
RoboPol. Consequently, Mrk~421 and Mrk~501 were excluded from our
sample, while for other four problematic cases we have observations taken with the Nordic Optical
Telescope (see below).

All the  non-TeV sources and some of the TeV sources were
already part of the RoboPol main monitoring sample (Pavlidou et
al. 2014), while others were added into the monitoring as a separate
TeV-source project. We only use data taken during the 2014 observing
season, which lasted from April to November. The mean number of
observations for the TeV sources is 9.6 and for the  non-TeV sources 10.8. These are also tabulated for each
individual source in Tables \ref{tevtable} and \ref{controltable}.

\subsection{NOT observations}\label{not}
For 13 TeV sources, we obtained observations also with the Nordic
Optical Telescope\footnote{\url{http://www.not.iac.es/}}. The observations of 1ES 0033+595 had to be excluded
from the analysis due to a close-by confusing source, bringing our final TeV-detected sample to
29 sources.

The observations were done with
ALFOSC\footnote{\url{http://www.not.iac.es/instruments/alfosc/}} in the
R-band using the standard setup for linear polarization observations
(lambda/2 retarder followed by a calcite). The observations were
performed twice per month from April 2014 to November 2014. In total we had 15 observing epochs. The exposure times varied from 10\,s to 360\,s depending on the source brightness.
Polarization standards were observed monthly to determine the zero
point of the position angle. The instrumental polarization was
determined using observations of zero polarization standard stars
and was found to be negligible. The observations were mostly conducted
under good ($\sim1\arcsec$) seeing conditions. 

The data were analyzed using the pipeline developed in Tuorla Observatory. It uses
standard procedures with semi-automatic software. The sky-subtracted
target counts were measured in the ordinary and extraordinary beams using aperture photometry.
The normalized Stokes parameters and the polarization fraction 
and position angle were calculated from the intensity ratios
of the two beams using standard formula \citep[e.g.][]{landi07}. 
As the data were taken under good seeing conditions, and the optics of
NOT are excellent, we were able to use aperture diameter of $3\arcsec$ to
minimize the contribution of the unpolarized host galaxy flux to our
measurements. 

Because the field-of-view in our observations is rather small, and in many cases includes no
comparison stars to be used for differential photometry, we were able
to perform photometry for the NOT data only for 1ES~2344+514.

\section{Results}\label{results}
We show the polarization time series of all sources in Appendix A. 
\subsection{Polarization fraction and its variability}\label{modindex}
Here we examine whether the mean polarization fraction and its variability
amplitude are different for the TeV and  non-TeV samples. We use a
maximum likelihood approach to
estimate the ``intrinsic'' mean polarization fraction and modulation
index (standard deviation over mean) of the source. The term ``intrinsic'' denotes values
we would expect if we had perfect sampling and no measurement
uncertainties. We assume the polarization fraction follows a Beta
distribution because Beta distribution is confined between 0 and 1 similarly as the
polarization fraction. The observational uncertainties are accounted
for by convolving the probability density of
the Beta distribution with a probability density of the Ricean
distribution (assumed distribution for a single polarization
measurement). This results in a probability density function
\begin{equation}
{\rm PDF}\left(p;\alpha, \beta\right)=\frac{p^{\alpha -1} \left( 1-p\right)^{\beta -1}}{B\left(\alpha, \beta\right)},
\end{equation}
where $p$ is the polarization fraction and $\alpha$ and $\beta$ determine
the shape of the Beta distribution $B\left(\alpha, \beta\right)$. If
the parameters $a, \beta$ of this distribution are known, the mean
polarization fraction and the intrinsic modulation index are then given by 
\begin{equation}
p_\mathrm{int}=\frac{\alpha}{\alpha + \beta}
\end{equation}
and
\begin{equation}
m_\mathrm{int}=\frac{\sqrt{Var}}{p_\mathrm{int}}=\frac{\sqrt{\frac{\alpha\beta}{\left(\alpha + \beta\right)^2 \left(\alpha + \beta
       +1 \right)}}}{\frac{\alpha}{\alpha + \beta}}, 
\end{equation}
where $Var$ is the variance of the distribution. 
Details of the method are described in Appendix A of
\cite{blinov16}.

The main advantage of this method is that it provides
uncertainties for both the mean polarization fraction and the modulation index, and when the
values cannot be constrained, it is possible to calculate a 2$\sigma$
upper limit. One important thing to note is that the method takes
as input the observed polarization fraction without debiasing, and
automatically and properly accounts for biasing \citep[for details on
why debiasing is typically applied, see e.g.,][]{simmons85}.
For this reason, the polarization curves presented in Appendix A do not have
debiasing applied.

\begin{figure}
   \centering
   \includegraphics[width=9cm]{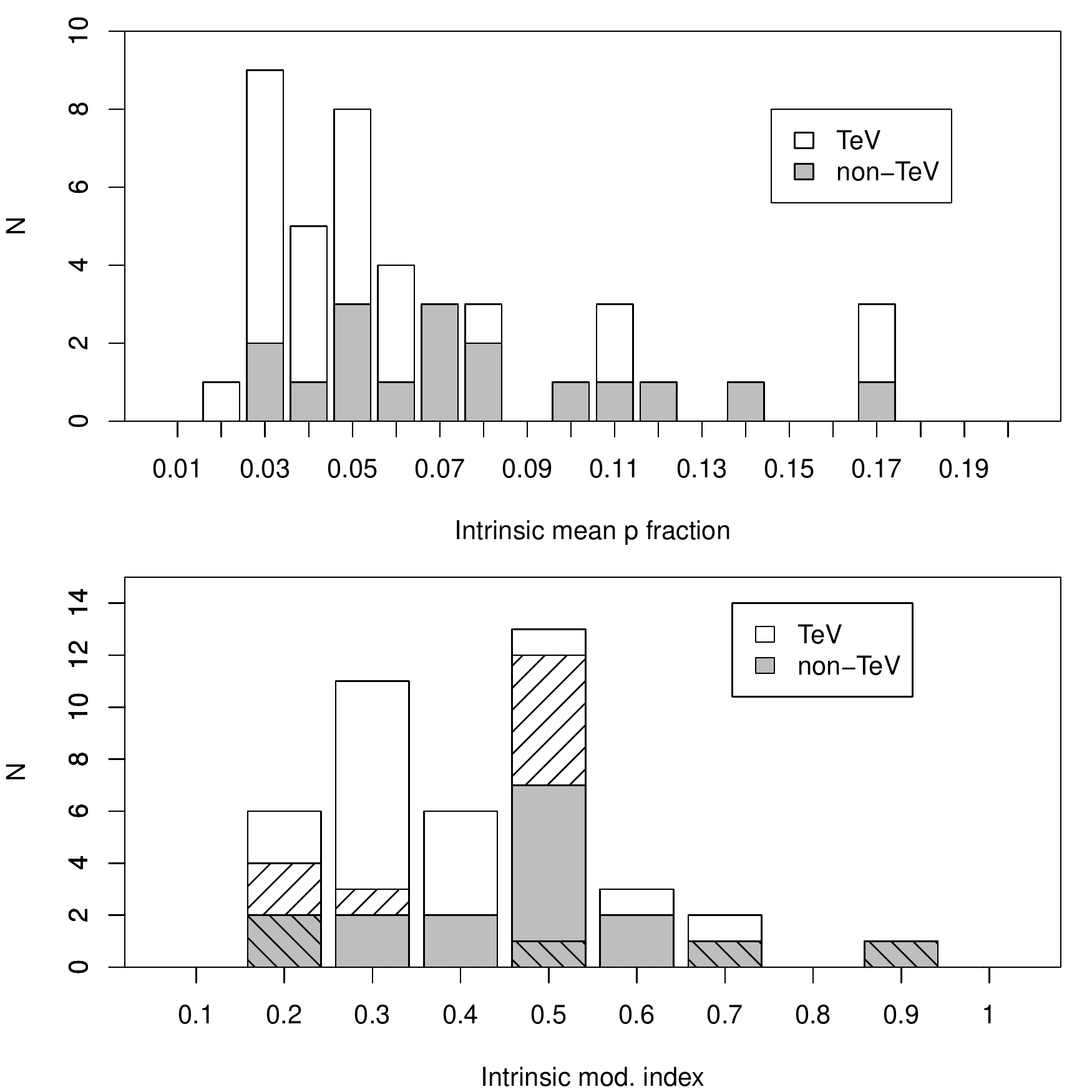}
   \caption{Top: Stacked histogram of the intrinsic mean polarization
     fraction in the TeV (white) and non-TeV (gray) samples. Bottom:
     Stacked histogram of the intrinsic modulation index in the TeV (white)
   and non-TeV (gray) samples. The hatched bars show the fraction of 
   TeV or non-TeV sources that have only 2$\sigma$ upper limits available.}
             \label{hist}%
    \end{figure}

The likelihood method is applicable to sources with at least three
observations out of which at least two have a
signal-to-noise ratio $\ge 3$. This results in a sample of
25 TeV and 17  non-TeV sources. The mean polarization fraction
and the intrinsic modulation index are tabulated in Table
\ref{tevtable} for the TeV sources and in Table \ref{controltable} for
the  non-TeV sources.

In Figure \ref{hist} (top panel) we show the distribution of the intrinsic mean
polarization fraction for the TeV and  non-TeV samples. The mean
polarization fraction for the TeV sources is $0.054\pm0.008$ and for the  non-TeV
sources $0.073\pm0.009$. A two-sample Kolmogorov-Smirnov (K-S) test
gives a p-value of 0.070 for the null hypothesis that the two samples
were drawn from the same distribution, so the null hypothesis cannot
be rejected on the grounds of that test.  However, a two-sided
Wilcoxon test gives a probability of $p=0.028$ for the means of the distributions to be the same, which indicates
that the  non-TeV sources have higher intrinsic polarization fraction
than the TeV sources.

The bottom panel of Fig. \ref{hist} shows the distributions for the
intrinsic modulation index. Because of the upper limits, we 
calculate the means of the distributions through a Kaplan-Meier
estimate, which gives $0.29\pm0.03$ for the TeV sources and
$0.38\pm0.04$ for the  non-TeV sources. Gehan's generalized Wilcoxon
test gives a probability of $p=0.031$ that the two distributions come
from the same population indicating that the  non-TeV sources are more
variable than the TeV sources. If we only consider the HSP sources,
the two samples can no longer be distinguished ($p=0.71$) and the mean
values are more similar ($0.31\pm0.03$ for the TeV and $0.36\pm0.09$
for the  non-TeV sources). However, our sample includes only five HSP
 non-TeV sources for which we obtained a modulation index so that the
result is affected by the small number of sources.

\subsection{Host galaxy contribution}\label{host}
The host galaxies of TeV blazars are known to contribute a significant
fraction to the total flux in some targets
\citep{nilsson07} and it is therefore possible that our polarization
fraction observations are affected by the unpolarized starlight from the
galaxy \citep[e.g.,][]{andruchow08, heidt11}. In order to test this, we collected host
galaxy magnitudes for the sources in our sample from the
literature. These
are tabulated in Tables \ref{tevtable} and \ref{controltable} along
with the observed and Galactic extinction--corrected magnitudes. We use
the re-calibrated dust maps of \cite{schlafly11} with
the reddening law of \cite{fitzpatrick99} extracted from NASA/IPAC
Extragalactic Database (NED) for the Galactic extinction
correction. 

Most of the host galaxy magnitude estimates in the literature are
obtained by modeling the core and galaxy emission using a De
Vaucouleurs intensity profile integrated to infinity
\citep[e.g.,][]{nilsson99}. In these cases, whenever the effective
radius of the galaxy was available in the literature, we estimate the
contribution of the host galaxy to our magnitude estimates by
integrating up to the aperture size used in our observations,
using the equations described in \cite{nilsson09}.  As explained
  in Sect.~\ref{obs}, we use a different aperture size for the TeV and
  non-TeV sources. Therefore, for the TeV
sources, we integrate up to a radius of $1.5\arcsec$, and
for the  non-TeV sources, we use an aperture radius of $2.6\arcsec$.

If the host galaxy magnitude or limit was not obtained using an R-band
filter, we convert between the magnitude systems by using the
following average color relations for 
elliptical galaxies from \cite{kotilainen98} and \cite{fukugita95}:
$R-H = 2.5$, $H-K=0.2$, and $R-I=0.7$. 
In \cite{shaw13}, the absolute magnitude of the host galaxy is
estimated from the spectra instead of fitting images. 
We convert their absolute magnitudes to apparent magnitudes using the cosmological parameters listed in their paper.
The values tabulated in Tables \ref{tevtable} and \ref{controltable}
are the R-band host magnitude values we use in our analysis while we give a reference to
the original paper where the host magnitude is given. We note that
ideally one should use the same aperture size and same calibration stars
as in the original derivation of the host magnitude in order to obtain
accurate results. As this is not possible for most of our sources, the
uncertainty in the host-corrected magnitudes is most likely quite
large and values for individual sources should be treated with caution.

In the following we only consider sources for which we were able to
determine a mean polarization fraction using the likelihood analysis. 
As explained in Sect.~\ref{not}, due to the small field of view
of the NOT polarimeter and the lack of calibrated standard stars in the field,
we were able to estimate photometry only for 1ES~2344+514 from our NOT observations. We tabulate the mean magnitudes for the sources without NOT
magnitudes using data from the Tuorla blazar monitoring
program\footnote{\url{http://users.utu.fi/kani/1m}} \citep{takalo08} taken
in 2014 but do not use them in the following analysis. 

\begin{figure}
  \centering
  \includegraphics[width=9cm]{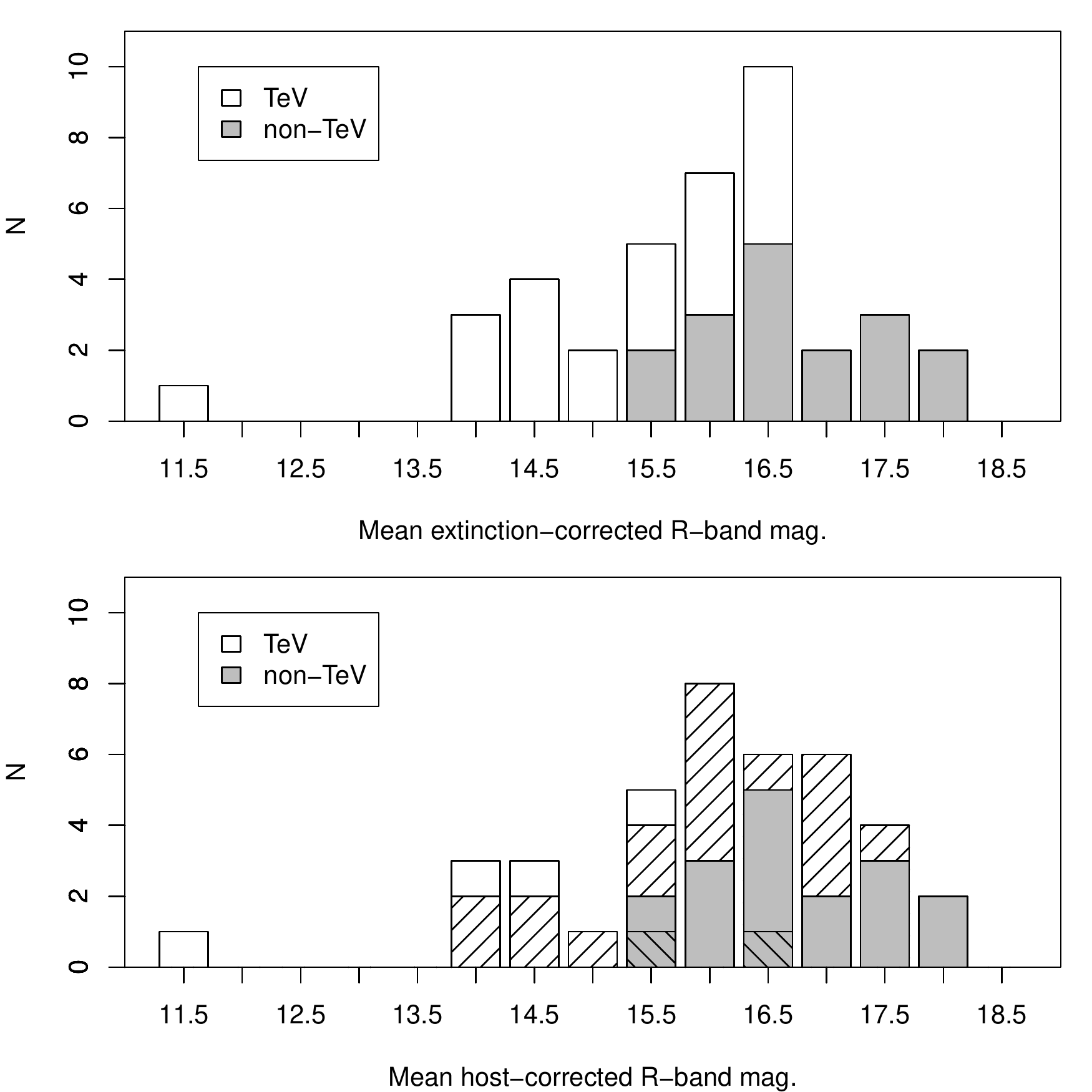}
  \caption{Top: Stacked histogram of the extinction-corrected mean magnitude in the TeV (white) and  non-TeV (gray) samples. Bottom:
    Stacked histogram of the host-corrected mean magnitude in the TeV
    (white hatched)  and  non-TeV (gray hatched) samples. For sources
     where a host correction is not available, we show the
     extinction-corrected mean magnitudes as in the top panel.}
   \label{mag}%
 \end{figure}

In Fig.~\ref{mag} top panel we show the extinction-corrected
magnitudes for the TeV and  non-TeV samples. The outlier TeV source in
the figure is the extreme HSP source HESS J1943+213
\citep{akiyama16,straal16} at a low Galactic latitude where the
extinction correction is uncertain. Therefore we exclude it from the
statistical tests (we note that the conclusion remains the same
regardless of its exclusion). The mean extinction-corrected magnitude
for the TeV sources is $15.2\pm0.2$ and for the  non-TeV sources
$16.4\pm0.2$. A K-S test gives $p=0.004$ for the distributions to
come from the same population, which indicates that our TeV and
 non-TeV sources have different magnitude distributions, which is not
surprising considering that the  non-TeV sources reside at higher
cosmological distances.  Because the TeV sources are much brighter
  in the optical than the non-TeV sources, they may also be brighter
  in the gamma-ray bands as the fluxes of BL~Lac objects in these two regimes are correlated \citep[e.g.,][]{bloom97,
    hovatta14a,wierzcholska15}. This may in part also explain why some
objects in our non-TeV sample are not detected at TeV energies with the current instruments.

The bottom panel of Fig.~\ref{mag} shows the host- and extinction--corrected magnitudes
in hatched for the sources where a host galaxy magnitude was found in
the literature. Combining the host- and extinction-only--corrected (where
host correction is not available) magnitudes, the TeV sources have a
mean magnitude of $15.4\pm0.2$ and the  non-TeV sources a mean
magnitude of $16.4\pm0.2$. A K-S test gives $p=0.043$, which indicates
that even when the host correction is accounted for, the  non-TeV
sources are fainter. We note that
we have a host magnitude estimate for only two  non-TeV sources, and in
both cases the magnitude changes only very little, so that the  non-TeV
sample mean is very similar to the un-corrected one. The lack of host
magnitude estimates for the  non-TeV sources is likely due to them
residing at higher redshifts than the TeV sources. In fact, many of them were observed by \cite{shaw13} but no host galaxy
contribution was seen in their spectra. Considering the strong dependence of the host galaxy
luminosity on redshift \citep{nilsson03}, we can expect the host
galaxy contribution in the remaining  non-TeV sources to be small.

We then proceed to estimate the amount of unpolarized
host-galaxy light on our polarization fraction estimates. We remove the host galaxy flux density from the mean
observed flux density and re-calculate the host-corrected polarization fraction
$p_\mathrm{corr}$ following,
\begin{equation}
p_\mathrm{corr} = \frac{p_\mathrm{int} I}{I - I_\mathrm{host}},
\end{equation}
where $p_\mathrm{int}$ is the mean polarization fraction from the likelihood
analysis, $I$ is the flux density calculated from the mean observed
magnitude, and $I_\mathrm{host}$ is the flux density of the host galaxy.

We find that the mean polarization fraction in the TeV sample is $0.068\pm0.010$,
which is similar (K-S test p=$0.345$) as in the  non-TeV sample
where the mean is $0.073\pm0.009$. This indicates that after
correcting for the host galaxy contribution, the polarization fraction
of the TeV and  non-TeV samples cannot be distinguished. 

\subsection{Polarization angle variability}\label{evpa}
In this section we examine whether there are differences in the
polarization angle variability of the TeV and  non-TeV samples. We do
this first by calculating the time-derivative of the electric vector
position angle (EVPA). In order to account for the n$\pi$~ambiguity in
the position angle, we smooth the EVPA curves by always requiring
that the difference between consecutive points is $<90^\circ$.
We only use observations with a signal-to-noise
ratio of at least three in the polarization fraction. In calculation of
the derivative, we further 
require that the change in the EVPA between consecutive points meets the following criteria
\begin{equation}
|\theta[i+1]-\theta[i]| > \sqrt{\sigma[i+1]^2+\sigma[i]^2},
\end{equation}
where $\theta$ is the EVPA and $\sigma$ its uncertainty. If this
criterion is not met, we can either set the derivative to zero or
ignore it. The first way is more appropriate when the sources do not
exhibit much variation, while the latter gives an estimate of the
typical rate in the polarization angle variability when they do change
significantly. In both cases we calculate the derivative between two
consecutive points and take the absolute value. For sources with a
redshift estimate or limit available, we multiply the observed
derivative by $(1+z)$ in order to look at the variations in the source
frame. We use the redshift limits as values when doing this. For
sources without redshift estimates, we do not correct the derivative.

\begin{figure}
   \centering
   \includegraphics[width=9cm]{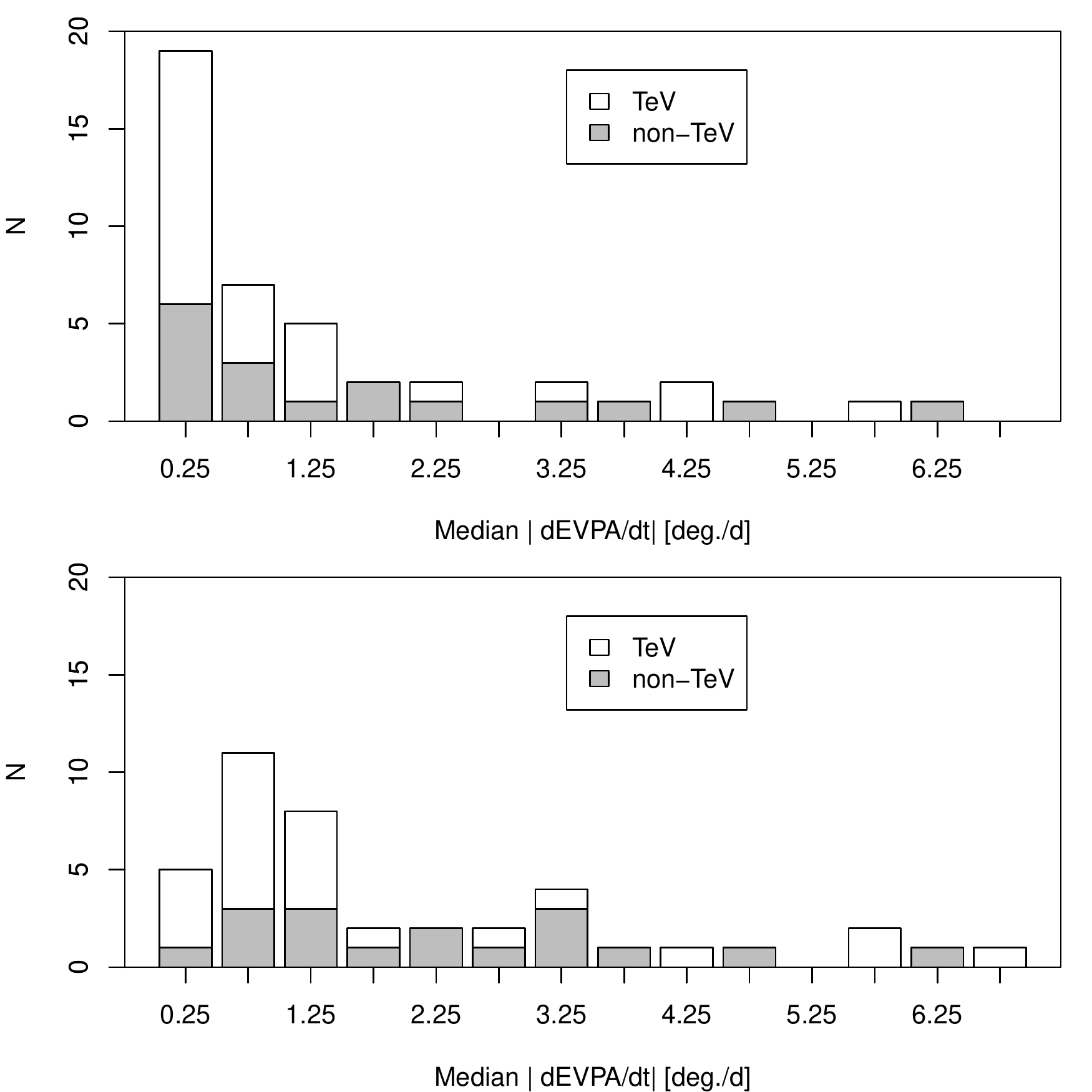}
   \caption{Stacked histogram of the median EVPA derivative for the TeV
     (white) and  non-TeV (gray) samples. Top panel shows the
     distribution when the derivative is set to zero for insignificant changes
   and the bottom panel shows the same when insignificant changes are
   ignored (see text for details).}
             \label{evpahist}%
     \end{figure}

The distribution of the median redshift-corrected absolute derivatives for each source are shown in
Fig.~\ref{evpahist}. The top panel shows the distributions when the
derivative is set to zero for insignificant changes. The mean values of the distributions are
similar for the TeV (mean $1.11\pm0.29$) and  non-TeV (mean $1.66\pm0.45$) samples. A
K-S test gives a probability $p=0.583$ for the them to come from the
same population. If we look at the HSP  non-TeV sources only (5
sources), their mean is $0.62\pm0.31$, which is similar to
the TeV mean.

The bottom panel shows the same when ignoring
derivatives between insignificant variations. As expected, the mean
values are now higher (TeV mean $1.81\pm0.39$ and  non-TeV mean
$2.38\pm0.40$) and in this case the K-S test gives a probability
$p=0.034$ indicating that when the sources show significant
variability between consecutive measurements, the magnitude is larger
in  non-TeV sources. We note that the true difference could be even
larger as many of the  non-TeV sources do not have redshift estimates
available. Again, if we only consider the HSP  non-TeV sources, the
mean is $1.71\pm0.55$, very close to the TeV source mean.

\begin{figure}
   \centering
   \includegraphics[width=9cm]{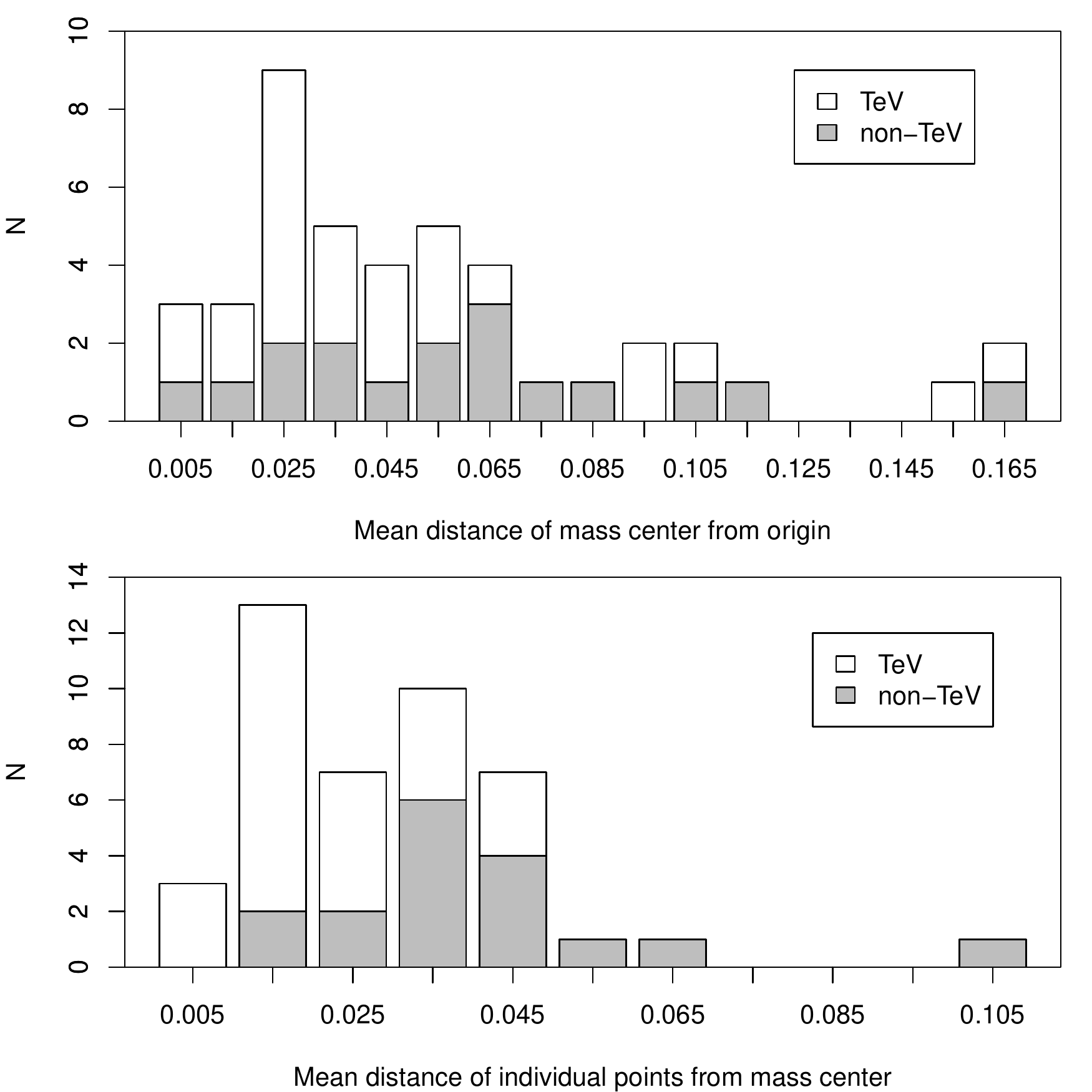}
   \caption{Top: Stacked histogram of the distance of the mass center in
     the $Q/I-U/I$-plane from the origin. Bottom:
Stacked histogram of the mean distance of the Q/I
     vs. U/I points from the mass center.  The TeV
     sources are shown in white and  non-TeV sources in gray.}
             \label{QUdist}%
    \end{figure}

Another way to study the polarization angle variability is to look at
the polarization in the $Q/I$--$U/I$ plane. The $Q/I$--$U/I$
plots for each individual source are shown in Appendix A. A larger
spread in the $Q/I-U/I$-plane suggests more EVPA variability, while a
clumped distribution away from the origin could be an indication of a preferred
polarization angle. In order to quantify these effects, we calculate the
weighted mean $Q/I$ and $U/I$ as the mass center of the points. The
distribution of the distance of the mass center from origin is shown
in Fig.~\ref{QUdist} top panel. The mean distance for the TeV sources
is $0.050\pm0.008$ and for the  non-TeV sources $0.060\pm0.010$. As
expected, these are similar to the intrinsic mean polarization degree
estimates. A K-S test gives $p=0.197$ for the distributions to come from the same
population, and we cannot reject the null hypothesis. 

We then estimate the distance of each individual measurement to this mass
center, and take a mean value to estimate the scatter in the $Q/I-U/I$-plane. The distributions of these mean
distances are shown in Fig.~\ref{QUdist} bottom panel for both the TeV and  non-TeV
samples. The mean value for the TeV sources is $0.021\pm0.003$ and for
the  non-TeV sources $0.041\pm0.005$. According to a K-S test, which
gives p=$0.003$, we can reject the null hypothesis that these come
from the same population, indicating that the TeV sources show less
spread in the $Q/I-U/I$-plane. This is in accordance with the
modulation index results where the TeV sources were found to show less
variability than the  non-TeV sources.

\begin{sidewaystable*}
\scriptsize
\caption{TeV sample sources and their variability properties.} \label{tevtable}      
\centering          
\begin{tabular}{l l l l l l l l l l l l l l l l r l}     % 7 columns 
\hline\hline       
Name & TeVCat name & 2FHL name & Redshift\tablefootmark{a} & SED & Intrinsic mean
& Intrinsic & median & N & host-corr & 
mean & de-red. & host-corr & host gal. & Ref. & Jet P.A. & Ref. \\ 
 & & & & class\tablefootmark{a} & p fraction & mod. index &
 |d$\theta$/d$t$| & obs & p fraction & R-mag.\tablefootmark{b} &
 R-mag. & R-mag. & R-mag.\tablefootmark{c} & host & (deg.) & jet. P.A.\\
\hline                    
J0136+3905 & RGB J0136+391 & 2FHL J0136.5+3906 & $>0.41$\tablefootmark{d} & HSP & $0.026\pm_{0.003}^{0.004}$  & $0.386\pm_{0.095}^{0.138}$ & 2.94 & 9 &
$-$ & $15.6\pm0.02$ & 15.5 & $-$ & $-$ & $-$ & $-$ & $-$ \\
J0152+0146 & RGB J0152+017 & 2FHL J0152.8+0146 & 0.082 & HSP &
$0.035\pm_{0.005}^{0.006}$  & $0.387\pm_{0.110}^{0.158}$ &
0.26 & 9 & 0.049  & $15.3\pm0.08$ & 15.2 & 15.6 & 16.7 & 1 & $-135.6$ & 8 \\
J0222+4302 & 3C 66A& 2FHL J0222.6+4301& $>0.335$\tablefootmark{e} & HSP & $0.055\pm_{0.007}^{0.008}$ & $0.293\pm_{0.074}^{0.118}$ & 1.05 & 7 & $-$ & $14.47\pm0.07$ & 14.29 & $-$ & $-$ & $-$ & $-$ & $-$\\
J0232+2017 & 1ES 0229+200 & $-$ & 0.139 & HSP & $0.024\pm_{0.010}^{0.019}$ & $0.474\pm_{0.181}^{0.380}$ & 0.41 & 3 &
$-$ & $16.1\pm0.01^*$ & 15.8 & 18.2 & 16.2 & 2 
& $-$ & $-$\\
J0319+1845 & RBS 0413 & 2FHL J0319.7+1849 & 0.192 & HSP & $-$ & $-$ &
$-$ & 3 & $-$ & $17.5\pm0.14$ & 17.3 & $-$ & $-$ & $-$ & $-$ & $-$\\
J0416+0105 & 1ES 0414+009 & 2FHL J0416.9+0105 & 0.287 & HSP & $-$ & $-$
& $-$ & 1 & $-$ & $16.55\pm0.04$ & 16.2 & $-$ & $-$ & $-$ & $-$ & $-$\\
J0507+6737 & 1ES 0502+675 & 2FHL J0507.9+6737& 0.341& HSP & $0.045\pm0.007$ & $0.240\pm_{0.095}^{0.169}$ & 0.80 & 5 & 0.062  & $16.6\pm0.11$ & 16.3 & 17.0 & 18.0 & 3& $-$ & $-$\\
J0521+2112 & VER J0521+211 & 2FHL J0521.7+2112 & 0.108 & ISP & $0.076\pm_{0.007}^{0.008}$ & $<0.435$ & 0.30 & 6 & $-$ & $15.4\pm0.01$ & 13.9 &
$-$ & $-$ & $-$ & $-$ & $-$\\
J0648+1516 & RX J0648.7+1516 & 2FHL J0648.6+1516 & 0.179 & HSP & $-$
& $-$ & $-$ & 1 & $-$ & $17.25\pm0.01$ & 16.94 & $-$ & $-$ & $-$ & $-$ & $-$\\
J0650+2502 & 1ES 0647+250 & 2FHL J0650.7+2502 & 0.410\tablefootmark{f}
& HSP & $-$ & $-$ & 
$-$ & 2 & $-$ & $15.6\pm0.04^*$ & 15.4 & 15.4 & 19.9 & 4 & 157.9 & 8\\
J0710+5908 & RGB J0710+591 & 2FHL J0710.5+5908 & 0.125 & HSP & $0.043\pm0.004$ & $<0.412$ & 0.33 & 6 & 0.060 & $16.4\pm0.03$ & 16.3 &
16.7 & 17.8 & 1 & $-145.3$ & 8 \\
J0809+5218 & 1ES 0806+524 & 2FHL J0809.7+5218 & 0.138 & HSP &
$0.051\pm0.002$ & $<0.205$ & 0.25 & 4 & 0.054 & $15.1\pm0.04$ & 15.0
& 15.1 & 18.0 & 6 & $-$ & $-$\\
J1136+7009 & Markarian 180 & 2FHL J1136.5+7009 & 0.045 & HSP &
$0.034\pm_{0.003}^{0.004}$ & $<0.437$ & 0.45 & 10 & 0.043 &
$14.9\pm0.03$ & 14.9 & 15.2 & 16.6 & 1 & $-$ & $-$\\
J1217+3007 & 1ES 1215+303 & 2FHL J1217.9+3006 & 0.130 & HSP &
$0.109\pm_{0.005}^{0.004}$ & $0.168\pm_{0.030}^{0.042}$ & 0.17 & 14 & 0.112 & $14.3\pm0.05$ & 14.2 & 14.3 & 18.1 & 1 & 143.6 & 9\\
J1221+3010 & 1ES 1218+304 & 2FHL J1221.3+3009 & 0.182 & HSP & $0.047\pm_{0.006}^{0.006}$ & $0.278\pm_{0.084}^{0.143}$ & 0.68 & 9 &
0.087 & $16.5\pm0.04$ & 16.5 & 17.2 & 17.4 & 3 & $-$ & $-$\\
J1221+2813 & W Comae & $-$ & 0.102 & ISP & $0.162\pm_{0.008}^{0.007}$ & $<0.192$ &0.47& 6 & 0.188 & $15.6\pm0.06$ & 15.6
& 15.8 & 17.9 & 1 & 114.6 & 9\\
J1224+2436 & MS 1221.8+2452 & 2FHL J1224.4+2435 & 0.218 & HSP &
$0.041\pm_{0.005}^{0.006}$ & $0.365\pm_{0.088}^{0.134}$ & 2.57 &
9 & 0.043 & $16.0\pm0.06$ & 15.9 & 16.0 & 19.5 & 2 & $-$ & $-$\\
J1427+2348 & PKS 1424+240 & 2FHL J1427.0+2348 & 0.6035\tablefootmark{g} & HSP & $0.039\pm_{0.006}^{0.008}$ & $0.544\pm_{0.100}^{0.136}$ & 3.59 & 12 &
0.039 & $14.1\pm0.05$ & 14.0 & 14.1 & $>21.0$ & 5 & $-$ & $-$\\
J1428+4240 & H 1426+428 & 2FHL J1428.5+4239 & 0.129 & HSP & $0.024\pm0.002$ & $<0.409$ & 0 & 7 & 0.038 & $16.5\pm0.02$ & 16.5 & 17.0
& 17.6 & 6 & $-11.9$ & 10\\
J1442+1200 & 1ES 1440+122 & 2FHL J1442.9+1159 & 0.163 & HSP & $0.028\pm0.003$ & $0.218\pm_{0.069}^{0.106}$ & 0.54 &7 & $-$ & $16.2\pm0.02^*$ & 16.2
& 16.5 & 17.5 & 1 & $-$ & $-$\\
J1555+1111 & PG 1553+113 & 2FHL J1555.7+1111 & $>0.395$\tablefootmark{h} & HSP &
$0.033\pm_{0.003}^{0.004}$ & $0.640\pm_{0.082}^{0.100}$ & 2.91
& 32 & 0.033 & $14.0\pm0.01$ & 13.9 & 14.00& $>21.6$ & 5 & $-$ & $-$\\
J1725+1152 & H 1722+119 & 2FHL J1725.1+1154 & 0.34\tablefootmark{i} & HSP &
$0.108\pm_{0.008}^{0.009}$ & $0.297\pm_{0.050 }^{0.067}$ &
0.84 & 16 & 0.108 & $14.8\pm0.05$ & 14.5 & 14.8 & $>21.4$ & 5 & $-$ & $-$\\
J1728+5013 & 1ES 1727+502 & 2FHL J1728.2+5013 & 0.055 & HSP &
$0.022\pm_{0.002}^{0.002}$ & $0.353\pm_{0.078}^{0.108}$ & 2.40
& 17 & 0.027 & $15.7\pm0.02$ & 15.6 & 15.9 & 17.5 & 6 & $-$ & $-$\\
J1743+1935 & 1ES 1741+196 & 2FHL J1743.9+1933 & 0.084 & HSP &
$0.048\pm0.005$ & $0.212\pm_{0.060}^{0.101}$ & 0.24 & 6 & $-$ & $15.0\pm0.01^*$
& 14.8 & 16.1 & 15.4 & 1 & 70.0 & 8\\
J1943+2118 & HESS J1943+213 & 2FHL J1944.1+2117 & $>0.03$\tablefootmark{j} & HSP &
$0.022\pm_{0.003}^{0.002}$ & $<0.417$ & 0  & 6 & $-$ & $17.15\pm0.01$ & 11.18\tablefootmark{l} & $-$ & $-$ & $-$ & $-149.2$ & 11\\
J1959+6508 & 1ES 1959+650 & 2FHL J2000.1+6508 & 0.048 & HSP & $0.058\pm0.002$ & $<0.160$ & 0.51 & 13 & 0.067 & $14.9\pm0.06$ & 14.5 &
15.0 & 17.0 & 5 & 152.7 & 10\\
J2001+4352 & MAGIC J2001+435 & 2FHL J2001.2+4352 & 0.190\tablefootmark{k} & ISP &
$0.167\pm0.008$ & $0.146\pm_{0.028}^{0.041}$ & 0.58 & 12 & 0.190 &
$16.8\pm0.04$ & 15.5 & 16.9 & 19.2 & 7 & $-$ & $-$\\
J2250+3825 & B3 2247+381 & 2FHL J2249.9+3826 & 0.119 & HSP & $0.020\pm0.002$ & $0.249\pm_{0.088}^{0.125}$ & 1.73 & 10 & 0.025 & $16.4\pm0.04$ & 16.1 & 16.6 & 18.2 & 1 & $-$ & $-$\\
J2347+5142 & 1ES 2344+514 & 2FHL J2347.1+5142 & 0.044 & HSP &
$0.026\pm0.004$ & $0.286\pm_{0.095}^{0.167}$ & 0 &
5 & 0.080 & $15.8\pm0.03$ & 15.3 & 17.0 & 16.2 & 6 & $-$ & $-$\\
\hline                  
\end{tabular}
\tablefoottext{a}{Taken from the 3rd {\it Fermi} AGN catalogue
  \citep{3LAC} unless otherwise stated.}
\tablefoottext{b}{Mean from RoboPol observations, except for sources
  with an asterisk for which this is from Tuorla monitoring programme. The quoted
  uncertainty is the standard error of the mean, unless N obs is one, in which case the uncertainty is the
  rms error of the measurement.}
\tablefoottext{c}{Host galaxy magnitude used in this work, converted
  to R-band magnitude where needed and with correct aperture size used
  in the estimation (3\arcsec for sources where RoboPol magnitudes
  were used and 1.5\arcsec for J2347+5142 where the NOT magnitude was used).}
\tablefoottext{d}{Lower limit based on non-detection of the host galaxy
\citep{nilsson12}.}
\tablefoottext{e}{Lower limit based on $Ly\alpha$ \citep{furniss13a}.}
\tablefoottext{f}{Based on a host galaxy detection
\citep{kotilainen11}.}
\tablefoottext{g}{Based on a lower limit on $Ly\alpha$
\citep{furniss13b}, and a detection of a group of galaxies at the same
redshift \citep{rovero16}.}
\tablefoottext{h}{Lower limit based on a confirmed  $Ly\alpha + O_{iv}$
absorber \citep{danforth10}.}
\tablefoottext{i}{Lower limit based on a featureless spectrum \citep{landoni14}.}
\tablefoottext{j}{Lower limit based on a host galaxy association
  \citep{peter14}}
\tablefoottext{k}{Based on a host galaxy detection
  \citep{aleksic14_J2001}.}
\tablefoottext{l}{Unreliable due to low Galactic latitude.}
\tablebib{(1)~\citet{nilsson03}; (2)~\citet{scarpa00};
  (3)~\citet{shaw13}; (4)~\citet{meisner10};
  (5)~\citet{urry00}; (6)~\citet{nilsson07};
  (7)~\citet{aleksic14_J2001}; (8)~\citet{piner14};
  (9)~\citet{lister13}; (10)~\citet{piner10}; (11)~\citet{akiyama16}} 

\end{sidewaystable*}

\begin{sidewaystable*}
\scriptsize
\caption{ Non-TeV sample sources and their variability properties.}             
\label{controltable}      
\centering          
\begin{tabular}{l l l l l l l l l l l l l l}     % 7 columns 
\hline\hline       
Name & Other name & 2FHL name & Redshift\tablefootmark{a} & SED & Intrinsic mean
& Intrinsic & median & N & 
host-corr & mean & de-red. & host-corr & host gal.\\ 
 & & & & class\tablefootmark{a} & p fraction & mod. index &
 |d$\theta$/d$t$| & obs & p fraction& R-mag.\tablefootmark{b} &
 R-mag. & R-mag. & R-mag.\tablefootmark{c} \\
\hline                    
J0114+1325 & GB6J0114+1325 & $-$ & $>0.26$\tablefootmark{d} & HSP & 0.072 $\pm$ $_{0.013}^{0.017}$ & 0.497
$\pm$ $_{0.113}^{0.166}$ & 2.43 & 8 & $-$ & $16.3\pm0.10$ & 16.2 &
$-$ & $-$ \\
J0848+6606 & GB6J0848+6605 & $-$ & $-$ & ISP & 0.063 $\pm$ $_{0.012}^{0.013}$ &
$<0.820$ & 0.73 & 4 & $-$ & 17.8 $\pm$ 0.08 & 17.6 & $-$ & $-$ \\
J1037+5711 & GB6J1037+5711 & 2FHL J1037.6+5710 & $>0.175$\tablefootmark{d} & ISP & 0.023 $\pm$ 0.005 & 0.499
$\pm$ $_{0.168}^{0.262}$ & 4.25 & 11 & $-$ & $15.6\pm0.05$ & 15.6 & $-$ & $-$ \\
J1203+6031 & SBS1200+608 & $-$ & 0.065 & ISP & $-$ & $-$ & $-$ & 9 & $-$ & $15.7\pm0.04$ &
15.7 & $-$ & $-$ \\
J1542+6129 & GB6J1542+6129 & 2FHL J1542.9+6129 & $-$ & ISP & 0.049 $\pm$ $_{0.004}^{0.005}$ &
0.401 $\pm$ $_{0.075}^{0.098}$ & 1.93 & 18 & $-$ & $15.3\pm0.02$ &
15.3 & $-$ & $-$ \\
J1558+5625 & TXS1557+565 & $-$ & 0.300 & ISP & 0.067 $\pm$ 0.008 &  0.447 $\pm$
$_{0.086}^{0.114}$ & 2.74 & 19 & $-$ & $17.3\pm0.05$ & 17.2 & $-$ & $-$ \\
J1649+5235 & 87GB164812.2+524023 & $-$ & $-$ & ISP & 0.028 $\pm$ $_{0.002}^{0.003}$ &
$<0.500$ & 0.27 & 16 & $-$ & $16.4\pm0.09$ & 16.3 & $-$ & $-$ \\
J1754+3212 & RXJ1754.1+3212 & 2FHL J1754.2+3211& $>0.54$\tablefootmark{e} & ISP &  0.047 $\pm$
$_{0.006}^{0.007}$ &  0.448 $\pm$ $_{0.103}^{0.142}$ & 0.24 & 13 & $-$ &
$16.7\pm0.07$ & 16.6 & $-$ & $-$ \\
J1809+2041 & RXJ1809.3+2041 & $-$ & $>0.28$\tablefootmark{f} & HSP & 0.038 $\pm$ $_{0.006}^{0.005}$ &
$<0.663$ & 1.42 & 12 & $-$ & $18.0\pm0.12$ & 17.8 & $-$ & $-$ \\
J1813+3144 & B21811+31 & $-$ & 0.117 & ISP & 0.058 $\pm$ $_{0.012}^{0.015}$ &
0.545 $\pm$ $_{0.129}^{0.195}$ & 0.61 & 8 & $0.066$ & $16.3\pm0.06$
& 16.2 & 16.5 & 18.6 \\
J1836+3136 & RXJ1836.2+3136 & $-$ & $-$ & ISP & 0.069 $\pm$ $_{0.011}^{0.013}$ & 0.472
$\pm$ $_{0.106}^{0.152}$ & 1.09 & 10 & $-$ & $17.5\pm0.05$ & 17.2 & $-$ & $-$ \\
J1838+4802 & GB6J1838+4802 & 2FHL J1838.9+4802 & 0.3\tablefootmark{g} & HSP & 0.093 $\pm$ 0.006 & 0.212 $\pm$
$_{0.043}^{0.062}$ & 0.41 & 11 & 0.095 & $15.2\pm0.06$ & 15.1 &
15.2 & 19.4\\
J1841+3218 & RXJ1841.7+3218 & $-$ & $-$ & HSP & $-$ & $-$ & $-$ & 8 & $-$ & $18.1\pm0.06$ & 17.9 & $-$ & $-$\\
J1903+5540 & TXS1902+556 & 2FHL J1903.2+5540 & $>0.727$\tablefootmark{f} & ISP & 0.08 $\pm$
$_{0.010}^{0.011}$ & 0.341 $\pm$ $_{0.075}^{0.110}$ & 1.31 & 9 & $-$ &
$15.7\pm0.07$ & 15.6 & $-$ & $-$\\
J2015$-$0137 & PKS2012-017 & $-$ & $-$ & ISP & 0.164 $\pm$ 0.006 & $<0.184$ & 0.27 &
10 & $-$ & $17.5\pm0.04$ & 17.2 & $-$ & $-$\\
J2022+7611 & S52023+760 & $-$ & 0.594 & ISP & 0.113 $\pm$ $_{0.011}^{0.012}$ &
0.302 $\pm$ $_{0.065}^{0.089}$ & 3.96 & 11 & $-$ & $16.5\pm0.16$ &
16.0 & $-$ & $-$\\
J2131$-$0915 & RBS1752 & 2FHL J2131.4-0914 & 0.449 & HSP &  0.103 $\pm$
$_{0.004}^{0.005}$ & $<0.168$ & 0 & 6 & $-$ & $17.0\pm0.02$ & 17.0
& $-$ & $-$\\
J2149+0322 & PKSB2147+031 & $-$ & $>0.724$\tablefootmark{f} & ISP &  0.131 $\pm$ $_{0.017}^{0.020}$ & 0.287
$\pm$ $_{0.088}^{0.131}$ & 0.24 & 7 & $-$ & $16.2\pm0.15$ & 16.0 & $-$ & $-$\\
J2340+8015 & 1RXSJ234051.4+801513 & 2FHL J2340.8+8014 & 0.274 & HSP & 0.041 $\pm$
$_{0.010}^{0.012}$ & 0.600 $\pm$ $_{0.236}^{0.380}$ & 0.28 & 11 & $-$
& $17.0\pm0.13$ & 16.5 & $-$ & $-$\\
\hline                  
\end{tabular}
\tablefoottext{a}{Taken from the 3rd {\it Fermi} AGN catalogue
  \citep{3LAC} unless otherwise stated.}
\tablefoottext{b}{Mean from RoboPol observations. The quoted
  uncertainty is the standard error of the mean.
\tablefoottext{c}{Host galaxy magnitude used in this work, using the
  correct aperture size of 5.2\arcsec for RoboPol observations. The original R-band
  values are taken from \citet{nilsson03}.}
\tablefoottext{d}{Lower limit based on non-detection of the host
  galaxy \citep{plotkin10}.}
\tablefoottext{e}{Lower limit based on non-detection of the host
  galaxy \citep{shaw13b}.}
\tablefoottext{f}{Lower limit based on intervening absorbers in the
  spectrum\citep{shaw13}.}
\tablefoottext{g}{Based on a host galaxy detection \citep{nilsson03}.}
  }
\end{sidewaystable*}

\section{Discussion}\label{discussion}
Our aim was to study the differences in the TeV-detected ISP and HSP
BL~Lac objects compared to non-TeV-detected objects. In this study we have compared the optical polarization
properties of a TeV-detected sample of 29 sources with a 
sample of 19 non-TeV objects. Our maximum likelihood analysis shows
that there are no differences in the mean polarization fraction in the
two samples, indicating that optical polarization
variability and the TeV emission are not directly related. 
  Instead, their redshift distributions, SED classifications and
  optical brightness are seen to differ significantly, which most
  likely explains why some of them are TeV detected while others are not.
In the following sections we will compare our results to earlier studies and analyze
the polarization angle behavior in more detail.

\subsection{Fraction of polarized sources and the duty cycle of high polarization}
Optical polarization of X-ray--selected BL~Lac objects was 
studied by \cite{jannuzi94} who examined 3 years of optical
polarization monitoring data. They detected significant polarization
in 28 out of 37 sources, out of which 19 showed significant
variability. We detected polarization at a level of signal-to-noise
greater than three in all but one TeV (RBS~0413)  and  non-TeV
(SBS~1200+608) source showing that our detection fraction is
higher (46 out of 48 sources). We detect significant variability in 17
TeV and 12  non-TeV sources, which makes the 
fraction of variable sources very similar to \cite{jannuzi94}.

\begin{figure}
   \centering
   \includegraphics[width=9cm]{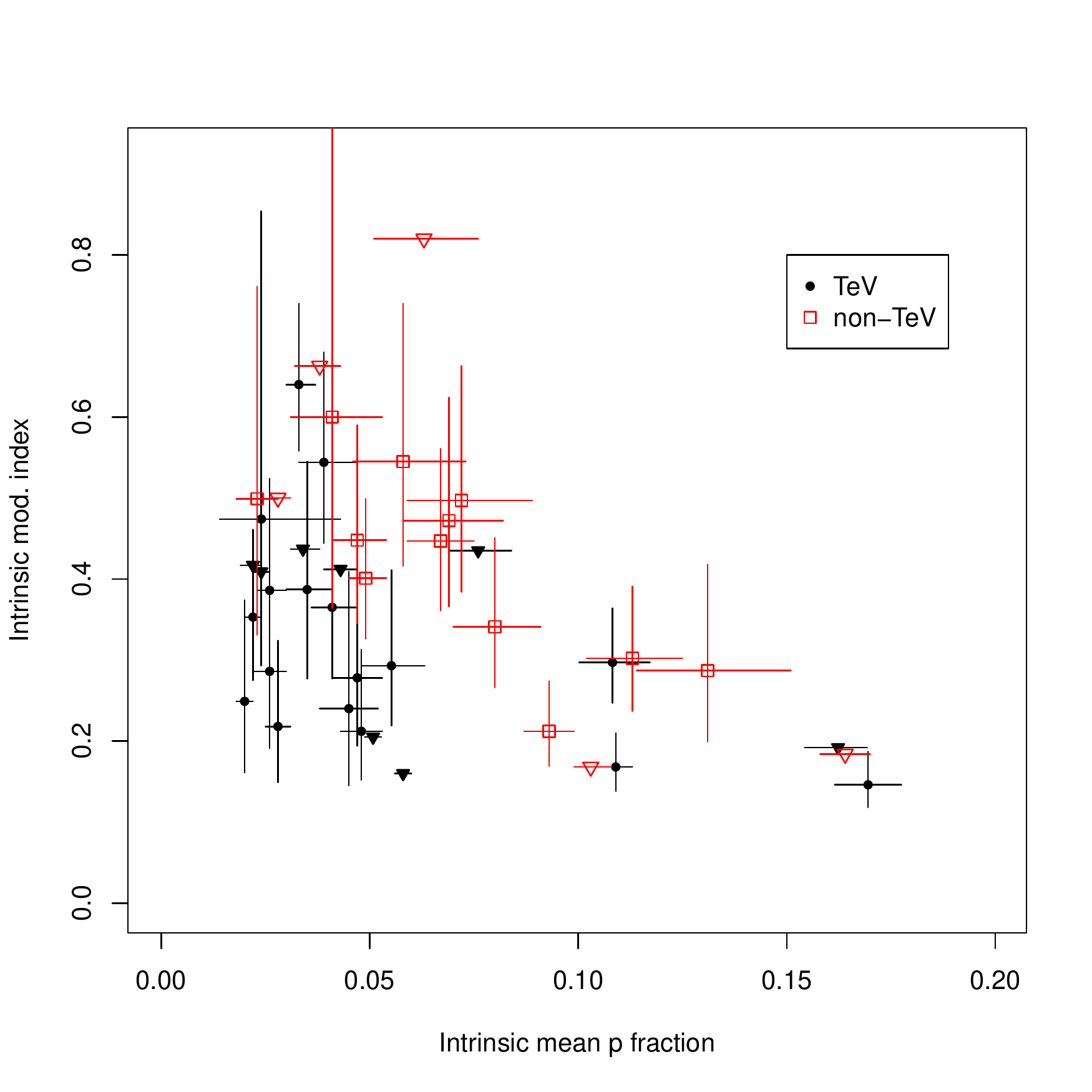}
   \caption{Intrinsic modulation index against the intrinsic mean
     polarization fraction for the TeV (black circles) and  non-TeV (red
     open squares) sources. Lower limits in modulation index are shown
   with downward triangles.}
             \label{m_vs_p}%
     \end{figure}

In Fig.~\ref{m_vs_p} we show the intrinsic modulation index against
the polarization fraction. There is a trend for sources with higher
mean polarization fraction to show smaller polarization variability
amplitudes. This could indicate that the sources with higher polarization
fractions had a more ordered dominating polarization
component. We note that a similar trend is seen when the full RoboPol
sample is examined \citep{angelakis16}.

\cite{jannuzi94} also find that the duty cycle for the fraction of
time these sources are highly polarized at $>4\%$ is 44\%. We
calculate the duty cycle in our sample in the same way as they do by taking the
first observation of each source and calculating the fraction of
sources that have polarization fraction $>4\%$. In the calculation of
the duty cycle, we account for the Ricean bias and de-bias our
polarization fraction observations similarly as in \cite{pavlidou14}. 
Accounting for the uncertainties in the measurements, we find the duty cycle in our TeV
objects to be $59^{+6.9}_{-15.1}\%$ and in the  non-TeV sources
$74^{+15.8}_{-21.1}\%$, which are consistent with each other within
uncertainties. They are higher than obtained by \cite{jannuzi94} but
similar to \cite{heidt11} who found a duty cycle of $66\%$ for HBL
BL~Lacs. 

\subsection{Host galaxy dilution}
\cite{heidt11} suggested that one reason why they obtained a higher
duty cycle than \cite{jannuzi94} is due to the larger aperture size
used by \cite{jannuzi94}, which would result in a larger fraction
of the host galaxy contaminating the polarization
results. \cite{heidt11} also found that sources with known redshifts are
less polarized than sources with unknown redshifts, and a trend
with sources at redshift of $\gtrsim 1$ to be more highly
polarized. They suggested this is also due to host galaxy dilution of
the polarization fraction as in lower redshift sources, for which the
redshift is also easier to determine, the host galaxy contribution
within the aperture is larger.

We examined this directly by collecting from the literature all the
available host galaxy magnitudes, and correcting our polarization
fraction by removing the contribution of the unpolarized host. In
Sect.~\ref{host}, we showed 
that this increases the polarization fraction in the TeV sources, and
reduces the difference between the TeV and  non-TeV sources. As only
13 of our 19  non-TeV sources have a redshift estimate or limit available,
this may also indicate that the redshifts of the remaining  non-TeV
sources are higher than in our TeV sources, in agreement with the
findings of \cite{heidt11}.

\subsection{Preferred polarization angle}
\cite{angel78} claimed that at least some BL~Lac objects have a
preferred polarization angle over several years of observations.
\cite{jannuzi94} found that 11 out of 13 sources in their well-studied
sample with a time span of at least 20 months between the first and
last observations have preferred polarization angles. 
In Sect.~\ref{evpa} we found that the distance of the mass center from
the origin was similar in the two samples, while
the scatter in the $Q/I-U/I$--plane
was significantly smaller for the TeV-detected than the  non-TeV
samples. If we look at the fraction of sources for which the scatter
is smaller than a third of the distance from the origin, i.e. sources
far away from the origin with small scatter (an indication of a
preferred polarization angle), the fraction of TeV sources is much
higher (11/26) than in the  non-TeV sources (3/17). 

\begin{figure}
   \centering
   \includegraphics[width=9cm]{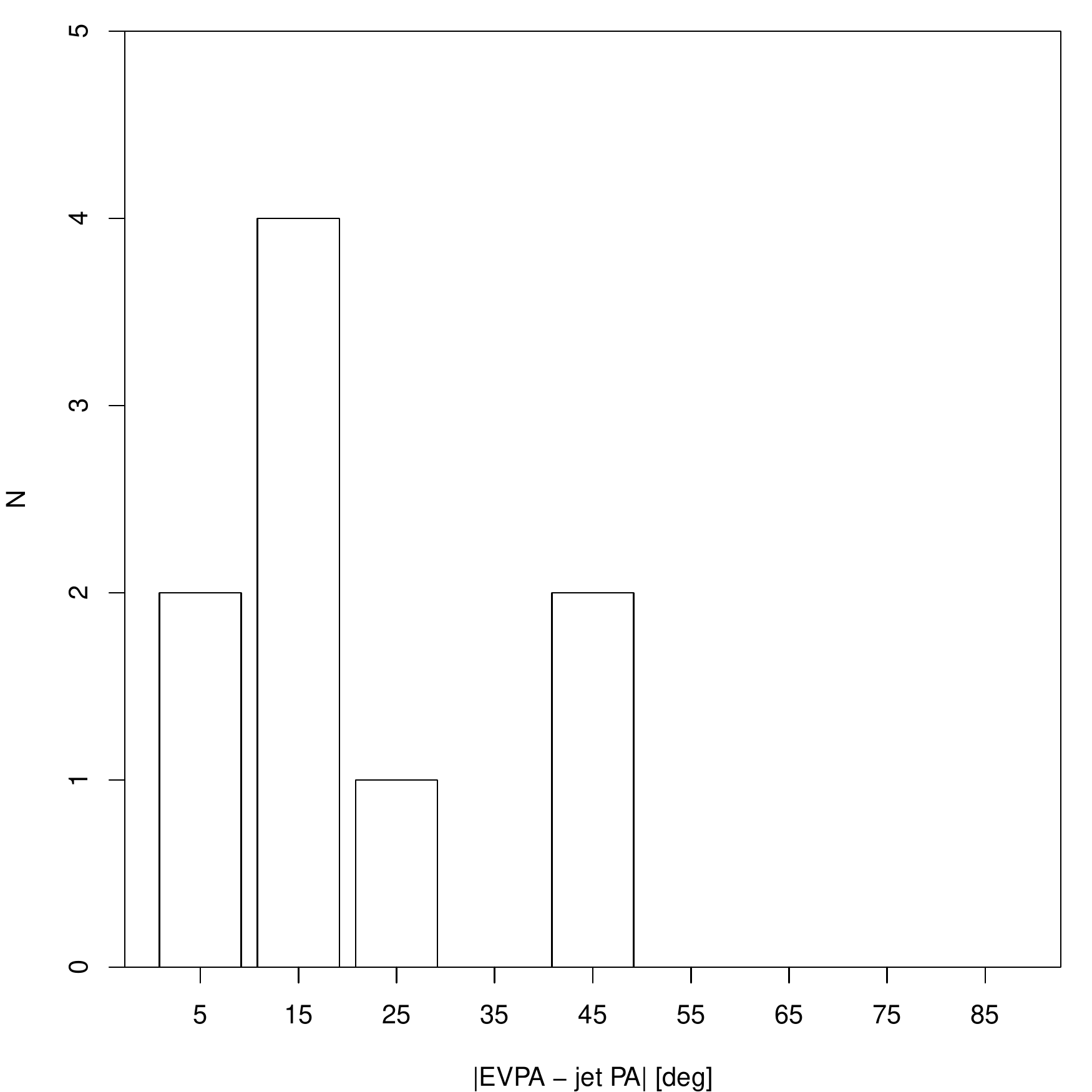}
   \caption{Stacked histogram of the difference between the  mean
     optical EVPA and inner jet position angle from VLBI observations for the TeV
     sources showing indications of a preferred polarization angle.}

             \label{jetpa}%
     \end{figure}

Assuming that these  14 sources do have a preferred polarization
angle, we can try to estimate the direction of the magnetic field relative to
the jet direction by comparing the  mean EVPA to the jet position angle. We
do this by collecting from the literature the jet position angles of
the innermost jet components obtained through Very Long Baseline
Interferometry (VLBI).   These are available for nine of the TeV
  sources and we tabulate them in
Table~\ref{tevtable}. We show the difference of the  mean EVPA and the jet
position angle in Fig.~\ref{jetpa}. As the optical emission is
optically thin, the projected magnetic field is perpendicular to the
observed EVPA
direction so that a small difference between EVPA and jet
position angle corresponds to magnetic field perpendicular to the jet
direction. We note that relativistic effects may alter the
appearance of this distribution, as discussed in \cite{lyutikov05}, so
the situation may not be as straightforward. 

 Figure~\ref{jetpa} shows that 67\% of the sources
for which both the EVPA and jet position angle are available (six out of
nine sources) show a difference less than 20 degrees, indicating that
the magnetic field is perpendicular to the jet direction.  A K-S test
gives a $p=0.0003$ for the sample to come from a uniform distribution.
The distribution
looks similar to a comparison between optical EVPA and the inner-jet
position angle at 43\,GHz for a sample of highly polarized quasars
\citep{lister00} and a sample of BL~Lac objects \citep{jorstad07}. 
We note a caveat that many of the jet position angle
observations are taken at fairly low radio frequencies (8 or 15\,GHz),
and therefore the position angle may not be representative of the jet
position angle in the optical band, as some blazars are known to show significant
curvature in the inner jets \citep[e.g.,][]{savolainen06}, although
for BL~Lac objects the alignment at least from 43\,GHz to optical seems to be better than in quasars
\citep{jorstad07}. Our analysis also relies on the assumption that the mean EVPA represents a
stable EVPA of the jet, which may not be the case considering the
fairly short time span of our observations. As discussed in e.g., 
\cite{villforth10} and \cite{sakimoto13}, it is also possible that the same sources
occasionally show a preferred polarization angle while at another time
they may not, so clearly long-term polarization observations are required to better
understand this in the HSP objects.

\subsection{Scatter in the $Q/I-U/I$-plane}

\begin{figure}
   \centering
   \includegraphics[width=4.5cm]{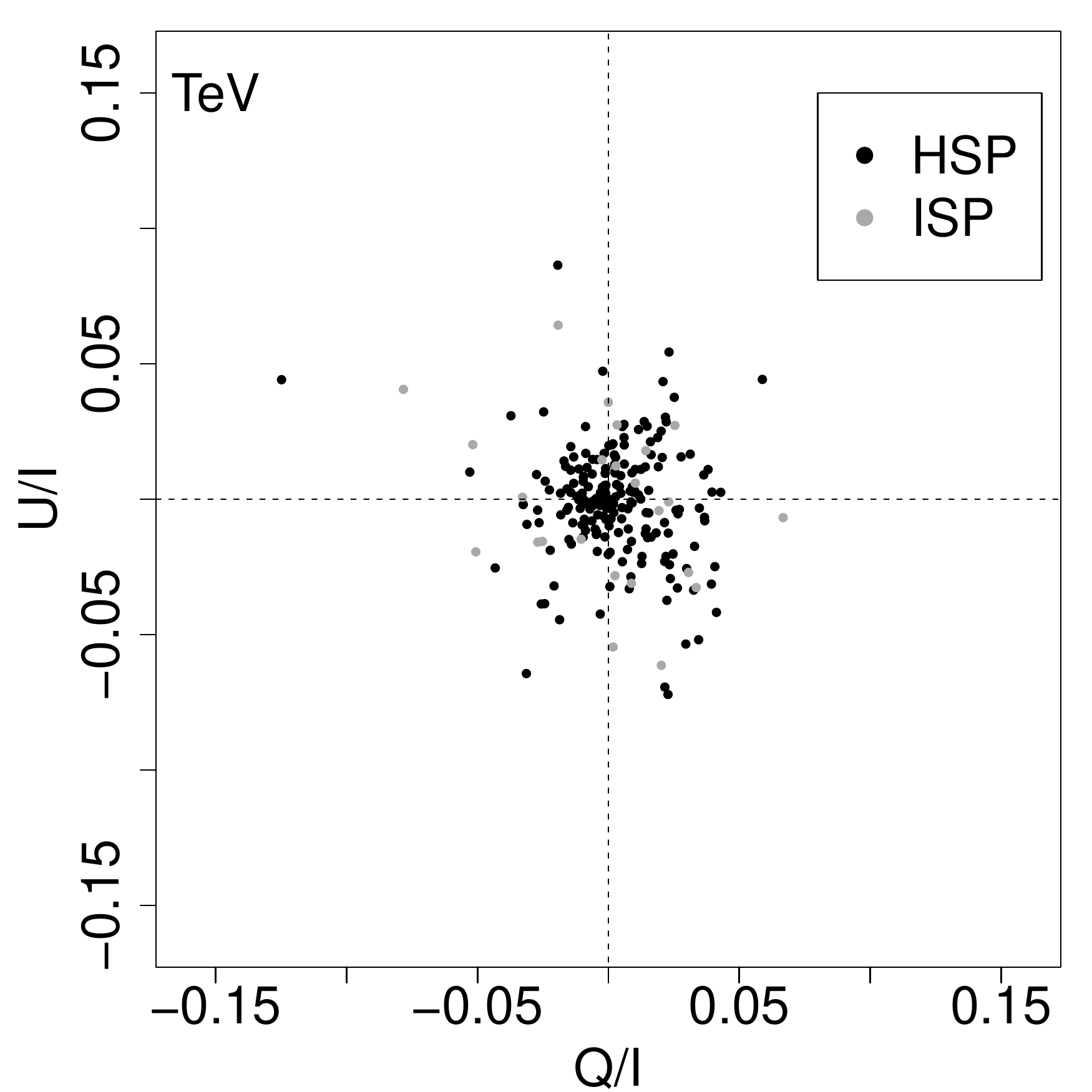}\includegraphics[width=4.5cm]{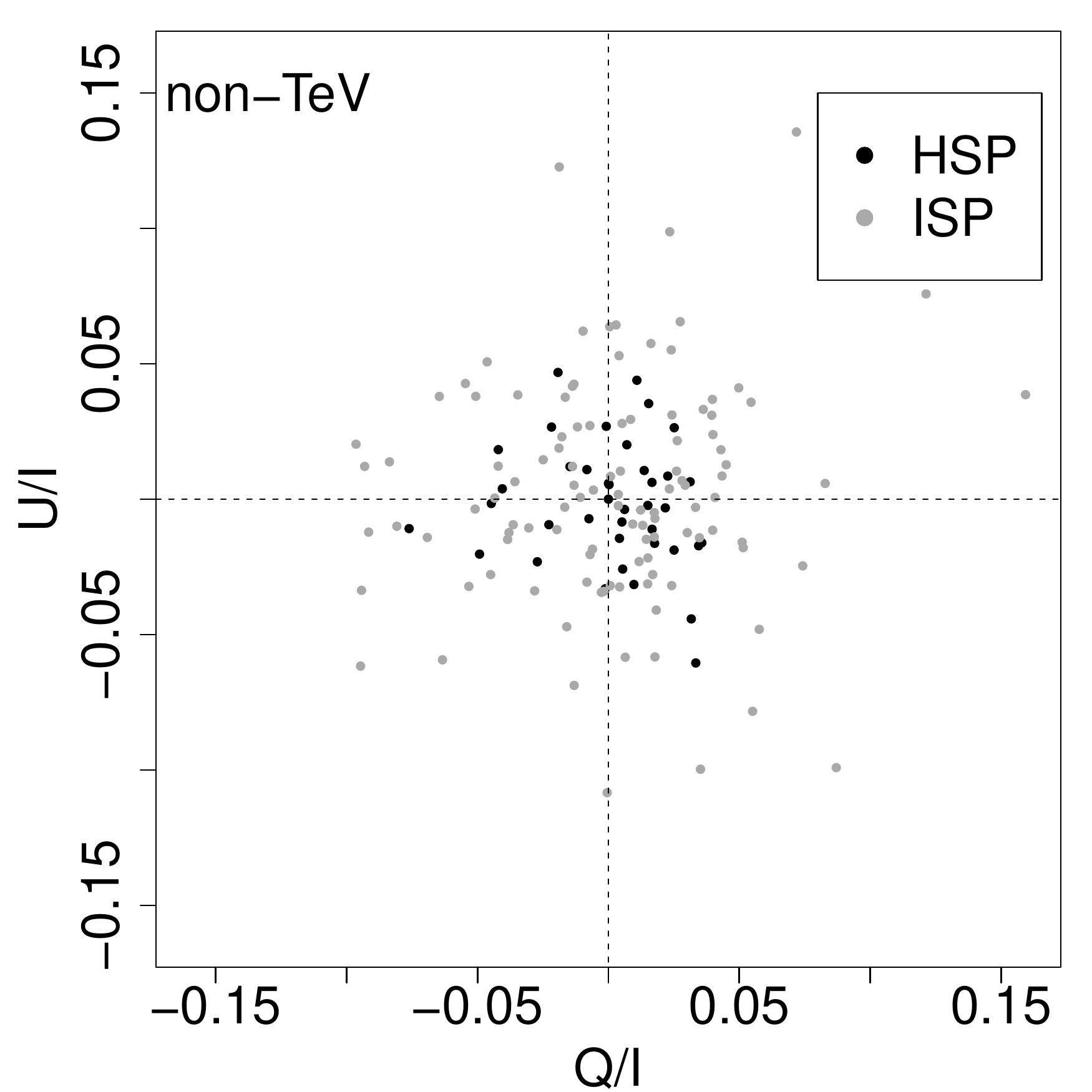}
\includegraphics[width=4.5cm]{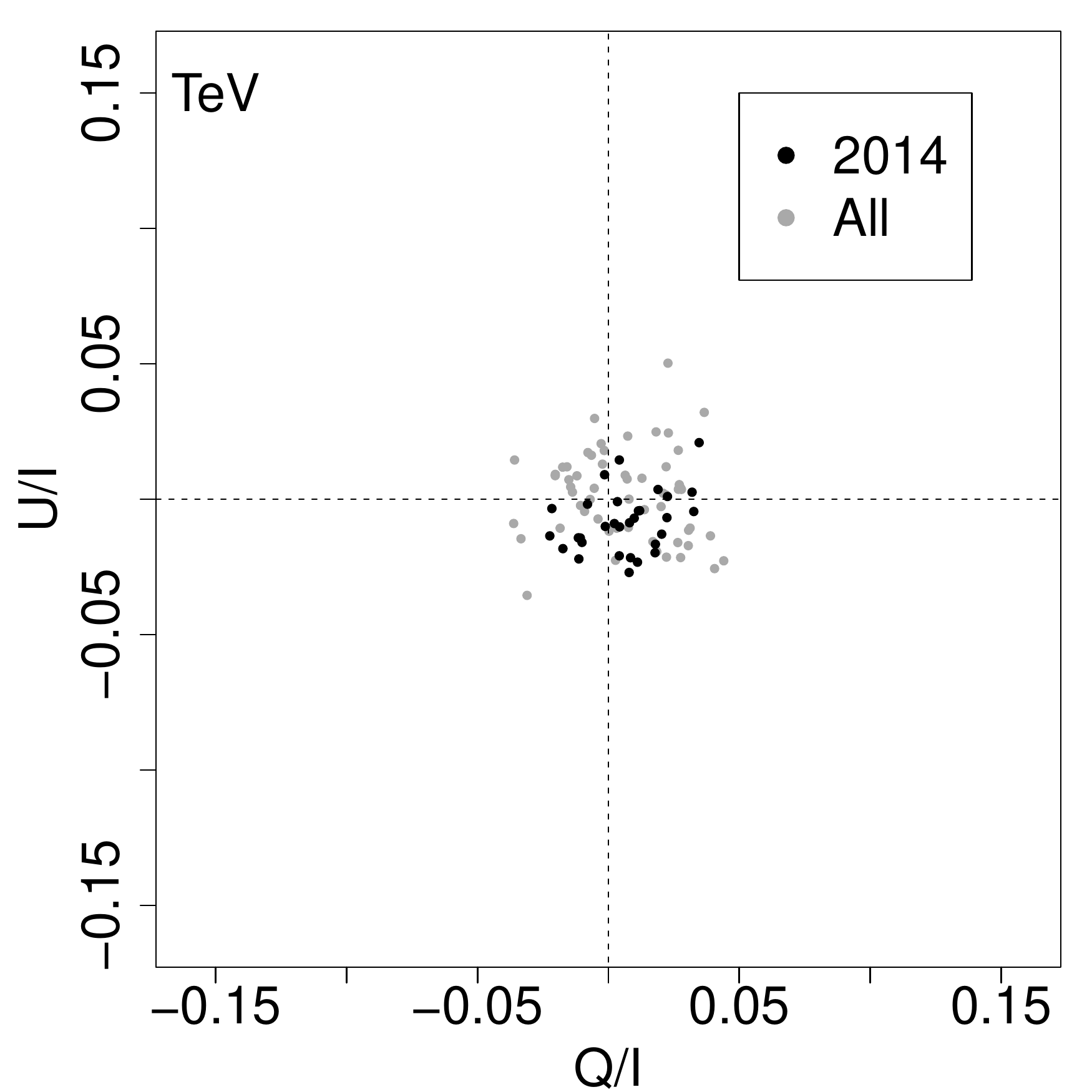}\includegraphics[width=4.5cm]{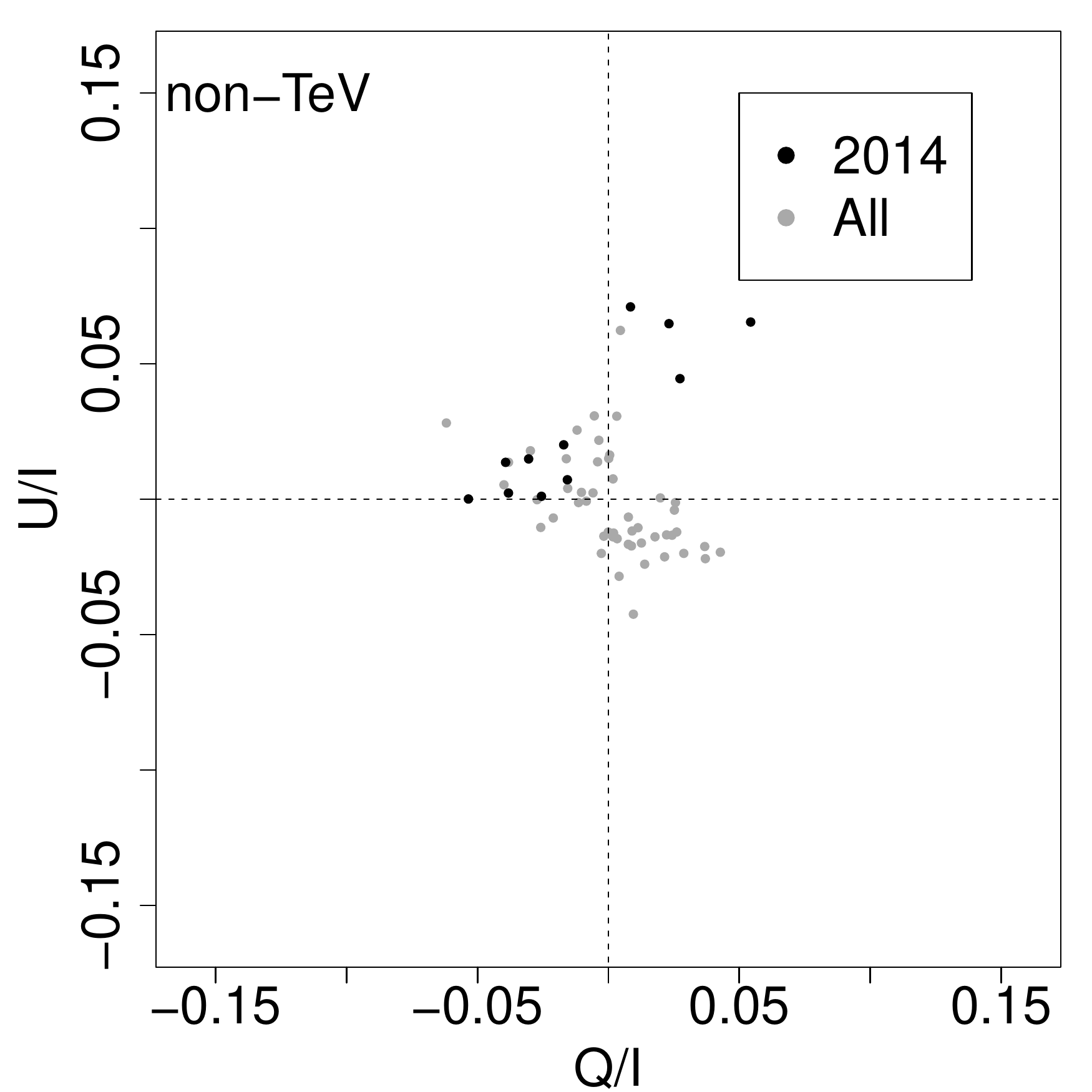}
   \caption{Top: Stacked Q/I vs. U/I plot for the TeV (left) and  non-TeV
     (right) samples. Bottom: Same for three TeV sources (left) and
     two  non-TeV sources (right) for which we have data over multiple
     seasons. The stacking is done by shifting the mass center of each
     source to the origin. In the bottom panel we have stacked the data based on
     the mass center of all data, which is why the top and bottom
     panels are not exactly the same for the 2014 data.}
             \label{QUstacked}%
    \end{figure}

In order to investigate the difference in the scatter between the two samples, in Fig.~\ref{QUstacked} we show stacked plots for all the TeV
(left) and  non-TeV (right) sources by shifting the mass center of the individual
sources to the origin. While the scatter is larger in the  non-TeV than in
the TeV sources, this seems to be a difference between HSP and ISP
type objects rather than TeV and  non-TeV sources.  As
 discussed in Sect.~\ref{SEDs}, the  non-TeV sample
contains a  much larger fraction of ISP objects than the TeV sample, and the
HSP sources in the  non-TeV sample have a similar range of scatter in
the $Q/I-U/I$-plane as the TeV sources. This agrees with our results
in Sect.~\ref{modindex} , where we found that the  non-TeV sample sources have higher polarization
fraction variability amplitudes, and Sect.~\ref{evpa}  where the rate
of EVPA change was found to be larger in the  non-TeV sample, but very similar to the TeV
sources if we only consider the HSP-type  non-TeV sources. Also, when
the full RoboPol sample is examined, a trend with higher synchrotron
peak sources having a more preferred EVPA distribution is seen \citep{angelakis16}.

There could be several causes for this. Because we are observing the
sources over a fixed band, we probe a different part of the SED in the
ISP and HSP sources. This results in larger total intensity
variability in the ISPs than in HSPs \citep[e.g.,][]{hovatta14a}
because the optical emission in ISPs is produced by
electrons with energies above the synchrotron break frequency while in 
HSPs the optical emission is produced by electrons with energies less than
the break frequency. Thus, any new emission component changes the
total intensity by a larger fraction in the ISPs, which could be
reflected in the polarization fraction observations, if the polarized
flux does not change at a same rate. In this case, we would expect to
see more scatter in the HSP sources when observed in X-ray bands, a
good test case for the future X-ray polarization missions. This effect
is discussed more in \cite{angelakis16} where the
polarization amplitude variability in the full RoboPol sample
including LSP, ISP and HSP sources is analyzed.

Another alternative could be lower optical Doppler beaming in the HSPs compared to ISPs as might be
expected based on radio observations of these objects
\citep[e.g.,][]{lister11}. If the Doppler factor in the HSP sources is
lower, it takes a longer time to probe the same range of
variability as in the ISP objects. Because we have only used one
season of data for these objects, it is possible that the true spread
in the $Q/I-U/I$--plane in the HSPs is larger, if we monitor them
longer. Some of the TeV sources in our sample are also in the main sample of
the RoboPol programme, and we can investigate whether inclusion of more
data changes the picture. We select three of the HSP TeV sources
(RGB\,J0710+591, Markarian\,180, and 1ES\,1959+650) with least
amount of scatter (mean distance from the mass center $<0.2$) that also
have data in 2013 (the first two sources) and in 2015 (the last
one). Similarly, we select two HSP  non-TeV sources (RBS~1752 and
1RXSJ~234051.4+801513) for which we have data from 2013--2015. 
In Fig.~\ref{QUstacked} lower panel we show
the $Q/I-U/I$ points for these sources with the 2014 points marked in
black symbols and the data from all seasons shown in gray symbols. 
We can see that the scatter increases when more data from the other seasons
are added, showing that longer monitoring time is required to draw
strong conclusions about the scatter.

\subsection{Rotations in the polarization plane}
It is clear that not all the HSP sources have a preferred angle in our
analysis and one reason for this could be rotations in polarization
plane. Even though EVPA rotations have been observed many decades ago
\citep[e.g.,][]{kikuchi88}, for a long time it was unclear whether these
are seen in all types of objects, and especially in the HSPs. During
the first observing season of RoboPol in 2013, we detected a 128 degree EVPA
rotation in the HSP source PG~1553+113 \citep{blinov15}. The same
source was seen to rotate during 2014 as well by about 145 degrees, as reported in
\cite{blinov16}, confirming the TeV HSPs as a class of objects with EVPA
rotations (see also \citealt{jermak16}).  

In this paper, following \cite{kiehlmann16} we define an EVPA
rotation as a period,  in which the EVPA continuously rotates in one
direction. Insignificant counter-rotations with 
\begin{equation}
\left| \theta_i - \theta_j \right| < 3 \sqrt{\sigma^2_i + \sigma^2_j},
\end{equation}
where $\theta_i$ and $\theta_j$ are the first and last data point of
the counter-rotation and $\sigma^2_i$, $\sigma^2_j$ the corresponding
uncertainties, are not considered to break a continuous rotation.
Additionally, we consider only smooth rotations, where each pair of
adjacent derivatives does not change by more 10 degrees per day. We
further require that the rotation consists of at least four
observations and that the corresponding polarization fraction has a
signal-to-noise ratio of at least three.

We find significant rotations in three of the HSP TeV sources
(RGB~J0136+391, PG~1553+113, and 1ES~1727+502), and nine rotations in
six  non-TeV sources (GB6\,J1037+5711,GB6\, J1542+6129, TXS\,1557+565,
87GB\,164812.2+524023, RXJ\,1809.3+2041, S5\,2023+760). These are shown as shaded regions
over the EVPA curves in Appendix A. The rotations in PG~1553+113,
GB6\,J1037+5711, and S5\,2023+760 were already reported in
\cite{blinov16}. This shows that EVPA rotations can occur also in ISP
and HSP
sources if they are observed at high enough cadence. We would not have
detected the rotations in RGB~J0136+391 and 1ES~1727+502 if we did not
have both RoboPol and NOT observations of them. The fraction of
rotating sources is much higher in the  non-TeV sample, although we
note that only one of them is an HSP type source (RXJ\,1809.3+2041), so this
could simply reflect the differences in the EVPA variations of the ISP
and HSP sources in the optical band. The differences in the number of
rotations in LSP, ISP and HSP sources in the RoboPol sample are
studied in \cite{blinov16b}

In order to investigate the physical nature behind these rotations, we
ran a set of simulations. We used the {\it simple $Q,U$ random walk process}
described in \cite{kiehlmann16}. Our jet consists of
$N_\mathrm{cells}$, each with a uniform magnetic field at a random
orientation. During each step of the simulation, we let
$n_\mathrm{var}$ cells change their magnetic field orientation. 
For details of the simulation steps, see \cite{kiehlmann16}. 

\begin{figure}
   \centering
   \includegraphics[width=9cm]{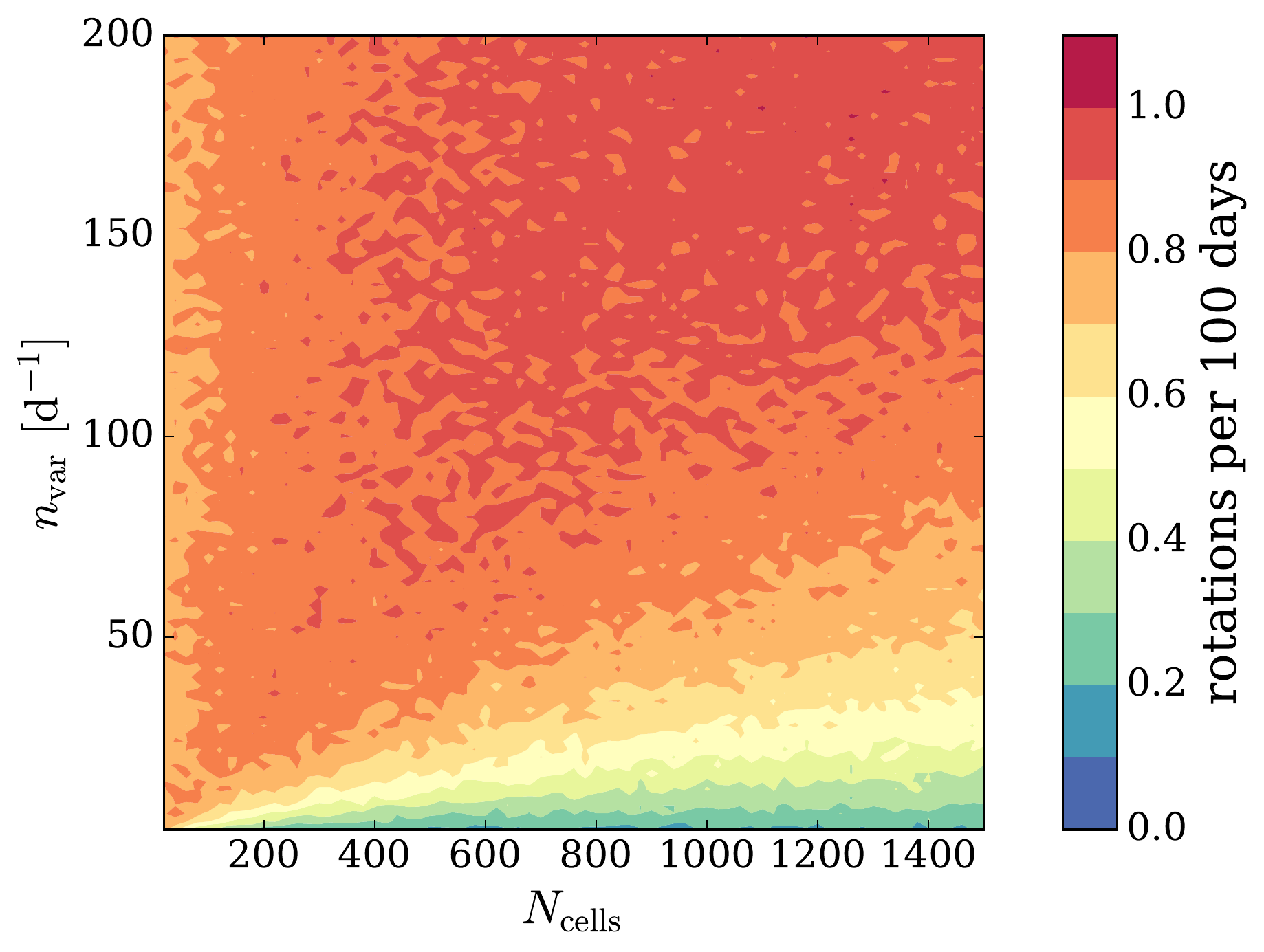}
   \caption{Frequency of rotations per 100 days for various
     combinations of the number of cells $N_\mathrm{cells}$ and number
     of them $n_\mathrm{var}$ that change during each step.}
             \label{fig:sim}%
     \end{figure}

In our simulations, we probe the range $N_\mathrm{cells} = [20, 40,
... , 1500]$ and $n_\mathrm{var} = [2, 4, ..., 200]$ and run 500
simulations for each combination. The time sampling of the light
curve, length of the season, and the $q$ and $u$ uncertainties are
taken from the cumulative distribution function of the real
observations. The rotations in the simulations are then identified in
the same manner as for the real data. This allows us to calculate the
frequency of rotations per 100 days for each parameter set, and the
result is shown in Fig.~\ref{fig:sim}. We can see that, as expected, the frequency
of rotations increases when a larger number of cells change per day,
and also that less total number of cells produces more frequently
rotations.

\begin{figure*}
   \centering
   \includegraphics[width=18cm]{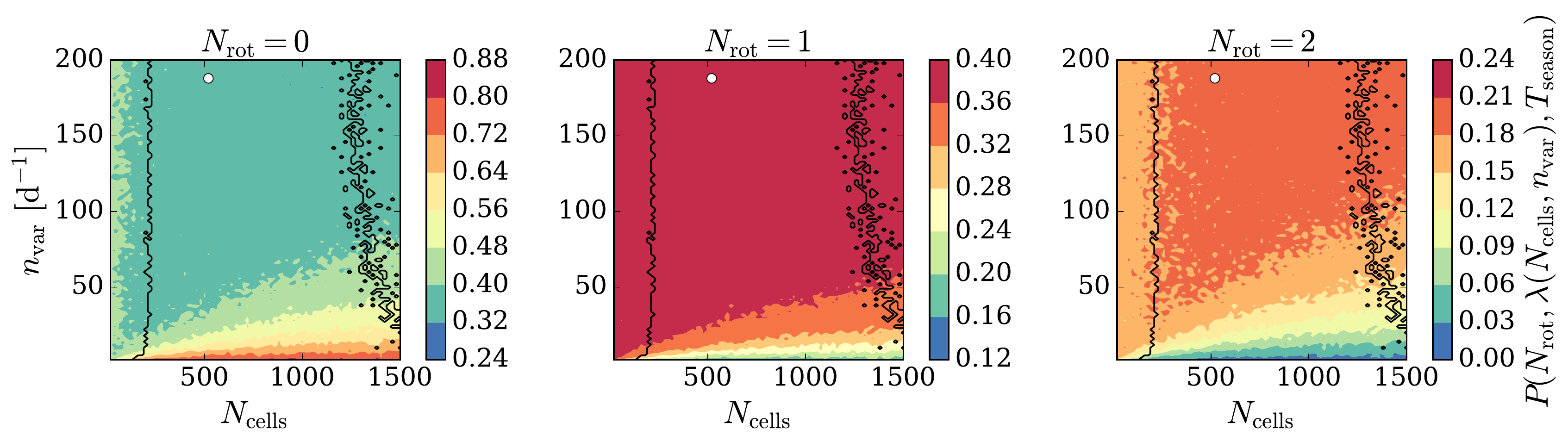}
   \includegraphics[width=18cm]{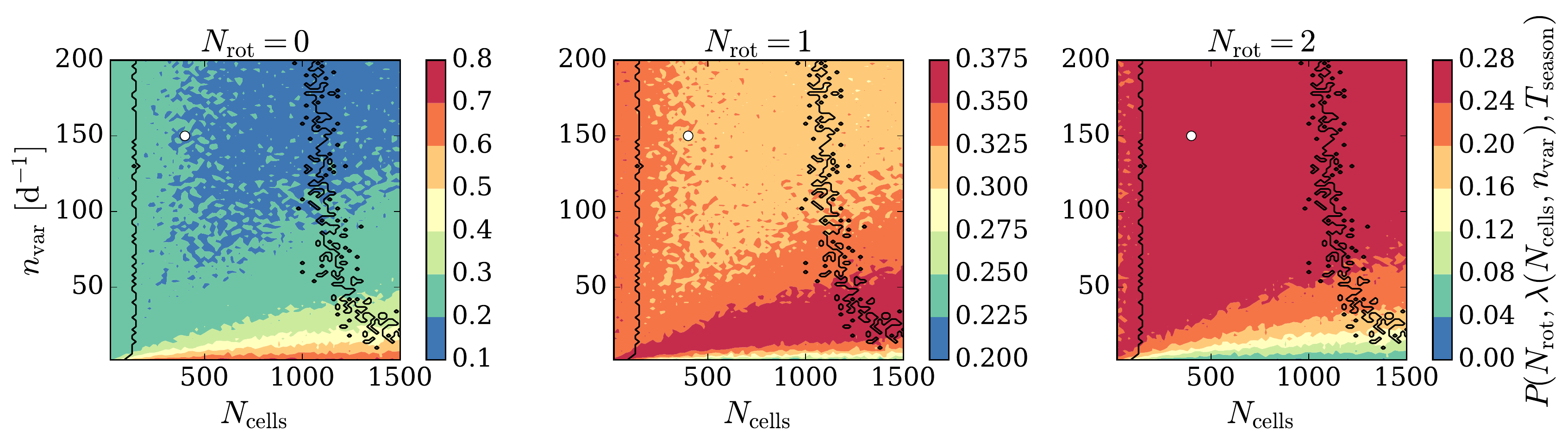}
   \caption{Probability to observe $n=0, 1, 2$ observations during our
   observing season for the TeV source RGB~J0136+391 (top) and the
    non-TeV source GB6\,J1037+5711 (bottom). The color scale shows the
   probability for a various set of parameters $n_\mathrm{var}$ and
   $N_\mathrm{cells}$. The white dot marks the best-fit region for
   obtaining a similar polarization fraction as in the observed light
   curve and the black curves show its 95\% confidence interval.}
             \label{simex}%
     \end{figure*}

We then examine how this relates to the rotations observed in the
individual objects. First, based on the expected frequency of
rotations $\lambda$, we calculate the probability of observing $n=0,1,2$
rotations over the model grid using Poisson statistics
\begin{equation*}
P(n,t,\lambda) = \frac{(\lambda t)^n}{n!}e^{-\lambda t},
\end{equation*}
where $t$ is the total length of the season for each source. The only
parameter that changes for each source is the season length $t$.
In Fig.~\ref{simex}, we show examples for two of the sources showing
rotations RGB~J0136+391 (TeV source) and GB6\,J1037+5711 ( non-TeV
source). Because the sampling is fairly uniform across the samples,
the results for all individual sources are very similar to the example
cases. 

We can see that the probability of observing 0, 1 or 2 rotations is
non-zero in all cases and at least some parameter combinations are
able to produce rotations during our season with high probability. In
order to further examine how well the simulated light curves match our
observed ones, we find all the simulations where the mean polarization
fraction of the simulated light curve is within uncertainties of the
observed mean polarization fraction. Here the uncertainties are
determined using a bootstrap approach, similarly to
\cite{kiehlmann16}. We then select the region of the parameter space
that most likely produces the observed mean polarization
fraction. The best-fit value is shown as a white dot in the the panels
in Fig.~\ref{simex}, and the black lines show the 95\% confidence intervals. 

For all the sources, the probability for observing a single rotation
is between 15 and 37\%, consistent with our detected number of
rotations especially in the  non-TeV sources. In TeV sources we see
less rotations than expected ($\sim 10$\%), which could be due to
them showing more often a preferred polarization angle, which would
hinder us to detect EVPA rotations. The probability to observe two rotations is much
less, although not small enough to rule out a random
walk process.  Even though it is not possible to rule it out for the
  individual sources, \cite{blinov15} showed that it is 
  unlikely that all rotations in the full RoboPol sample are caused by
  a random
  walk process. We also note that we have not examined the
characteristics of the rotations (e.g., their smoothness), which may
further restrict the parameter space where rotations can be observed,  and
  help to distinguish between deterministic and stochastic rotations
(\citealt{kiehlmann16}, Kiehlmann et al. 2016b in prep.).
Detailed modeling of these rotations along with
multifrequency data will be presented elsewhere.

\section{Conclusions}\label{conclusions}
We have studied the optical polarization variability in a sample of
TeV-detected and non-detected ISP and HSP type BL~Lac objects using
data from the RoboPol and NOT instruments. Our main conclusions can be
summarized as follows:
\begin{enumerate}
\item The mean polarization fraction of the TeV-detected BL~Lacs is 5\%
     while the non-TeV sources show a higher mean polarization fraction of 7\%. This difference in polarization fraction disappears when the
     dilution by the unpolarized light of the host galaxy is accounted
     for.  This is a similar polarization fraction as in
       optically-selected BL~Lac objects \citep[e.g.,][]{smith07},
       although an analysis of the full RoboPol sample reveals a
       negative trend in the optical polarization fraction as a
       function of the synchrotron peak with LSP sources showing
       typically a higher mean polarization fraction than HSP sources
       \citep{angelakis16}.
\item When the polarization variations are studied, the rate of EVPA change
   is similar in both samples. The fraction of sources with a
   smaller spread in the $Q/I - U/I$--plane along with a clumped
   distribution of points away from the origin, possibly indicating a
   preferred polarization angle, is larger in the TeV than in the  non-TeV
   sources. We also find the  non-TeV sources to show larger
   polarization fraction variability amplitudes than the TeV sources. This difference between TeV and and non-TeV samples seems
   to arise from differences between the ISP and HSP type sources instead of
   the TeV detection. 
\item We detect significant EVPA rotations in both TeV and  non-TeV sources, showing that rotations can occur in high
 spectral peaking BL Lac objects when the monitoring cadence is dense
 enough. Our simulations show that we cannot exclude a random walk
 origin for these rotations.
   \end{enumerate}
We conclude that TeV loudness is more likely connected to general
flaring activity,  redshift, and the location of the synchrotron peak rather than the polarization properties of these sources.

\begin{acknowledgements}
We thank the referee, P. Giommi, for the careful reading of the paper and
the suggestions that improved it significantly. T.~H. was supported by the Academy of Finland project number
     267324. Part of this work was supported by the COST Action MP1104
     "Polarization as a tool to study the Solar System and
     beyond". D.~B. acknowledges support from the St. Petersburg
     University research grant 6.38.335.2015. I.M. was supported for this research through a stipend from the International Max Planck Research School (IMPRS) for Astronomy and Astrophysics at the Universities of Bonn and Cologne.
The data presented here were obtained in part with
     ALFOSC, which is provided by the Instituto de Astrofisica de
     Andalucia (IAA) under a joint agreement with the University of
     Copenhagen and NOTSA. The RoboPol project is a collaboration between the University
of Crete/FORTH in Greece, Caltech in the USA, MPIfR
in Germany, IUCAA in India and Tor\'un Centre for Astronomy
in Poland. The U. of Crete group acknowledges support
by the “RoboPol” project, which is co-funded by the European
Social Fund (ESF) and Greek National Resources, and
by the European Commission Seventh Framework Programme
(FP7) through grants PCIG10-GA-2011-304001 “JetPop”
and PIRSES-GA-2012-31578 “EuroCal”. This research was
supported in part by NASA grant NNX11A043G and NSF
grant AST-1109911, and by the Polish National Science Centre,
grant number 2011/01/B/ST9/04618. This research has made use of the NASA/IPAC Extragalactic Database (NED),
which is operated by the Jet Propulsion Laboratory, California Institute of Technology,
under contract with the National Aeronautics and Space Administration.
\end{acknowledgements}

% WARNING
%-------------------------------------------------------------------
% Please note that we have included the references to the file aa.dem in
% order to compile it  but we ask you to:
%
% - use BibTeX with the regular commands:
   \bibliographystyle{aa} % style aa.bst
   \bibliography{thbib.bib} % your references Yourfile.bib

\begin{thebibliography}{80}
\expandafter\ifx\csname natexlab\endcsname\relax\def\natexlab#1{#1}\fi

\bibitem[{{Ackermann} {et~al.}(2011){Ackermann}, {Ajello}, {Allafort},
  {Antolini}, {Atwood}, {Axelsson}, {Baldini}, {Ballet}, {Barbiellini},
  {Bastieri}, {Bechtol}, {Bellazzini}, {Berenji}, {Blandford}, {Bloom}, \&
  et~al.}]{ackermann11}
{Ackermann}, M., {Ajello}, M., {Allafort}, A., {et~al.} 2011, \apj, 743, 171

\bibitem[{{Ackermann} {et~al.}(2015){Ackermann}, {Ajello}, {Atwood}, {Baldini},
  {Ballet}, {Barbiellini}, {Bastieri}, {Becerra Gonzalez}, {Bellazzini},
  {Bissaldi}, {Blandford}, {Bloom}, {Bonino}, {Bottacini}, {Brandt}, {Bregeon},
  {Britto}, {Bruel}, {Buehler}, {Buson}, {Caliandro}, {Cameron}, {Caragiulo},
  {Caraveo}, {Carpenter}, {Casandjian}, {Cavazzuti}, {Cecchi}, {Charles},
  {Chekhtman}, {Cheung}, {Chiang}, {Chiaro}, {Ciprini}, {Claus},
  {Cohen-Tanugi}, {Cominsky}, {Conrad}, {Cutini}, {D'Abrusco}, {D'Ammando}, {de
  Angelis}, {Desiante}, {Digel}, {Di Venere}, {Drell}, {Favuzzi}, {Fegan},
  {Ferrara}, {Finke}, {Focke}, {Franckowiak}, {Fuhrmann}, {Fukazawa},
  {Furniss}, {Fusco}, {Gargano}, {Gasparrini}, {Giglietto}, {Giommi},
  {Giordano}, {Giroletti}, {Glanzman}, {Godfrey}, {Grenier}, {Grove},
  {Guiriec}, {Hewitt}, {Hill}, {Horan}, {Itoh}, {J{\'o}hannesson}, {Johnson},
  {Johnson}, {Kataoka}, {Kawano}, {Krauss}, {Kuss}, {La Mura}, {Larsson},
  {Latronico}, {Leto}, {Li}, {Li}, {Longo}, {Loparco}, {Lott}, {Lovellette},
  {Lubrano}, {Madejski}, {Mayer}, {Mazziotta}, {McEnery}, {Michelson},
  {Mizuno}, {Moiseev}, {Monzani}, {Morselli}, {Moskalenko}, {Murgia}, {Nuss},
  {Ohno}, {Ohsugi}, {Ojha}, {Omodei}, {Orienti}, {Orlando}, {Paggi}, {Paneque},
  {Perkins}, {Pesce-Rollins}, {Piron}, {Pivato}, {Porter}, {Rain{\`o}},
  {Rando}, {Razzano}, {Razzaque}, {Reimer}, {Reimer}, {Romani}, {Salvetti},
  {Schaal}, {Schinzel}, {Schulz}, {Sgr{\`o}}, {Siskind}, {Sokolovsky}, {Spada},
  {Spandre}, {Spinelli}, {Stawarz}, {Suson}, {Takahashi}, {Takahashi},
  {Tanaka}, {Thayer}, {Thayer}, {Tibaldo}, {Torres}, {Torresi}, {Tosti},
  {Troja}, {Uchiyama}, {Vianello}, {Winer}, {Wood}, \& {Zimmer}}]{3LAC}
{Ackermann}, M., {Ajello}, M., {Atwood}, W.~B., {et~al.} 2015, \apj, 810, 14

\bibitem[{{Ahnen} {et~al.}(2015){Ahnen}, {Ansoldi}, {Antonelli}, {Antoranz},
  {Babic}, {Banerjee}, {Bangale}, {Barres de Almeida}, {Barrio}, {Bednarek}, \&
  et~al.}]{ahnen15_1441}
{Ahnen}, M.~L., {Ansoldi}, S., {Antonelli}, L.~A., {et~al.} 2015, \apjl, 815,
  L23

\bibitem[{{Ahnen} {et~al.}(2016){Ahnen}, {Ansoldi}, {Antonelli}, {Antoranz},
  {Babic}, {Banerjee}, {Bangale}, {de Almeida}, {Barrio}, {Gonz{\'a}lez},
  {Bednarek}, {Bernardini}, {Biasuzzi}, {Biland}, {Blanch}, {Bonnefoy},
  {Bonnoli}, {Borracci}, {Bretz}, {Carmona}, {Carosi}, {Chatterjee}, {Clavero},
  {Colin}, {Colombo}, {Contreras}, {Cortina}, {Covino}, {Da Vela}, {Dazzi}, {De
  Angelis}, {De Caneva}, {De Lotto}, {de O{\~n}a Wilhelmi}, {Mendez}, {Pierro},
  {Prester}, {Dorner}, {Doro}, {Einecke}, {Elsaesser}, {Fern{\'a}ndez-Barral},
  {Fidalgo}, {Fonseca}, {Font}, {Frantzen}, {Fruck}, {Galindo}, {L{\'o}pez},
  {Garczarczyk}, {Terrats}, {Gaug}, {Giammaria}, {(Eisenacher)},
  {Godinovi{\'c}}, {Mu{\~n}oz}, {Guberman}, {Hanabata}, {Hayashida}, {Herrera},
  {Hose}, {Hrupec}, {Hughes}, {Idec}, {Kodani}, {Konno}, {Kubo}, {Kushida},
  {Barbera}, {Lelas}, {Lindfors}, {Lombardi}, {Longo}, {L{\'o}pez},
  {L{\'o}pez-Coto}, {L{\'o}pez-Oramas}, {Lorenz}, {Majumdar}, {Makariev},
  {Mallot}, {Maneva}, {Manganaro}, {Mannheim}, {Maraschi}, {Marcote},
  {Mariotti}, {Mart{\'{\i}}nez}, {Mazin}, {Menzel}, {Miranda}, {Mirzoyan},
  {Moralejo}, {Nakajima}, {Neustroev}, {Niedzwiecki}, {Rosillo}, {Nilsson},
  {Nishijima}, {Noda}, {Orito}, {Overkemping}, {Paiano}, {Palacio},
  {Palatiello}, {Paneque}, {Paoletti}, {Paredes}, {Paredes-Fortuny}, {Persic},
  {Poutanen}, {Moroni}, {Prandini}, {Puljak}, {Reinthal}, {Rhode}, {Rib{\'o}},
  {Rico}, {Garcia}, {R{\"u}gamer}, {Saito}, {Satalecka}, {Scapin}, {Schultz},
  {Schweizer}, {Shore}, {Sillanp{\"a}{\"a}}, {Sitarek}, {Snidaric},
  {Sobczynska}, {Stamerra}, {Steinbring}, {Strzys}, {Takalo}, {Takami},
  {Tavecchio}, {Temnikov}, {Terzi{\'c}}, {Tescaro}, {Teshima}, {Thaele},
  {Torres}, {Toyama}, {Treves}, {Verguilov}, {Vovk}, {Ward}, {Will}, {Wu},
  {Zanin}, {Lucarelli}, {Pittori}, {Vercellone}, {Berdyugin}, {Carini},
  {L{\"a}hteenm{\"a}ki}, {Pasanen}, {Pease}, {Sainio}, {Tornikoski}, \&
  {Walters}}]{ahnen16_1011}
{Ahnen}, M.~L., {Ansoldi}, S., {Antonelli}, L.~A., {et~al.} 2016, \mnras, 459,
  2286

\bibitem[{{Akiyama} {et~al.}(2016){Akiyama}, {Stawarz}, {Tanaka}, {Nagai},
  {Giroletti}, \& {Honma}}]{akiyama16}
{Akiyama}, K., {Stawarz}, {\L}., {Tanaka}, Y.~T., {et~al.} 2016, ApJL submitted
  [\eprint[arXiv]{1603.00877}]

\bibitem[{{Aleksi{\'c}} {et~al.}(2014){Aleksi{\'c}}, {Ansoldi}, {Antonelli},
  {Antoranz}, {Babic}, {Bangale}, {Barres de Almeida}, {Barrio}, {Becerra
  Gonz{\'a}lez}, {Bednarek}, {Bernardini}, {Biland}, {Blanch}, {Bonnefoy},
  {Bonnoli}, {Borracci}, {Bretz}, {Carmona}, {Carosi}, {Carreto Fidalgo},
  {Colin}, {Colombo}, {Contreras}, {Cortina}, {Covino}, {da Vela}, {Dazzi}, {de
  Angelis}, {de Caneva}, {de Lotto}, {Delgado Mendez}, {Doert},
  {Dom{\'{\i}}nguez}, {Dominis Prester}, {Dorner}, {Doro}, {Einecke},
  {Eisenacher}, {Elsaesser}, {Farina}, {Ferenc}, {Fonseca}, {Font}, {Frantzen},
  {Fruck}, {Garc{\'{\i}}a L{\'o}pez}, {Garczarczyk}, {Garrido Terrats}, {Gaug},
  {Godinovi{\'c}}, {Gonz{\'a}lez Mu{\~n}oz}, {Gozzini}, {Hadasch}, {Hayashida},
  {Herrera}, {Herrero}, {Hildebrand}, {Hose}, {Hrupec}, {Idec}, {Kadenius},
  {Kellermann}, {Kodani}, {Konno}, {Krause}, {Kubo}, {Kushida}, {La Barbera},
  {Lelas}, {Lewandowska}, {Lindfors}, {Lombardi}, {L{\'o}pez},
  {L{\'o}pez-Coto}, {L{\'o}pez-Oramas}, {Lorenz}, {Lozano}, {Makariev},
  {Mallot}, {Maneva}, {Mankuzhiyil}, {Mannheim}, {Maraschi}, {Marcote},
  {Mariotti}, {Mart{\'{\i}}nez}, {Mazin}, {Menzel}, {Meucci}, {Miranda},
  {Mirzoyan}, {Moralejo}, {Munar-Adrover}, {Nakajima}, {Niedzwiecki},
  {Nilsson}, {Nishijima}, {Noda}, {Nowak}, {Orito}, {Overkemping}, {Paiano},
  {Palatiello}, {Paneque}, {Paoletti}, {Paredes}, {Paredes-Fortuny}, {Partini},
  {Persic}, {Prada}, {Prada Moroni}, {Prandini}, {Preziuso}, {Puljak},
  {Reinthal}, {Rhode}, {Rib{\'o}}, {Rico}, {Rodriguez Garcia}, {R{\"u}gamer},
  {Saggion}, {Saito}, {Saito}, {Satalecka}, {Scalzotto}, {Scapin}, {Schultz},
  {Schweizer}, {Shore}, {Sillanp{\"a}{\"a}}, {Sitarek}, {Snidaric},
  {Sobczynska}, {Spanier}, {Stamatescu}, {Stamerra}, {Steinbring}, {Storz},
  {Strzys}, {Sun}, {Suri{\'c}}, {Takalo}, {Takami}, {Tavecchio}, {Temnikov},
  {Terzi{\'c}}, {Tescaro}, {Teshima}, {Thaele}, {Tibolla}, {Torres}, {Toyama},
  {Treves}, {Uellenbeck}, {Vogler}, {Wagner}, {Zandanel}, {Zanin}, \& {MAGIC
  Collaboration}}]{aleksic14_J2001}
{Aleksi{\'c}}, J., {Ansoldi}, S., {Antonelli}, L.~A., {et~al.} 2014, \aap, 572,
  A121

\bibitem[{{Aleksi{\'c}} {et~al.}(2015){Aleksi{\'c}}, {Ansoldi}, {Antonelli},
  {Antoranz}, {Babic}, {Bangale}, {Barrio}, {Becerra Gonz{\'a}lez}, {Bednarek},
  {Bernardini}, {Biasuzzi}, {Biland}, {Blanch}, {Bonnefoy}, {Bonnoli},
  {Borracci}, {Bretz}, {Carmona}, {Carosi}, {Colin}, {Colombo}, {Contreras},
  {Cortina}, {Covino}, {Da Vela}, {Dazzi}, {De Angelis}, {De Caneva}, {De
  Lotto}, {de O{\~n}a Wilhelmi}, {Delgado Mendez}, {Di Pierro}, {Dominis
  Prester}, {Dorner}, {Doro}, {Einecke}, {Eisenacher}, {Elsaesser},
  {Fern{\'a}ndez-Barral}, {Fidalgo}, {Fonseca}, {Font}, {Frantzen}, {Fruck},
  {Galindo}, {Garc{\'{\i}}a L{\'o}pez}, {Garczarczyk}, {Garrido Terrats},
  {Gaug}, {Godinovi{\'c}}, {Gonz{\'a}lez Mu{\~n}oz}, {Gozzini}, {Hadasch},
  {Hanabata}, {Hayashida}, {Herrera}, {Hose}, {Hrupec}, {Idec}, {Kadenius},
  {Kellermann}, {Knoetig}, {Kodani}, {Konno}, {Krause}, {Kubo}, {Kushida}, {La
  Barbera}, {Lelas}, {Lewandowska}, {Lindfors}, {Lombardi}, {Longo},
  {L{\'o}pez}, {L{\'o}pez-Coto}, {L{\'o}pez-Oramas}, {Lorenz}, {Lozano},
  {Makariev}, {Mallot}, {Maneva}, {Mannheim}, {Maraschi}, {Marcote},
  {Mariotti}, {Mart{\'{\i}}nez}, {Mazin}, {Menzel}, {Miranda}, {Mirzoyan},
  {Moralejo}, {Munar-Adrover}, {Nakajima}, {Neustroev}, {Niedzwiecki}, {Nievas
  Rosillo}, {Nilsson}, {Nishijima}, {Noda}, {Orito}, {Overkemping}, {Paiano},
  {Palatiello}, {Paneque}, {Paoletti}, {Paredes}, {Paredes-Fortuny}, {Persic},
  {Poutanen}, {Prada Moroni}, {Prandini}, {Puljak}, {Reinthal}, {Rhode},
  {Rib{\'o}}, {Rico}, {Rodriguez Garcia}, {Saito}, {Saito}, {Satalecka},
  {Scalzotto}, {Scapin}, {Schultz}, {Schweizer}, {Shore}, {Sillanp{\"a}{\"a}},
  {Sitarek}, {Snidaric}, {Sobczynska}, {Stamerra}, {Steinbring}, {Strzys},
  {Takalo}, {Takami}, {Tavecchio}, {Temnikov}, {Terzi{\'c}}, {Tescaro},
  {Teshima}, {Thaele}, {Torres}, {Toyama}, {Treves}, {Vogler}, {Will}, {Zanin},
  {Berger}, {Buson}, {D'Ammando}, {Gasparrini}, {Hovatta}, {Max-Moerbeck},
  {Readhead}, \& {Richards}}]{aleksic15_0806}
{Aleksi{\'c}}, J., {Ansoldi}, S., {Antonelli}, L.~A., {et~al.} 2015, \mnras,
  451, 739

\bibitem[{{Andruchow} {et~al.}(2008){Andruchow}, {Cellone}, \&
  {Romero}}]{andruchow08}
{Andruchow}, I., {Cellone}, S.~A., \& {Romero}, G.~E. 2008, \mnras, 388, 1766

\bibitem[{{Angel} {et~al.}(1978){Angel}, {Boroson}, {Adams}, {Duerr},
  {Glampapa}, {Cresham}, {Gural}, {Hubbard}, {Kopriva}, \& {Moore}}]{angel78}
{Angel}, J.~R.~P., {Boroson}, T.~A., {Adams}, M.~T., {et~al.} 1978, in BL Lac
  Objects, ed. A.~M. {Wolfe}, 117--146

\bibitem[{{Angel} \& {Stockman}(1980)}]{angel80}
{Angel}, J.~R.~P. \& {Stockman}, H.~S. 1980, \araa, 18, 321

\bibitem[{Angelakis {et~al.}(2016)Angelakis, Hovatta, Blinov, Pavlidou,
  Kiehlmann, Myserlis, \& Boettcher}]{angelakis16}
Angelakis, E., Hovatta, T., Blinov, D., {et~al.} 2016, MNRAS in press

\bibitem[{{Archambault} {et~al.}(2016){Archambault}, {Archer}, {Benbow},
  {Bird}, {Biteau}, {Buchovecky}, {Buckley}, {Bugaev}, {Byrum}, {Cerruti},
  {Chen}, {Ciupik}, {Connolly}, {Cui}, {Eisch}, {Errando}, {Falcone}, {Feng},
  {Finley}, {Fleischhack}, {Fortin}, {Fortson}, {Furniss}, {Gillanders},
  {Griffin}, {Grube}, {Gyuk}, {H{\"u}tten}, {Hakansson}, {Hanna}, {Holder},
  {Humensky}, {Johnson}, {Kaaret}, {Kar}, {Kelley-Hoskins}, {Kertzman},
  {Kieda}, {Krause}, {Krennrich}, {Kumar}, {Lang}, {Maier}, {McArthur},
  {McCann}, {Meagher}, {Moriarty}, {Mukherjee}, {Nguyen}, {Nieto},
  {O'Faol{\'a}in De Bhr{\'o}ithe}, {Ong}, {Otte}, {Park}, {Perkins}, {Pichel},
  {Pohl}, {Popkow}, {Pueschel}, {Quinn}, {Ragan}, {Reynolds}, {Richards},
  {Roache}, {Rovero}, {Santander}, {Sembroski}, {Shahinyan}, {Smith},
  {Staszak}, {Telezhinsky}, {Tucci}, {Tyler}, {Vincent}, {Wakely}, {Weiner},
  {Weinstein}, {Williams}, {Zitzer}, {Fumagalli}, \&
  {Prochaska}}]{veritas16_UL}
{Archambault}, S., {Archer}, A., {Benbow}, W., {et~al.} 2016, ArXiv e-prints
  [\eprint[arXiv]{1603.02410}]

\bibitem[{{Blinov} {et~al.}(2016){Blinov}, {Pavlidou}, {Papadakis}, {Hovatta},
  {Liodakis}, {Panopoulou}, {Angelakis}, {Balokovi{\'c}}, {Das}, {Khodade},
  {Kiehlmann}, {King}, {Kus}, {Kylafis}, {Mahabal}, {Marecki}, {Modi},
  {Myserlis}, {Paleologou}, {Papamastorakis}, Pazderska, Pazderski, {Pearson},
  {Rajarshi}, {Ramaprakash}, {Readhead}, {Reig}, {Tassis}, \&
  {Zensus}}]{blinov16}
{Blinov}, D., {Pavlidou}, V., {Papadakis}, I., {et~al.} 2016, \mnras, 457, 2252

\bibitem[{{Blinov} {et~al.}(2016b){Blinov}, {Pavlidou}, {Papadakis}, Kiehlmann,
  Liodakis, Panopoulou, Angelakis, \& et~al.}]{blinov16b}
{Blinov}, D., {Pavlidou}, V., {Papadakis}, I., {et~al.} 2016b, \mnras, in
  Press. arXiv:1607.04292

\bibitem[{{Blinov} {et~al.}(2015){Blinov}, {Pavlidou}, {Papadakis},
  {Kiehlmann}, {Panopoulou}, {Liodakis}, {King}, {Angelakis}, {Balokovi{\'c}},
  {Das}, {Feiler}, {Fuhrmann}, {Hovatta}, {Khodade}, {Kus}, {Kylafis},
  {Mahabal}, {Myserlis}, {Modi}, {Pazderska}, {Pazderski}, {Papamastorakis},
  {Pearson}, {Rajarshi}, {Ramaprakash}, {Reig}, {Readhead}, {Tassis}, \&
  {Zensus}}]{blinov15}
{Blinov}, D., {Pavlidou}, V., {Papadakis}, I., {et~al.} 2015, \mnras, 453, 1669

\bibitem[{{Bloom} {et~al.}(1997){Bloom}, {Bertsch}, {Hartman}, {Sreekumar},
  {Thompson}, {Balonek}, {Beckerman}, {Davis}, {Whitman}, {Miller}, {Nair},
  {Roberts}, {Tosti}, {Massaro}, {Nesci}, {Maesano}, {Montagni}, {Jang},
  {Bock}, {Dietrich}, {Herter}, {Otterbein}, {Pfeiffer}, {Seitz}, \&
  {Wagner}}]{bloom97}
{Bloom}, S.~D., {Bertsch}, D.~L., {Hartman}, R.~C., {et~al.} 1997, \apjl, 490,
  L145

\bibitem[{{B{\"o}ttcher} {et~al.}(2013){B{\"o}ttcher}, {Reimer}, {Sweeney}, \&
  {Prakash}}]{bottcher13}
{B{\"o}ttcher}, M., {Reimer}, A., {Sweeney}, K., \& {Prakash}, A. 2013, \apj,
  768, 54

\bibitem[{{Danforth} {et~al.}(2010){Danforth}, {Keeney}, {Stocke}, {Shull}, \&
  {Yao}}]{danforth10}
{Danforth}, C.~W., {Keeney}, B.~A., {Stocke}, J.~T., {Shull}, J.~M., \& {Yao},
  Y. 2010, \apj, 720, 976

\bibitem[{{Dom{\'{\i}}nguez} {et~al.}(2011){Dom{\'{\i}}nguez}, {Primack},
  {Rosario}, {Prada}, {Gilmore}, {Faber}, {Koo}, {Somerville},
  {P{\'e}rez-Torres}, {P{\'e}rez-Gonz{\'a}lez}, {Huang}, {Davis},
  {Guhathakurta}, {Barmby}, {Conselice}, {Lozano}, {Newman}, \&
  {Cooper}}]{dominguez11}
{Dom{\'{\i}}nguez}, A., {Primack}, J.~R., {Rosario}, D.~J., {et~al.} 2011,
  \mnras, 410, 2556

\bibitem[{{Fitzpatrick}(1999)}]{fitzpatrick99}
{Fitzpatrick}, E.~L. 1999, \pasp, 111, 63

\bibitem[{{Franceschini} {et~al.}(2008){Franceschini}, {Rodighiero}, \&
  {Vaccari}}]{franceschini08}
{Franceschini}, A., {Rodighiero}, G., \& {Vaccari}, M. 2008, \aap, 487, 837

\bibitem[{{Fukugita} {et~al.}(1995){Fukugita}, {Shimasaku}, \&
  {Ichikawa}}]{fukugita95}
{Fukugita}, M., {Shimasaku}, K., \& {Ichikawa}, T. 1995, \pasp, 107, 945

\bibitem[{{Furniss} {et~al.}(2013{\natexlab{a}}){Furniss}, {Fumagalli},
  {Danforth}, {Williams}, \& {Prochaska}}]{furniss13a}
{Furniss}, A., {Fumagalli}, M., {Danforth}, C., {Williams}, D.~A., \&
  {Prochaska}, J.~X. 2013{\natexlab{a}}, \apj, 766, 35

\bibitem[{{Furniss} {et~al.}(2013{\natexlab{b}}){Furniss}, {Williams},
  {Danforth}, {Fumagalli}, {Prochaska}, {Primack}, {Urry}, {Stocke},
  {Filippenko}, \& {Neely}}]{furniss13b}
{Furniss}, A., {Williams}, D.~A., {Danforth}, C., {et~al.} 2013{\natexlab{b}},
  \apjl, 768, L31

\bibitem[{{Giommi} {et~al.}(2000){Giommi}, {Padovani}, \& {Perlman}}]{giommi00}
{Giommi}, P., {Padovani}, P., \& {Perlman}, E. 2000, \mnras, 317, 743

\bibitem[{{Giommi} {et~al.}(2012){Giommi}, {Padovani}, {Polenta}, {Turriziani},
  {D'Elia}, \& {Piranomonte}}]{giommi12}
{Giommi}, P., {Padovani}, P., {Polenta}, G., {et~al.} 2012, \mnras, 420, 2899

\bibitem[{{Heidt} \& {Nilsson}(2011)}]{heidt11}
{Heidt}, J. \& {Nilsson}, K. 2011, \aap, 529, A162

\bibitem[{{Hovatta} {et~al.}(2014){Hovatta}, {Pavlidou}, {King}, {Mahabal},
  {Sesar}, {Dancikova}, {Djorgovski}, {Drake}, {Laher}, {Levitan},
  {Max-Moerbeck}, {Ofek}, {Pearson}, {Prince}, {Readhead}, {Richards}, \&
  {Surace}}]{hovatta14a}
{Hovatta}, T., {Pavlidou}, V., {King}, O.~G., {et~al.} 2014, \mnras, 439, 690

\bibitem[{{Jannuzi} {et~al.}(1994){Jannuzi}, {Smith}, \& {Elston}}]{jannuzi94}
{Jannuzi}, B.~T., {Smith}, P.~S., \& {Elston}, R. 1994, \apj, 428, 130

\bibitem[{Jermak {et~al.}(2016)Jermak, Steele, Lindfors, Hovatta, Nilsson,
  Lamb, Mundell, Barres~de Almeida, Berdyugin, Kadenius, Reinthal, \&
  Takalo}]{jermak16}
Jermak, H., Steele, I.~A., Lindfors, E., {et~al.} 2016, MNRAS in press

\bibitem[{{Jorstad} {et~al.}(2007){Jorstad}, {Marscher}, {Stevens}, {Smith},
  {Forster}, {Gear}, {Cawthorne}, {Lister}, {Stirling}, {G{\'o}mez}, {Greaves},
  \& {Robson}}]{jorstad07}
{Jorstad}, S.~G., {Marscher}, A.~P., {Stevens}, J.~A., {et~al.} 2007, \aj, 134,
  799

\bibitem[{{Kiehlmann} {et~al.}(2016){Kiehlmann}, {Savolainen}, {Jorstad},
  {Sokolovsky}, {Schinzel}, {Marscher}, {Larionov}, {Agudo}, {Akitaya},
  {Ben{\'{\i}}tez}, {Berdyugin}, {Blinov}, {Bochkarev}, {Borman}, {Burenkov},
  {Casadio}, {Doroshenko}, {Efimova}, {Fukazawa}, {G{\'o}mez}, {Grishina},
  {Hagen-Thorn}, {Heidt}, {Hiriart}, {Itoh}, {Joshi}, {Kawabata}, {Kimeridze},
  {Kopatskaya}, {Korobtsev}, {Krajci}, {Kurtanidze}, {Kurtanidze}, {Larionova},
  {Larionova}, {Lindfors}, {L{\'o}pez}, {McHardy}, {Molina}, {Moritani},
  {Morozova}, {Nazarov}, {Nikolashvili}, {Nilsson}, {Pulatova}, {Reinthal},
  {Sadun}, {Sasada}, {Savchenko}, {Sergeev}, {Sigua}, {Smith}, {Sorcia},
  {Spiridonova}, {Takaki}, {Takalo}, {Taylor}, {Troitsky}, {Uemura},
  {Ugolkova}, {Ui}, {Yoshida}, {Zensus}, \& {Zhdanova}}]{kiehlmann16}
{Kiehlmann}, S., {Savolainen}, T., {Jorstad}, S.~G., {et~al.} 2016, A\&A in
  press [\eprint[arXiv]{1603.00249}]

\bibitem[{{Kikuchi} {et~al.}(1988){Kikuchi}, {Mikami}, {Inoue}, {Tabara}, \&
  {Kato}}]{kikuchi88}
{Kikuchi}, S., {Mikami}, Y., {Inoue}, M., {Tabara}, H., \& {Kato}, T. 1988,
  \aap, 190, L8

\bibitem[{{King} {et~al.}(2014){King}, {Blinov}, {Ramaprakash}, {Myserlis},
  {Angelakis}, {Balokovi{\'c}}, {Feiler}, {Fuhrmann}, {Hovatta}, {Khodade},
  {Kougentakis}, {Kylafis}, {Kus}, {Modi}, {Paleologou}, {Panopoulou},
  {Papadakis}, {Papamastorakis}, {Paterakis}, {Pavlidou}, {Pazderska},
  {Pazderski}, {Pearson}, {Rajarshi}, {Readhead}, {Reig}, {Steiakaki},
  {Tassis}, \& {Zensus}}]{king14}
{King}, O.~G., {Blinov}, D., {Ramaprakash}, A.~N., {et~al.} 2014, \mnras, 442,
  1706

\bibitem[{{Kotilainen} {et~al.}(1998){Kotilainen}, {Falomo}, \&
  {Scarpa}}]{kotilainen98}
{Kotilainen}, J.~K., {Falomo}, R., \& {Scarpa}, R. 1998, \aap, 332, 503

\bibitem[{{Kotilainen} {et~al.}(2011){Kotilainen}, {Hyv{\"o}nen}, {Falomo},
  {Treves}, \& {Uslenghi}}]{kotilainen11}
{Kotilainen}, J.~K., {Hyv{\"o}nen}, T., {Falomo}, R., {Treves}, A., \&
  {Uslenghi}, M. 2011, \aap, 534, L2

\bibitem[{{Landi Degl'Innocenti} {et~al.}(2007){Landi Degl'Innocenti},
  {Bagnulo}, \& {Fossati}}]{landi07}
{Landi Degl'Innocenti}, E., {Bagnulo}, S., \& {Fossati}, L. 2007, in
  Astronomical Society of the Pacific Conference Series, Vol. 364, The Future
  of Photometric, Spectrophotometric and Polarimetric Standardization, ed.
  C.~{Sterken}, 495

\bibitem[{{Landoni} {et~al.}(2014){Landoni}, {Falomo}, {Treves}, \&
  {Sbarufatti}}]{landoni14}
{Landoni}, M., {Falomo}, R., {Treves}, A., \& {Sbarufatti}, B. 2014, \aap, 570,
  A126

\bibitem[{{Lavalley} {et~al.}(1992){Lavalley}, {Isobe}, \&
  {Feigelson}}]{lavalley92}
{Lavalley}, M., {Isobe}, T., \& {Feigelson}, E. 1992, in Astronomical Society
  of the Pacific Conference Series, Vol.~25, Astronomical Data Analysis
  Software and Systems I, ed. D.~M. {Worrall}, C.~{Biemesderfer}, \&
  J.~{Barnes}, 245

\bibitem[{{Lister} {et~al.}(2011){Lister}, {Aller}, {Aller}, {Hovatta},
  {Kellermann}, {Kovalev}, {Meyer}, {Pushkarev}, {Ros}, {MOJAVE Collaboration},
  {Ackermann}, {Antolini}, {Baldini}, {Ballet}, {Barbiellini}, {Bastieri},
  {Bechtol}, {Bellazzini}, {Berenji}, {Blandford}, {Bloom}, {Boeck},
  {Bonamente}, {Borgland}, {Bregeon}, {Brigida}, {Bruel}, {Buehler}, {Buson},
  {Caliandro}, {Cameron}, {Caraveo}, {Casandjian}, {Cavazzuti}, {Cecchi},
  {Chang}, {Charles}, {Chekhtman}, {Cheung}, {Chiang}, {Ciprini}, {Claus},
  {Cohen-Tanugi}, {Conrad}, {Cutini}, {de Palma}, {Dermer}, {Silva}, {Drell},
  {Drlica-Wagner}, \& et~al.}]{lister11}
{Lister}, M.~L., {Aller}, M., {Aller}, H., {et~al.} 2011, \apj, 742, 27

\bibitem[{{Lister} {et~al.}(2013){Lister}, {Aller}, {Aller}, {Homan},
  {Kellermann}, {Kovalev}, {Pushkarev}, {Richards}, {Ros}, \&
  {Savolainen}}]{lister13}
{Lister}, M.~L., {Aller}, M.~F., {Aller}, H.~D., {et~al.} 2013, \aj, 146, 120,
  Paper X

\bibitem[{{Lister} \& {Smith}(2000)}]{lister00}
{Lister}, M.~L. \& {Smith}, P.~S. 2000, \apj, 541, 66

\bibitem[{{Lyutikov} {et~al.}(2005){Lyutikov}, {Pariev}, \&
  {Gabuzda}}]{lyutikov05}
{Lyutikov}, M., {Pariev}, V.~I., \& {Gabuzda}, D.~C. 2005, \mnras, 360, 869

\bibitem[{{MAGIC Collaboration} {et~al.}(2016){MAGIC Collaboration}, {Ahnen},
  {Ansoldi}, {Antonelli}, {Antoranz}, {Babic}, {Banerjee}, {Bangale}, {Barres
  de Almeida}, {Barrio}, {Becerra Gonz{\'a}lez}, {Bednarek}, {Bernardini},
  {Biasuzzi}, {Biland}, {Blanch}, {Bonnefoy}, {Bonnoli}, {Borracci}, {Bretz},
  {Buson}, {Carosi}, {Chatterjee}, {Clavero}, {Colin}, {Colombo}, {Contreras},
  {Cortina}, {Covino}, {Da Vela}, {Dazzi}, {De Angelis}, {De Lotto}, {de Ona
  Wilhelmi}, {Di Pierro}, {Doert}, {Dom{\'{\i}}nguez}, {Dominis Prester},
  {Dorner}, {Doro}, {Einecke}, {Eisenacher Glawion}, {Elsaesser}, {Fallah
  Ramazani}, {Fern{\'a}ndez-Barral}, {Fidalgo}, {Fonseca}, {Font}, {Frantzen},
  {Fruck}, {Galindo}, {Garc{\'{\i}}a L{\'o}pez}, {Garczarczyk}, {Garrido
  Terrats}, {Gaug}, {Giammaria}, {Godinovi{\'c}}, {Gonz{\'a}lez Munoz}, {Gora},
  {Guberman}, {Hadasch}, {Hahn}, {Hanabata}, {Hayashida}, {Herrera}, {Hose},
  {Hrupec}, {Hughes}, {Idec}, {Kodani}, {Konno}, {Kubo}, {Kushida}, {La
  Barbera}, {Lelas}, {Lindfors}, {Lombardi}, {Longo}, {L{\'o}pez},
  {L{\'o}pez-Coto}, {Majumdar}, {Makariev}, {Mallot}, {Maneva}, {Manganaro},
  {Mannheim}, {Maraschi}, {Marcote}, {Mariotti}, {Mart{\'{\i}}nez}, {Mazin},
  {Menzel}, {Miranda}, {Mirzoyan}, {Moralejo}, {Moretti}, {Nakajima},
  {Neustroev}, {Niedzwiecki}, {Nievas Rosillo}, {Nilsson}, {Nishijima}, {Noda},
  {Nogu{\'e}s}, {Orito}, {Overkemping}, {Paiano}, {Palacio}, {Palatiello},
  {Paneque}, {Paoletti}, {Paredes}, {Paredes-Fortuny}, {Pedaletti}, {Perri},
  {Persic}, {Poutanen}, {Prada Moroni}, {Prandini}, {Puljak}, {Rhode},
  {Rib{\'o}}, {Rico}, {Rodriguez Garcia}, {Saito}, {Satalecka}, {Schultz},
  {Schweizer}, {Sillanp{\"a}{\"a}}, {Sitarek}, {Snidaric}, {Sobczynska},
  {Stamerra}, {Steinbring}, {Strzys}, {Takalo}, {Takami}, {Tavecchio},
  {Temnikov}, {Terzi{\'c}}, {Tescaro}, {Teshima}, {Thaele}, {Torres}, {Toyama},
  {Treves}, {Verguilov}, {Vovk}, {Ward}, {Will}, {Wu}, {Zanin}, {D'Ammando},
  {Hovatta}, {Max-Moerbeck}, {Raiteri}, {Readhead}, {Reinthal}, {Richards},
  {Verrecchia}, \& {Villata}}]{ahnen16_1722}
{MAGIC Collaboration}, {Ahnen}, M.~L., {Ansoldi}, S., {et~al.} 2016, ArXiv
  e-prints [\eprint[arXiv]{1603.06523}]

\bibitem[{{Meisner} \& {Romani}(2010)}]{meisner10}
{Meisner}, A.~M. \& {Romani}, R.~W. 2010, \apj, 712, 14

\bibitem[{Monet {et~al.}(2003)}]{Monet2003}
Monet, D.~G. {et~al.} 2003, AJ, 125, 984

\bibitem[{{Nieppola} {et~al.}(2006){Nieppola}, {Tornikoski}, \&
  {Valtaoja}}]{nieppola06}
{Nieppola}, E., {Tornikoski}, M., \& {Valtaoja}, E. 2006, \aap, 445, 441

\bibitem[{{Nilsson} {et~al.}(2007){Nilsson}, {Pasanen}, {Takalo}, {Lindfors},
  {Berdyugin}, {Ciprini}, \& {Pforr}}]{nilsson07}
{Nilsson}, K., {Pasanen}, M., {Takalo}, L.~O., {et~al.} 2007, \aap, 475, 199

\bibitem[{{Nilsson} {et~al.}(2003){Nilsson}, {Pursimo}, {Heidt}, {Takalo},
  {Sillanp{\"a}{\"a}}, \& {Brinkmann}}]{nilsson03}
{Nilsson}, K., {Pursimo}, T., {Heidt}, J., {et~al.} 2003, \aap, 400, 95

\bibitem[{{Nilsson} {et~al.}(1999){Nilsson}, {Pursimo}, {Takalo},
  {Sillanp{\"a}{\"a}}, {Pietil{\"a}}, \& {Heidt}}]{nilsson99}
{Nilsson}, K., {Pursimo}, T., {Takalo}, L.~O., {et~al.} 1999, \pasp, 111, 1223

\bibitem[{{Nilsson} {et~al.}(2009){Nilsson}, {Pursimo}, {Villforth},
  {Lindfors}, \& {Takalo}}]{nilsson09}
{Nilsson}, K., {Pursimo}, T., {Villforth}, C., {Lindfors}, E., \& {Takalo},
  L.~O. 2009, \aap, 505, 601

\bibitem[{{Nilsson} {et~al.}(2012){Nilsson}, {Pursimo}, {Villforth},
  {Lindfors}, {Takalo}, \& {Sillanp{\"a}{\"a}}}]{nilsson12}
{Nilsson}, K., {Pursimo}, T., {Villforth}, C., {et~al.} 2012, \aap, 547, A1

\bibitem[{{Ofek} {et~al.}(2012){Ofek}, {Laher}, {Surace}, {Levitan}, {Sesar},
  {Horesh}, {Law}, {van Eyken}, {Kulkarni}, {Prince}, {Nugent}, {Sullivan},
  {Yaron}, {Pickles}, {Ag{\"u}eros}, {Arcavi}, {Bildsten}, {Bloom}, {Cenko},
  {Gal-Yam}, {Grillmair}, {Helou}, {Kasliwal}, {Poznanski}, \&
  {Quimby}}]{ofek12b}
{Ofek}, E.~O., {Laher}, R., {Surace}, J., {et~al.} 2012, \pasp, 124, 854

\bibitem[{{Padovani} \& {Giommi}(1995)}]{padovani95}
{Padovani}, P. \& {Giommi}, P. 1995, \apj, 444, 567

\bibitem[{{Pavlidou} {et~al.}(2014){Pavlidou}, {Angelakis}, {Myserlis},
  {Blinov}, {King}, {Papadakis}, {Tassis}, {Hovatta}, {Pazderska},
  {Paleologou}, {Balokovi{\'c}}, {Feiler}, {Fuhrmann}, {Khodade}, {Kus},
  {Kylafis}, {Modi}, {Panopoulou}, {Papamastorakis}, {Pazderski}, {Pearson},
  {Rajarshi}, {Ramaprakash}, {Readhead}, {Reig}, \& {Zensus}}]{pavlidou14}
{Pavlidou}, V., {Angelakis}, E., {Myserlis}, I., {et~al.} 2014, \mnras, 442,
  1693

\bibitem[{{Peter} {et~al.}(2014){Peter}, {Domainko}, {Sanchez}, {van der Wel},
  \& {G{\"a}ssler}}]{peter14}
{Peter}, D., {Domainko}, W., {Sanchez}, D.~A., {van der Wel}, A., \&
  {G{\"a}ssler}, W. 2014, \aap, 571, A41

\bibitem[{{Pian} {et~al.}(1998){Pian}, {Vacanti}, {Tagliaferri}, {Ghisellini},
  {Maraschi}, {Treves}, {Urry}, {Fiore}, {Giommi}, {Palazzi}, {Chiappetti}, \&
  {Sambruna}}]{pian98}
{Pian}, E., {Vacanti}, G., {Tagliaferri}, G., {et~al.} 1998, \apjl, 492, L17

\bibitem[{{Piner} \& {Edwards}(2014)}]{piner14}
{Piner}, B.~G. \& {Edwards}, P.~G. 2014, \apj, 797, 25

\bibitem[{{Piner} {et~al.}(2010){Piner}, {Pant}, \& {Edwards}}]{piner10}
{Piner}, B.~G., {Pant}, N., \& {Edwards}, P.~G. 2010, \apj, 723, 1150

\bibitem[{{Plotkin} {et~al.}(2010){Plotkin}, {Anderson}, {Brandt},
  {Diamond-Stanic}, {Fan}, {Hall}, {Kimball}, {Richmond}, {Schneider},
  {Shemmer}, {Voges}, {York}, {Bahcall}, {Snedden}, {Bizyaev}, {Brewington},
  {Malanushenko}, {Malanushenko}, {Oravetz}, {Pan}, \& {Simmons}}]{plotkin10}
{Plotkin}, R.~M., {Anderson}, S.~F., {Brandt}, W.~N., {et~al.} 2010, \aj, 139,
  390

\bibitem[{{Reinthal} {et~al.}(2012){Reinthal}, {Lindfors}, {Mazin}, {Nilsson},
  {Takalo}, {Sillanp{\"a}{\"a}}, {Berdyugin}, \& {MAGIC
  Collaboration}}]{reinthal12}
{Reinthal}, R., {Lindfors}, E.~J., {Mazin}, D., {et~al.} 2012, Journal of
  Physics Conference Series, 355, 012013

\bibitem[{{Rovero} {et~al.}(2016){Rovero}, {Muriel}, {Donzelli}, \&
  {Pichel}}]{rovero16}
{Rovero}, A.~C., {Muriel}, H., {Donzelli}, C., \& {Pichel}, A. 2016, ArXiv
  e-prints [\eprint[arXiv]{1602.08364}]

\bibitem[{{Sakimoto} {et~al.}(2013){Sakimoto}, {Uemura}, {Sasada}, {Kawabata},
  {Fukazawa}, {Yamanaka}, {Itoh}, {Ohsugi}, {Yoshida}, {Akitaya}, {Sato}, \&
  {Kino}}]{sakimoto13}
{Sakimoto}, K., {Uemura}, M., {Sasada}, M., {et~al.} 2013, \pasj, 65

\bibitem[{{Savolainen} {et~al.}(2006){Savolainen}, {Wiik}, {Valtaoja},
  {Kadler}, {Ros}, {Tornikoski}, {Aller}, \& {Aller}}]{savolainen06}
{Savolainen}, T., {Wiik}, K., {Valtaoja}, E., {et~al.} 2006, \apj, 647, 172

\bibitem[{{Scarpa} {et~al.}(2000){Scarpa}, {Urry}, {Falomo}, {Pesce}, \&
  {Treves}}]{scarpa00}
{Scarpa}, R., {Urry}, C.~M., {Falomo}, R., {Pesce}, J.~E., \& {Treves}, A.
  2000, \apj, 532, 740

\bibitem[{{Schlafly} \& {Finkbeiner}(2011)}]{schlafly11}
{Schlafly}, E.~F. \& {Finkbeiner}, D.~P. 2011, \apj, 737, 103

\bibitem[{{Shaw} {et~al.}(2013{\natexlab{a}}){Shaw}, {Filippenko}, {Romani},
  {Cenko}, \& {Li}}]{shaw13b}
{Shaw}, M.~S., {Filippenko}, A.~V., {Romani}, R.~W., {Cenko}, S.~B., \& {Li},
  W. 2013{\natexlab{a}}, \aj, 146, 127

\bibitem[{{Shaw} {et~al.}(2013{\natexlab{b}}){Shaw}, {Romani}, {Cotter},
  {Healey}, {Michelson}, {Readhead}, {Richards}, {Max-Moerbeck}, {King}, \&
  {Potter}}]{shaw13}
{Shaw}, M.~S., {Romani}, R.~W., {Cotter}, G., {et~al.} 2013{\natexlab{b}},
  \apj, 764, 135

\bibitem[{{Simmons} \& {Stewart}(1985)}]{simmons85}
{Simmons}, J.~F.~L. \& {Stewart}, B.~G. 1985, \aap, 142, 100

\bibitem[{{Sitarek} {et~al.}(2015){Sitarek}, {Becerra Gonz{\'a}lez}, {Dominis
  Prester}, {Lindfors}, {Manganaro}, {Mazin}, {Nievas Rosillo}, {Stamerra},
  {Ievgen Vovk for the MAGIC Collaboration}, \& {Sara Buson for the Fermi-LAT
  Collaboration}}]{sitarek15}
{Sitarek}, J., {Becerra Gonz{\'a}lez}, J., {Dominis Prester}, D., {et~al.}
  2015, Proceedings of the 34th International Cosmic Ray Conference
  [\eprint[arXiv]{1508.04580}]

\bibitem[{{Smith} {et~al.}(2007){Smith}, {Williams}, {Schmidt},
  {Diamond-Stanic}, \& {Means}}]{smith07}
{Smith}, P.~S., {Williams}, G.~G., {Schmidt}, G.~D., {Diamond-Stanic}, A.~M.,
  \& {Means}, D.~L. 2007, \apj, 663, 118

\bibitem[{{Stickel} {et~al.}(1991){Stickel}, {Fried}, {Kuehr}, {Padovani}, \&
  {Urry}}]{stickel91}
{Stickel}, M., {Fried}, J.~W., {Kuehr}, H., {Padovani}, P., \& {Urry}, C.~M.
  1991, \apj, 374, 431

\bibitem[{{Stocke} {et~al.}(1985){Stocke}, {Liebert}, {Schmidt}, {Gioia},
  {Maccacaro}, {Schild}, {Maccagni}, \& {Arp}}]{stocke85}
{Stocke}, J.~T., {Liebert}, J., {Schmidt}, G., {et~al.} 1985, \apj, 298, 619

\bibitem[{{Stocke} {et~al.}(1991){Stocke}, {Morris}, {Gioia}, {Maccacaro},
  {Schild}, {Wolter}, {Fleming}, \& {Henry}}]{stocke91}
{Stocke}, J.~T., {Morris}, S.~L., {Gioia}, I.~M., {et~al.} 1991, \apjs, 76, 813

\bibitem[{{Straal} {et~al.}(2016){Straal}, {Gabanyi}, {van Leeuwen}, {Clarke},
  {Dubner}, {Frey}, {Giacani}, \& {Paragi}}]{straal16}
{Straal}, S.~M., {Gabanyi}, K.~E., {van Leeuwen}, J., {et~al.} 2016, ArXiv
  e-prints [\eprint[arXiv]{1603.01226}]

\bibitem[{{Takalo} {et~al.}(2008){Takalo}, {Nilsson}, {Lindfors},
  {Sillanp{\"a}{\"a}}, {Berdyugin}, \& {Pasanen}}]{takalo08}
{Takalo}, L.~O., {Nilsson}, K., {Lindfors}, E., {et~al.} 2008, in American
  Institute of Physics Conference Series, Vol. 1085, American Institute of
  Physics Conference Series, ed. F.~A. {Aharonian}, W.~{Hofmann}, \&
  F.~{Rieger}, 705--707

\bibitem[{{The Fermi-LAT Collaboration}(2015)}]{2FHL2015}
{The Fermi-LAT Collaboration}. 2015, ArXiv e-prints
  [\eprint[arXiv]{1508.04449}]

\bibitem[{{Urry} {et~al.}(2000){Urry}, {Scarpa}, {O'Dowd}, {Falomo}, {Pesce},
  \& {Treves}}]{urry00}
{Urry}, C.~M., {Scarpa}, R., {O'Dowd}, M., {et~al.} 2000, \apj, 532, 816

\bibitem[{{Villforth} {et~al.}(2010){Villforth}, {Nilsson}, {Heidt}, {Takalo},
  {Pursimo}, {Berdyugin}, {Lindfors}, {Pasanen}, {Winiarski}, {Drozdz},
  {Ogloza}, {Kurpinska-Winiarska}, {Siwak}, {Koziel-Wierzbowska}, {Porowski},
  {Kuzmicz}, {Krzesinski}, {Kundera}, {Wu}, {Zhou}, {Efimov}, {Sadakane},
  {Kamada}, {Ohlert}, {Hentunen}, {Nissinen}, {Dietrich}, {Assef}, {Atlee},
  {Bird}, {Depoy}, {Eastman}, {Peeples}, {Prieto}, {Watson}, {Yee}, {Liakos},
  {Niarchos}, {Gazeas}, {Dogru}, {Donmez}, {Marchev}, {Coggins-Hill},
  {Mattingly}, {Keel}, {Haque}, {Aungwerojwit}, \& {Bergvall}}]{villforth10}
{Villforth}, C., {Nilsson}, K., {Heidt}, J., {et~al.} 2010, \mnras, 402, 2087

\bibitem[{{Wierzcholska} {et~al.}(2015){Wierzcholska}, {Ostrowski}, {Stawarz},
  {Wagner}, \& {Hauser}}]{wierzcholska15}
{Wierzcholska}, A., {Ostrowski}, M., {Stawarz}, {\L}., {Wagner}, S., \&
  {Hauser}, M. 2015, \aap, 573, A69

\end{thebibliography}
%
% - join the .bib files when you upload your source files
%-------------------------------------------------------------------

\clearpage
\appendix
\section{Polarization curves of all the sources}
 In this appendix we show plots of the polarization fraction, EVPA, and corresponding Stokes parameters for all sources discussed in this paper.
\begin{figure}
\includegraphics[width=0.45\textwidth]{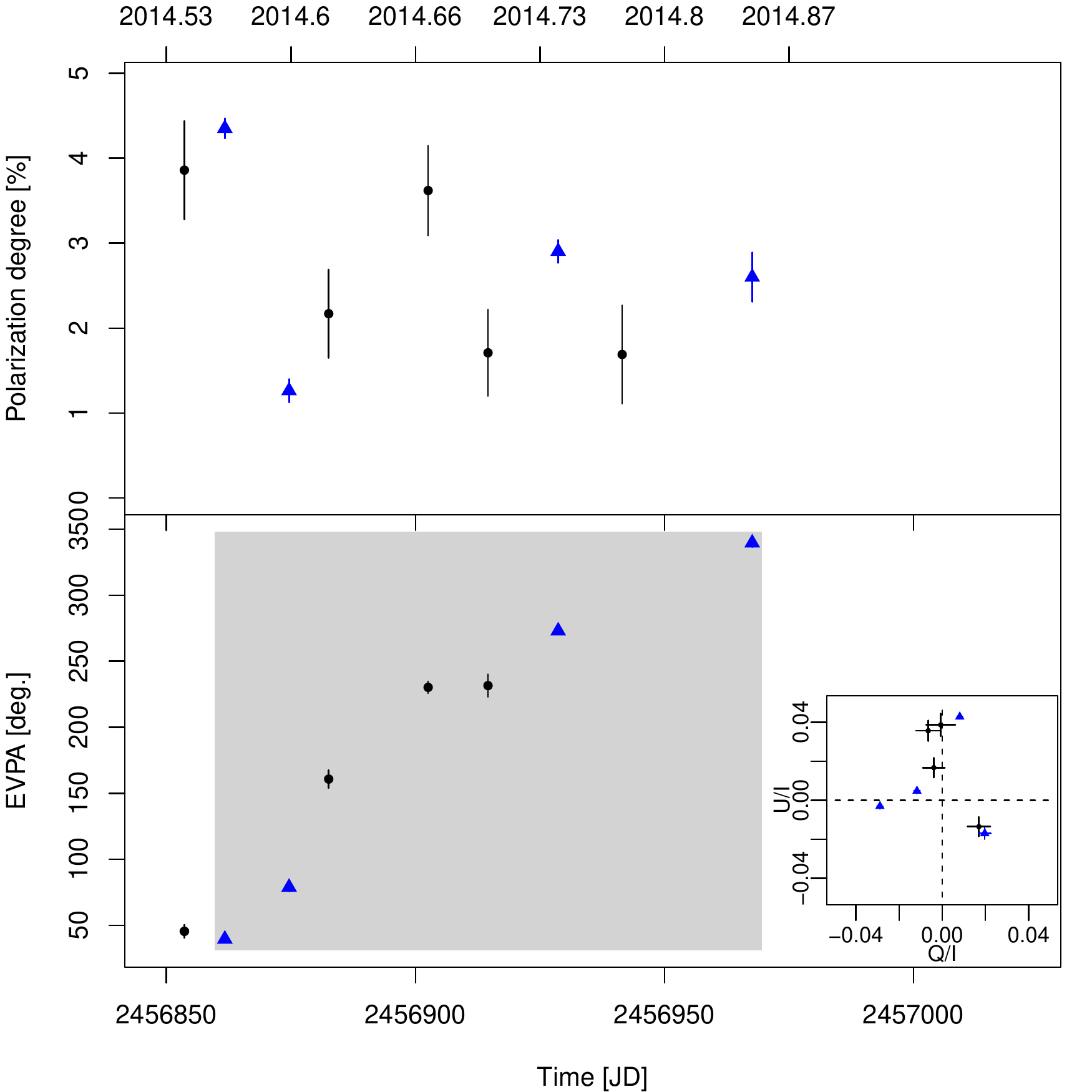}
\caption{Fractional polarization (top) and EVPA (bottom) of the TeV
  source J0136+3905. Black circles are RoboPol data and blue
  triangles NOT data. The shaded region shows the period of a
  significant EVPA rotation. The inset in the lower panel shows the
  $Q/I$ vs. $U/I$. EVPA data are shown only for
  observations where the signal-to-noise in the polarization fraction
  $\ge 3$.}
\label{Fig:lc1}
\end{figure}

\begin{figure}
\includegraphics[width=0.45\textwidth]{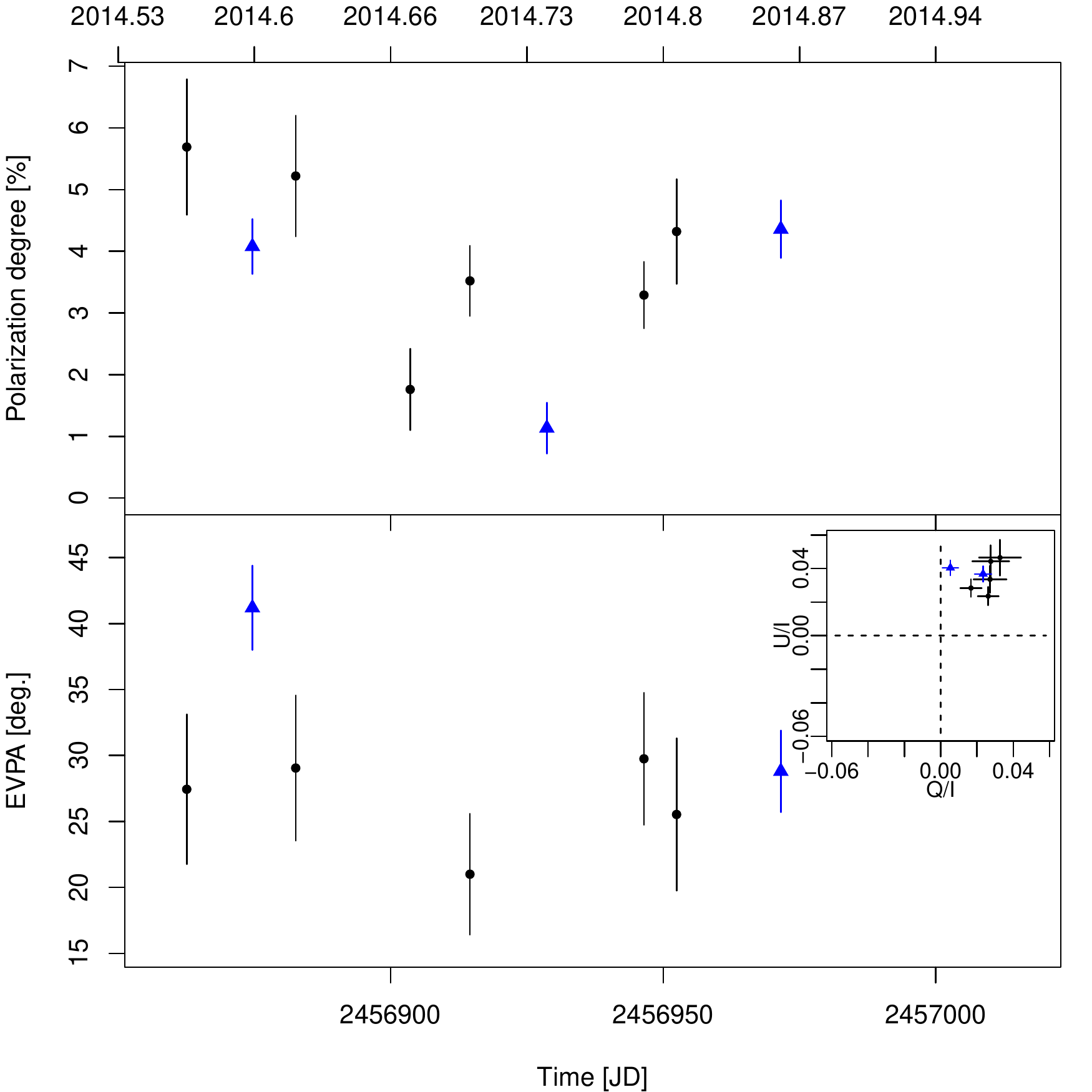}
\caption{Fractional polarization (top) and EVPA (bottom) of the TeV
  source J0152+0146. Black circles are RoboPol data and blue
  triangles NOT data. The inset in the lower panel shows the
  $Q/I$ vs. $U/I$. EVPA data are shown only for
  observations where the signal-to-noise in the polarization fraction
  $\ge 3$.}
\label{Fig:lc2}
\end{figure}

\begin{figure}
\includegraphics[width=0.45\textwidth]{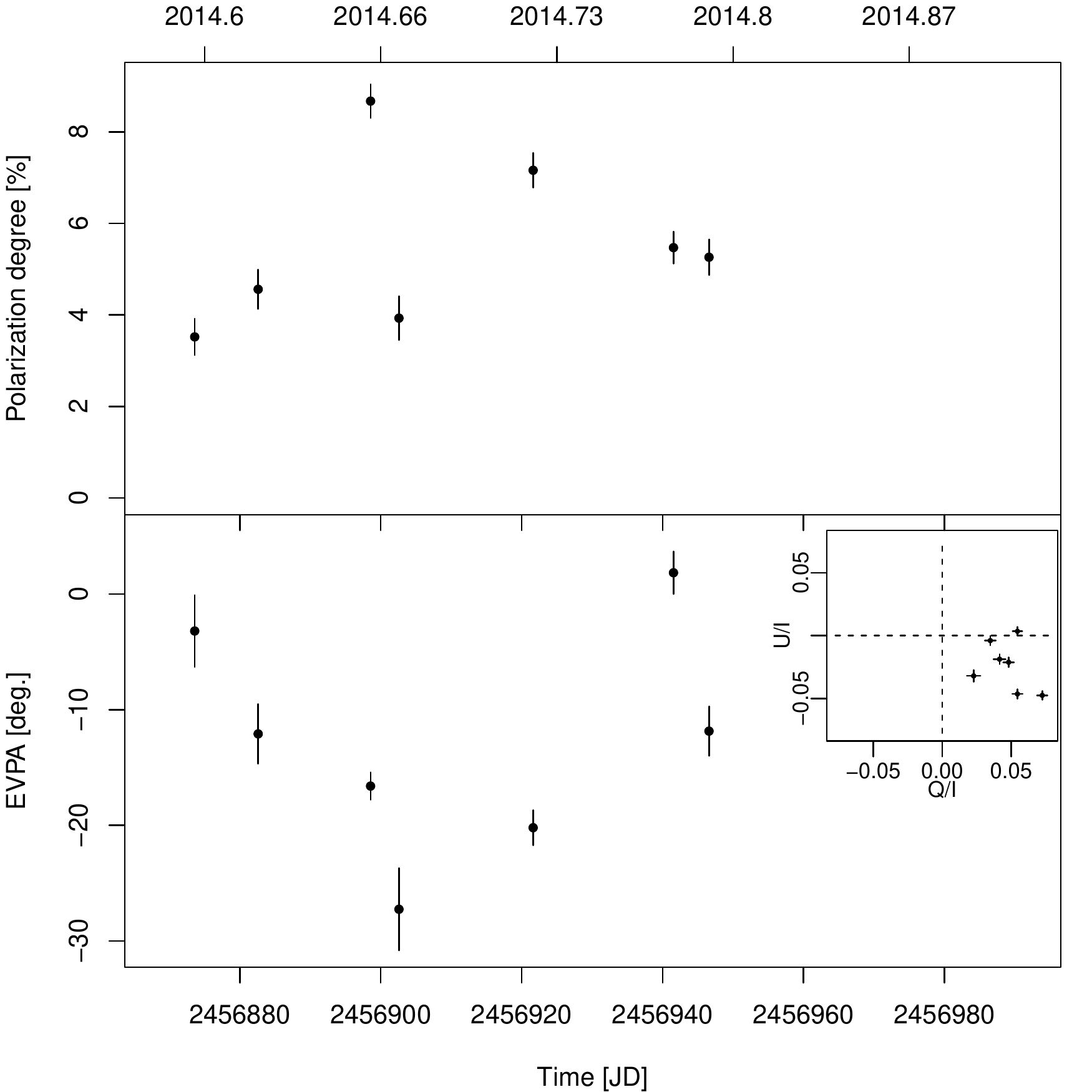}
\caption{Fractional polarization (top) and EVPA (bottom) of the TeV
  source J0222+4302. The inset in the lower panel shows 
  $Q/I$ vs. $U/I$. EVPA data are shown only for
  observations where the signal-to-noise in the polarization fraction
  $\ge 3$.}
\label{Fig:lc3}
\end{figure}

\begin{figure}
\includegraphics[width=0.45\textwidth]{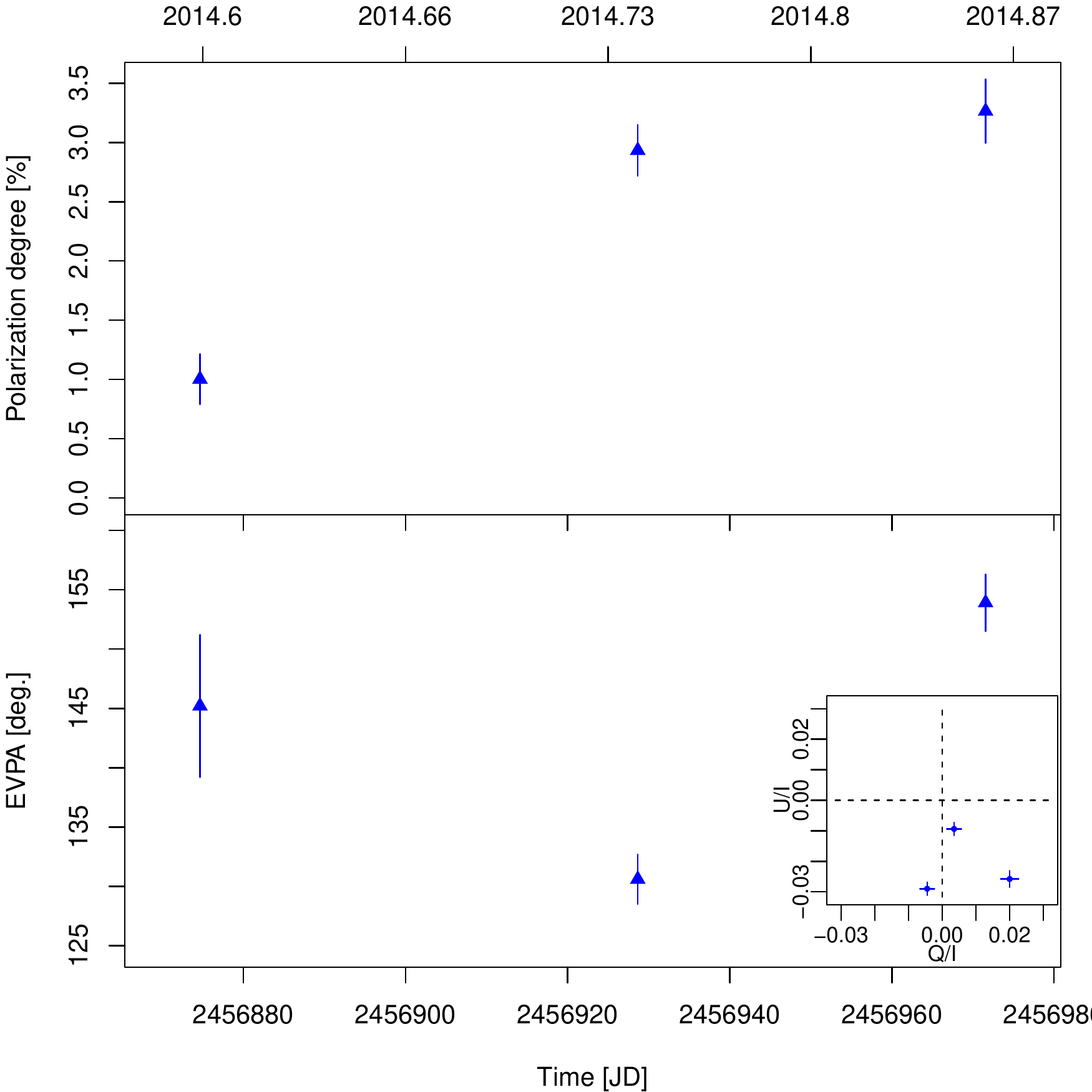}
\caption{Fractional polarization (top) and EVPA (bottom) of the TeV
  source J0232+2017. The inset in the lower panel shows 
  $Q/I$ vs. $U/I$. EVPA data are shown only for
  observations where the signal-to-noise in the polarization fraction
  $\ge 3$. }
\label{Fig:lc4}
\end{figure}

\begin{figure}
\includegraphics[width=0.45\textwidth]{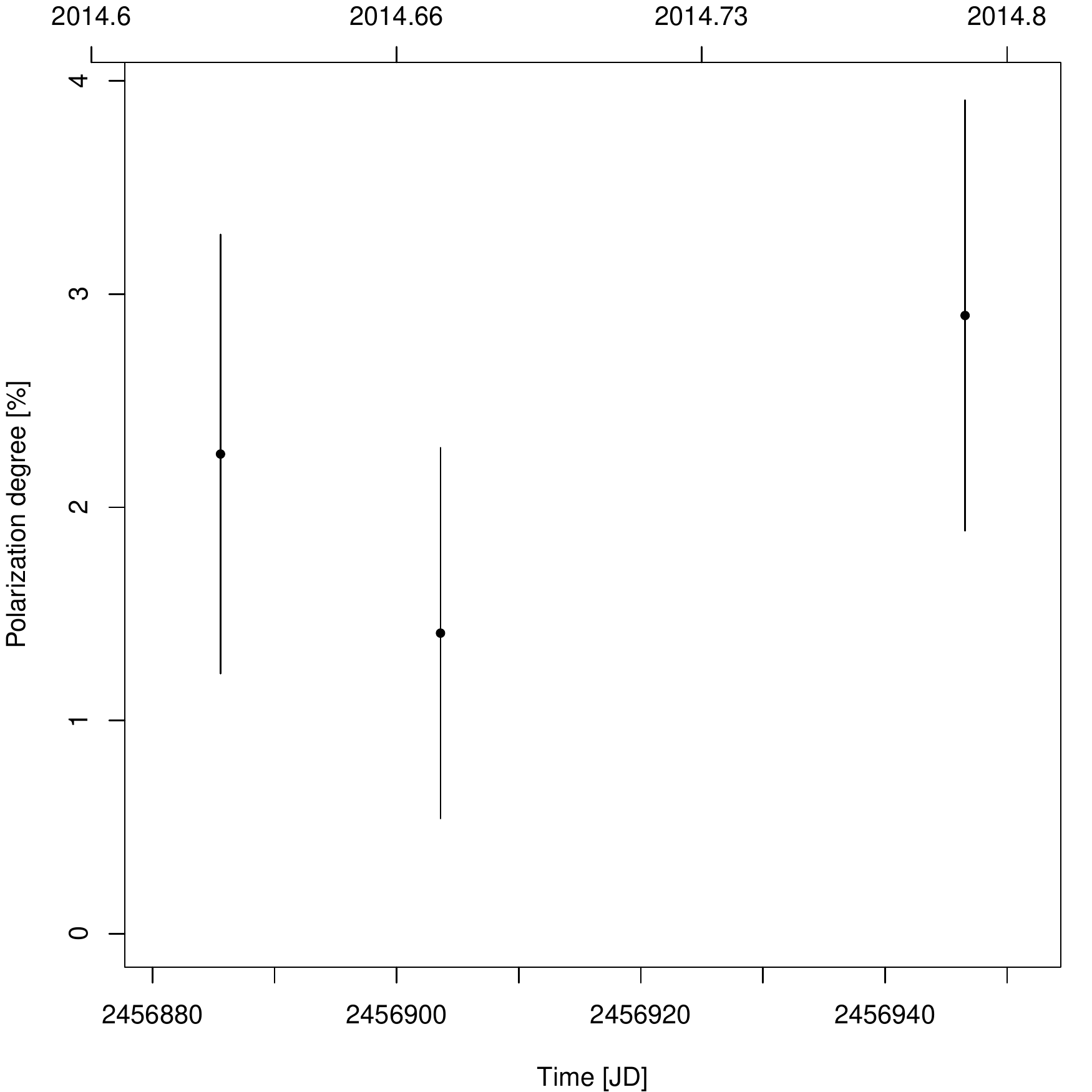}
\caption{Fractional polarization (top) and EVPA (bottom) of the TeV
  source J0319+1845. None of the observations have polarization fraction
  $\ge 3$ and no EVPA data are shown.}
\label{Fig:lc5}
\end{figure}

\begin{figure}
\includegraphics[width=0.45\textwidth]{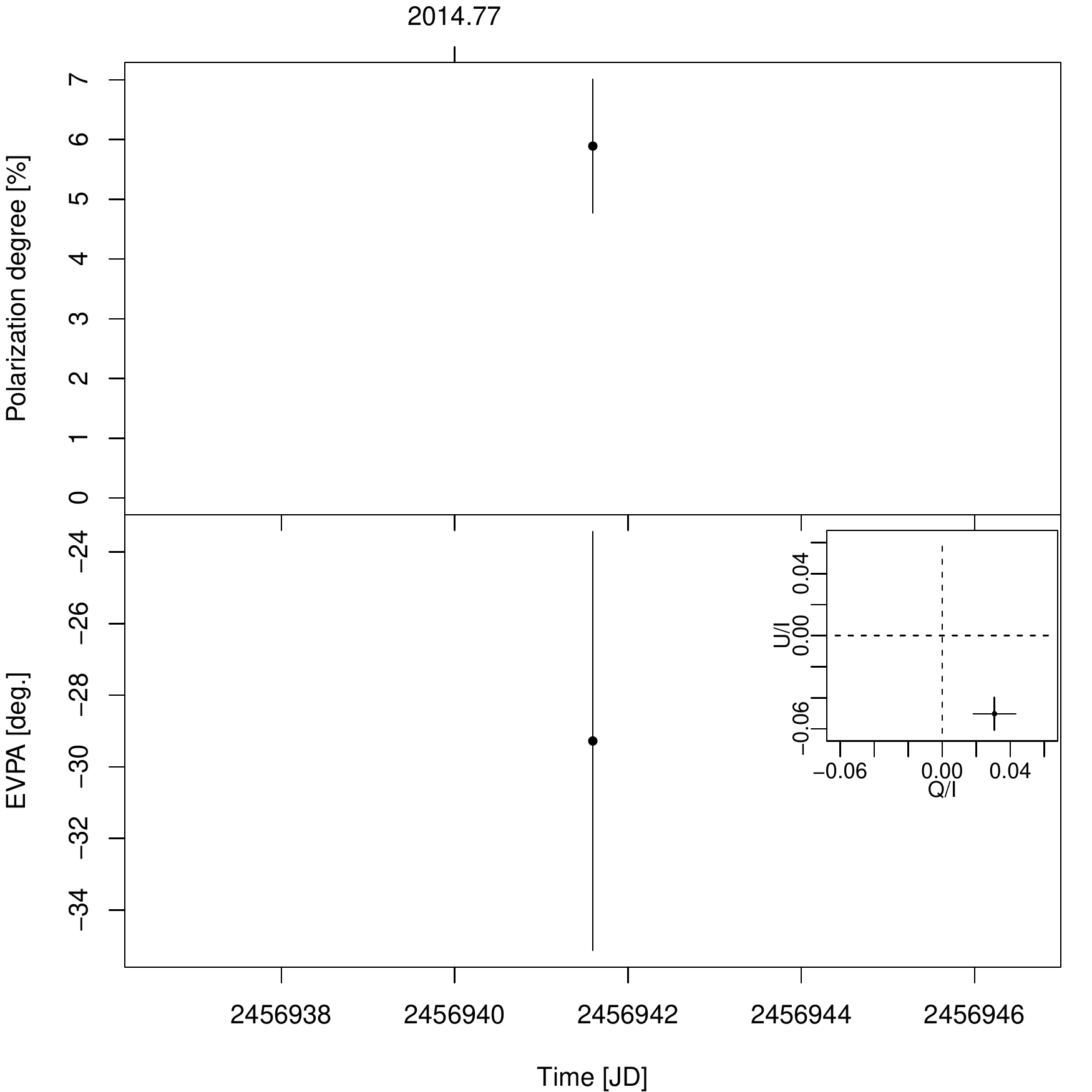}
\caption{Fractional polarization (top) and EVPA (bottom) of the TeV
  source J0416+0105. The inset in the lower panel shows
  $Q/I$ vs. $U/I$. EVPA data are shown only for
  observations where the signal-to-noise in the polarization fraction
  $\ge 3$.}
\label{Fig:lc6}
\end{figure}

\begin{figure}
\includegraphics[width=0.45\textwidth]{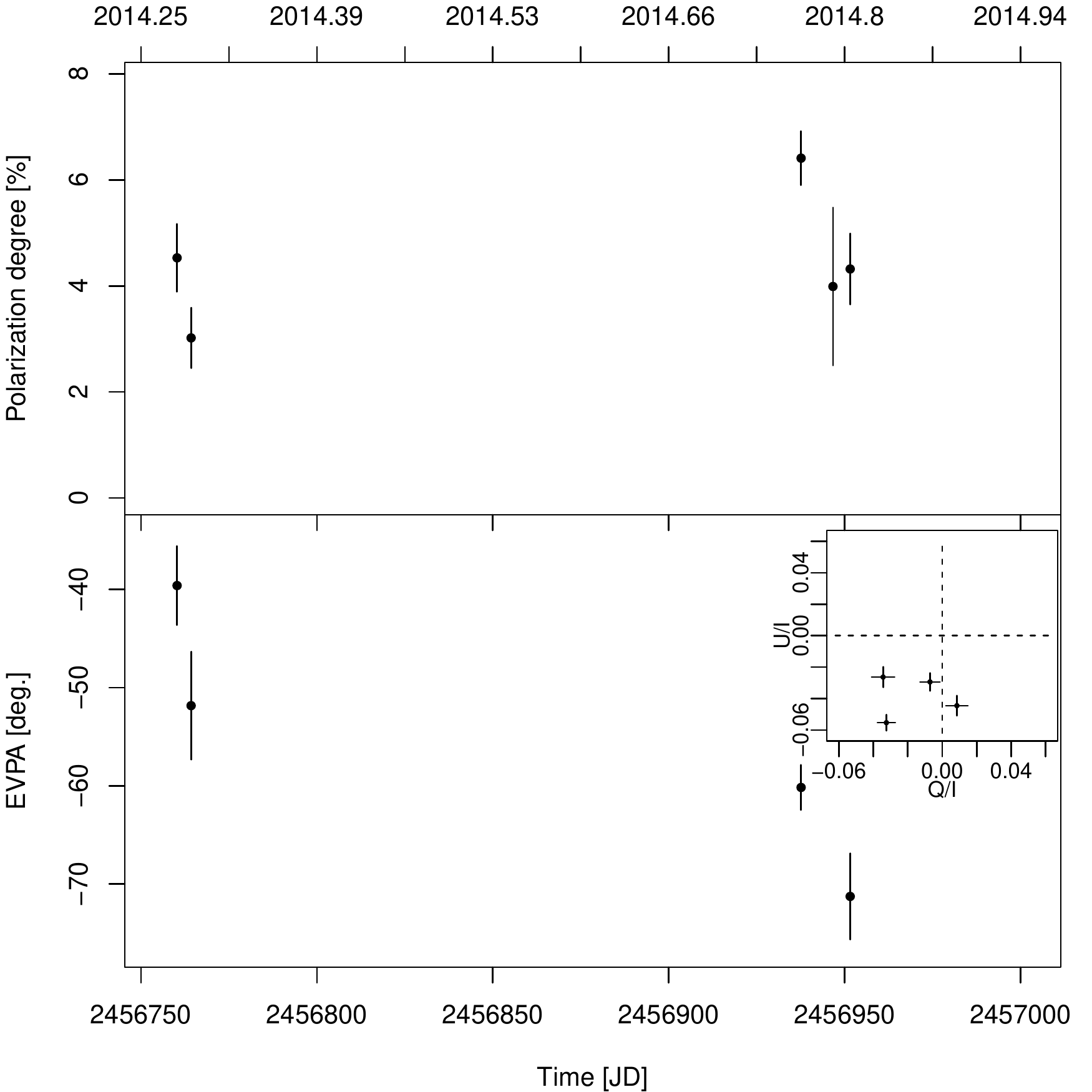}
\caption{Fractional polarization (top) and EVPA (bottom) of the TeV
  source J0507+6737. The inset in the lower panel shows 
  $Q/I$ vs. $U/I$. EVPA data are shown only for
  observations where the signal-to-noise in the polarization fraction
  $\ge 3$.}
\label{Fig:lc7}
\end{figure}

\begin{figure}
\includegraphics[width=0.45\textwidth]{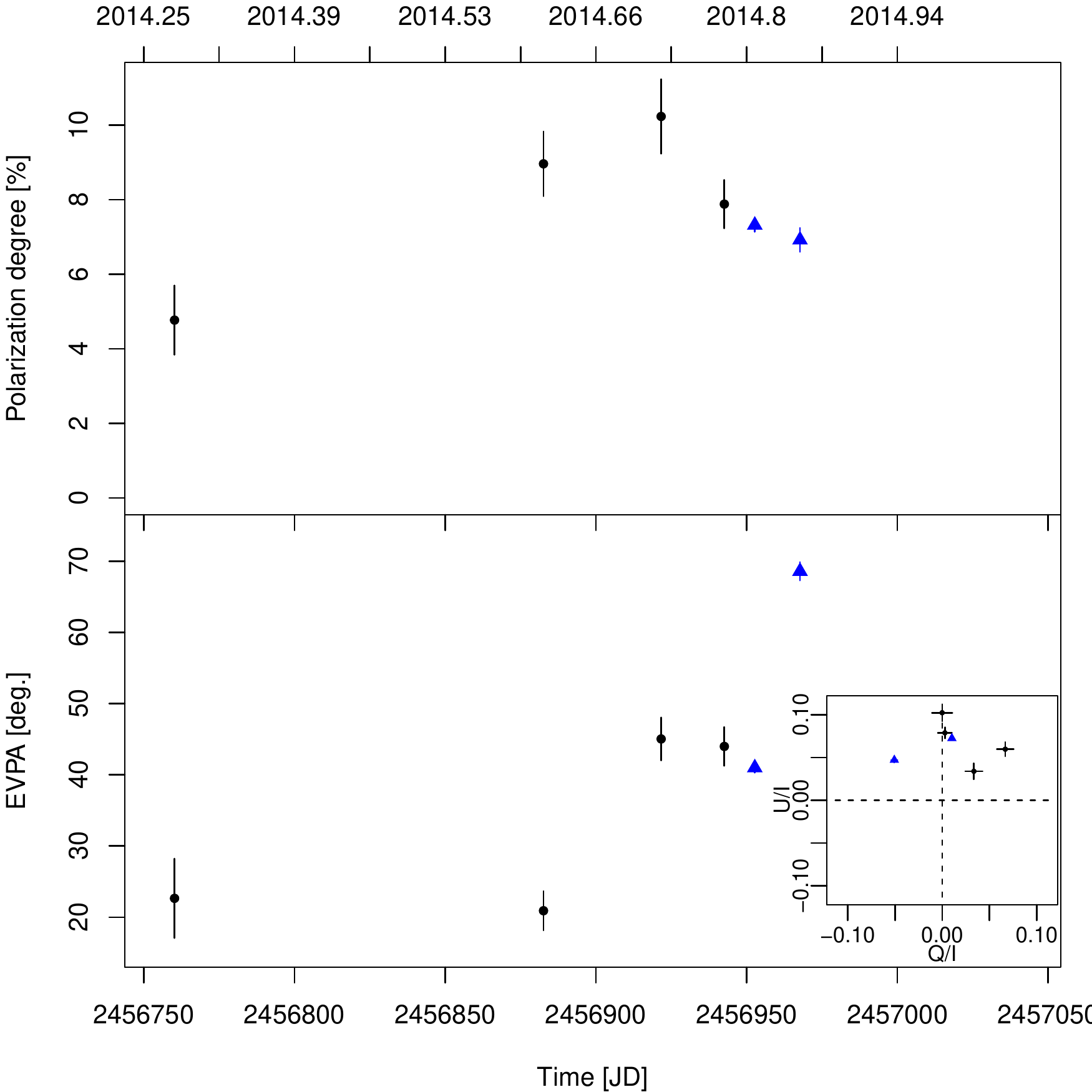}
\caption{Fractional polarization (top) and EVPA (bottom) of the TeV
  source J0521+2112. The inset in the lower panel shows 
  $Q/I$ vs. $U/I$. EVPA data are shown only for
  observations where the signal-to-noise in the polarization fraction
  $\ge 3$.}
\label{Fig:lc8}
\end{figure}

\begin{figure}
\includegraphics[width=0.45\textwidth]{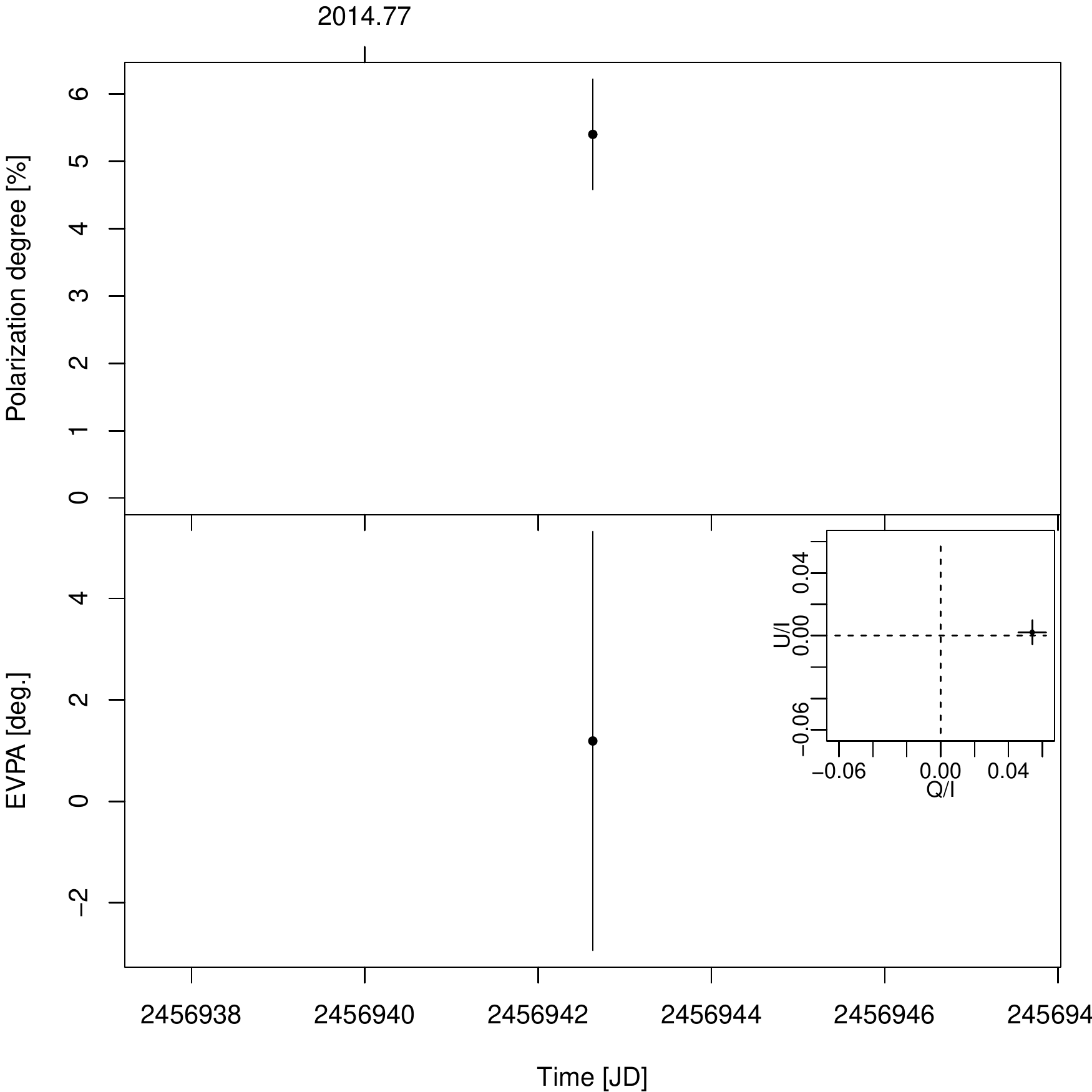}
\caption{Fractional polarization (top) and EVPA (bottom) of the TeV
  source J0648+1516. The inset in the lower panel shows 
  $Q/I$ vs. $U/I$. EVPA data are shown only for
  observations where the signal-to-noise in the polarization fraction
  $\ge 3$.}
\label{Fig:lc9}
\end{figure}

\begin{figure}
\includegraphics[width=0.45\textwidth]{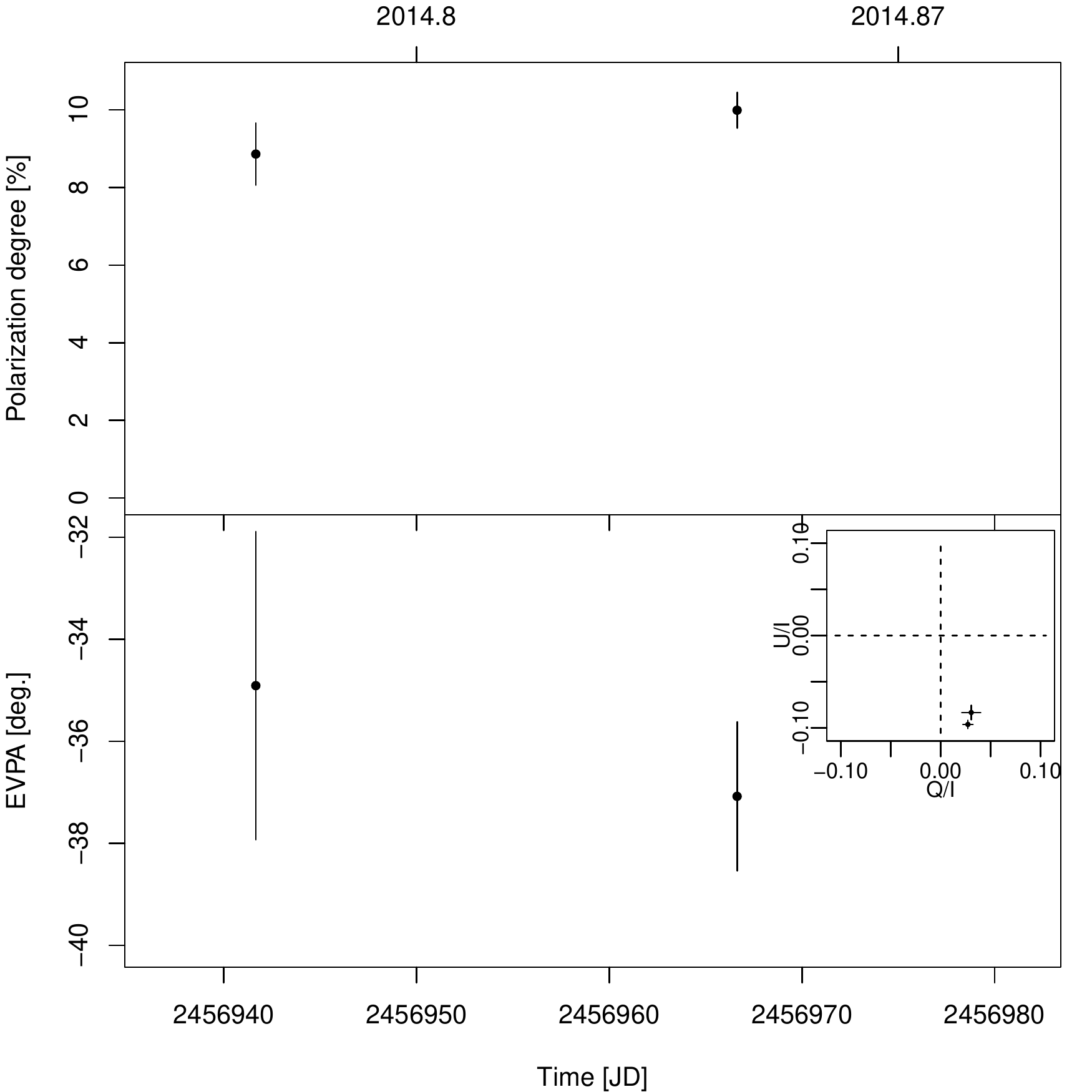}
\caption{Fractional polarization (top) and EVPA (bottom) of the TeV
  source J0650+2502. The inset in the lower panel shows 
  $Q/I$ vs. $U/I$. EVPA data are shown only for
  observations where the signal-to-noise in the polarization fraction
  $\ge 3$.}
\label{Fig:lc10}
\end{figure}

\begin{figure}
\includegraphics[width=0.45\textwidth]{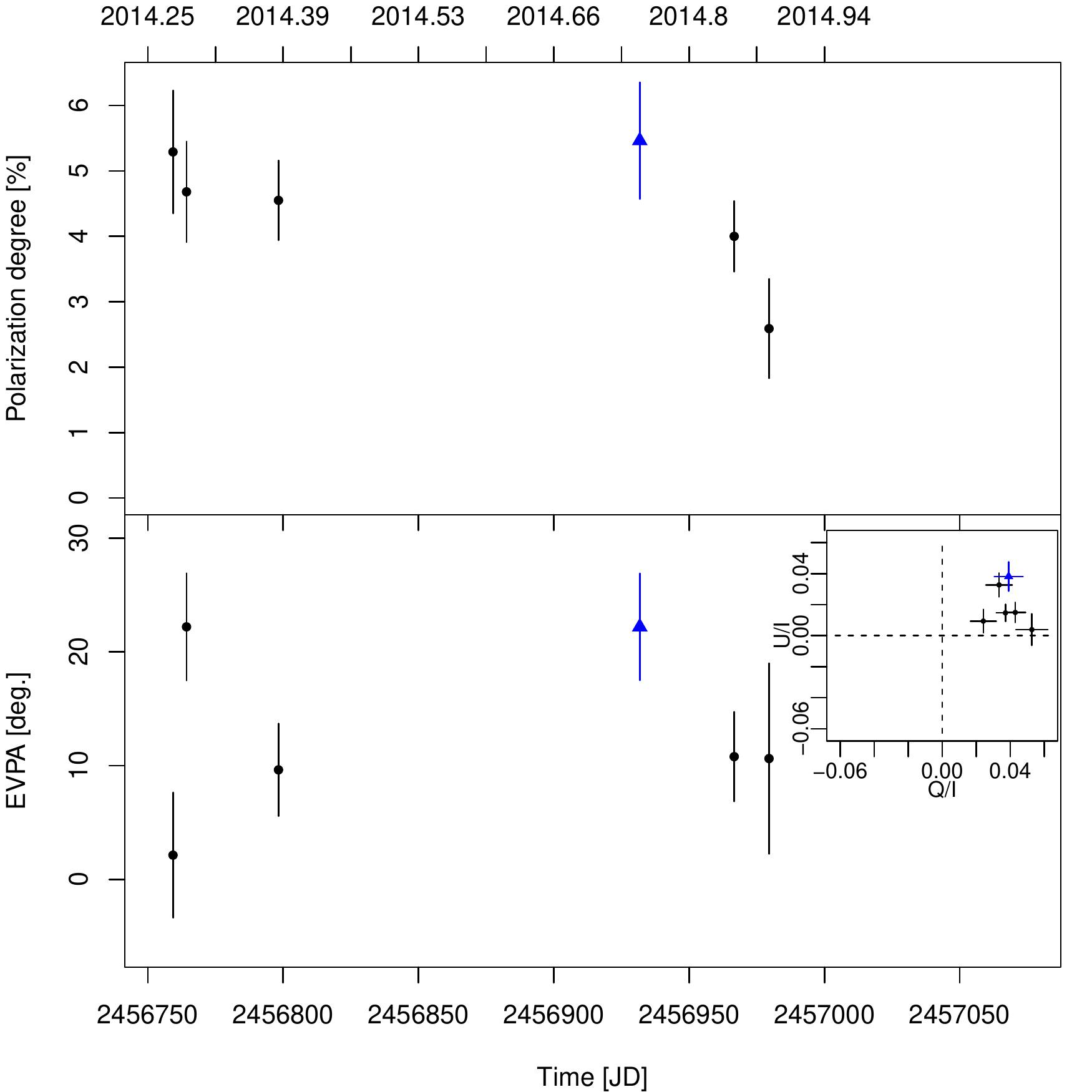}
\caption{Fractional polarization (top) and EVPA (bottom) of the TeV
  source J0710+5908. The inset in the lower panel shows 
  $Q/I$ vs. $U/I$. EVPA data are shown only for
  observations where the signal-to-noise in the polarization fraction
  $\ge 3$.}
\label{Fig:lc11}
\end{figure}

\begin{figure}
\includegraphics[width=0.45\textwidth]{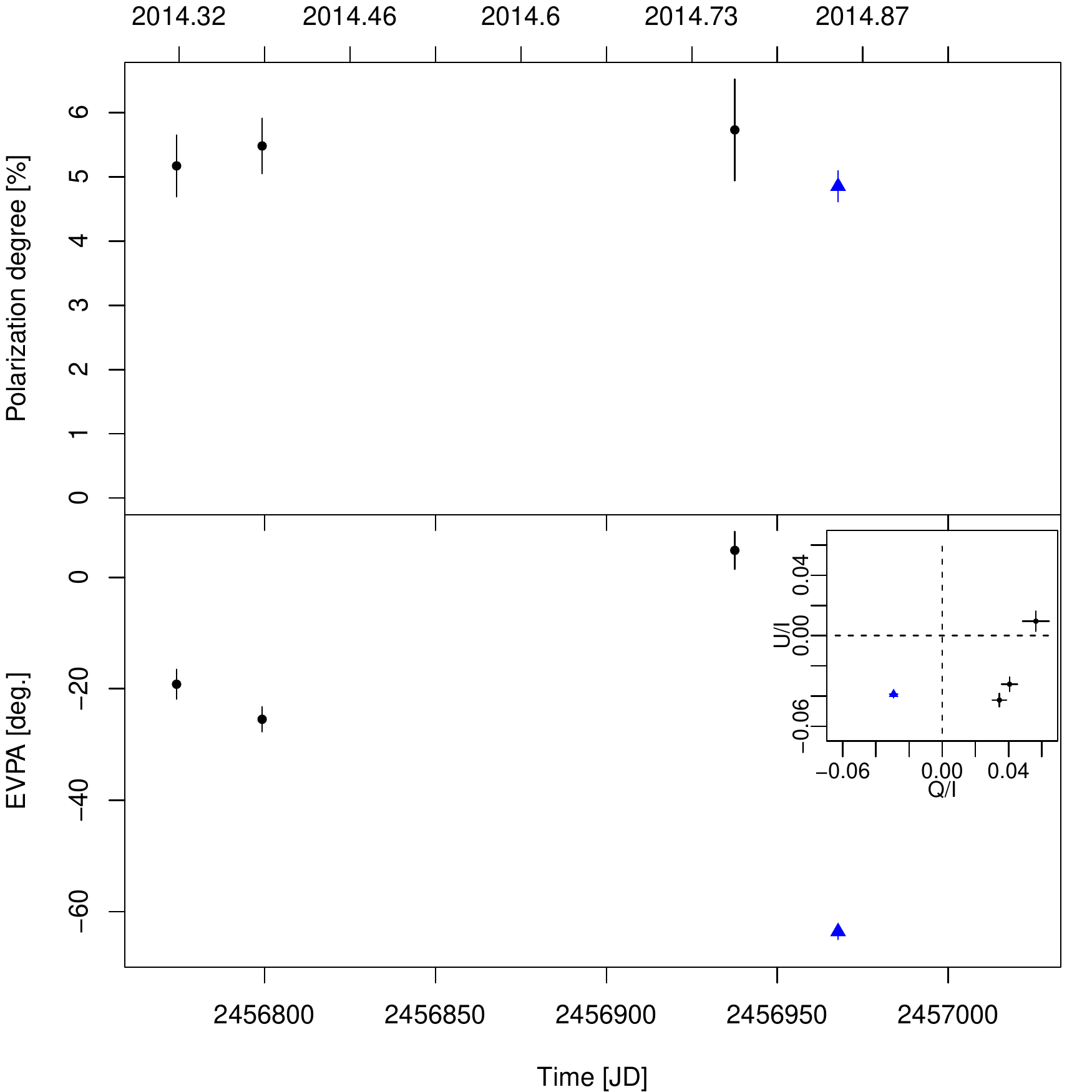}
\caption{Fractional polarization (top) and EVPA (bottom) of the TeV
  source J0809+5218. The inset in the lower panel shows 
  $Q/I$ vs. $U/I$. EVPA data are shown only for
  observations where the signal-to-noise in the polarization fraction
  $\ge 3$.}
\label{Fig:lc12}
\end{figure}

\begin{figure}
\includegraphics[width=0.45\textwidth]{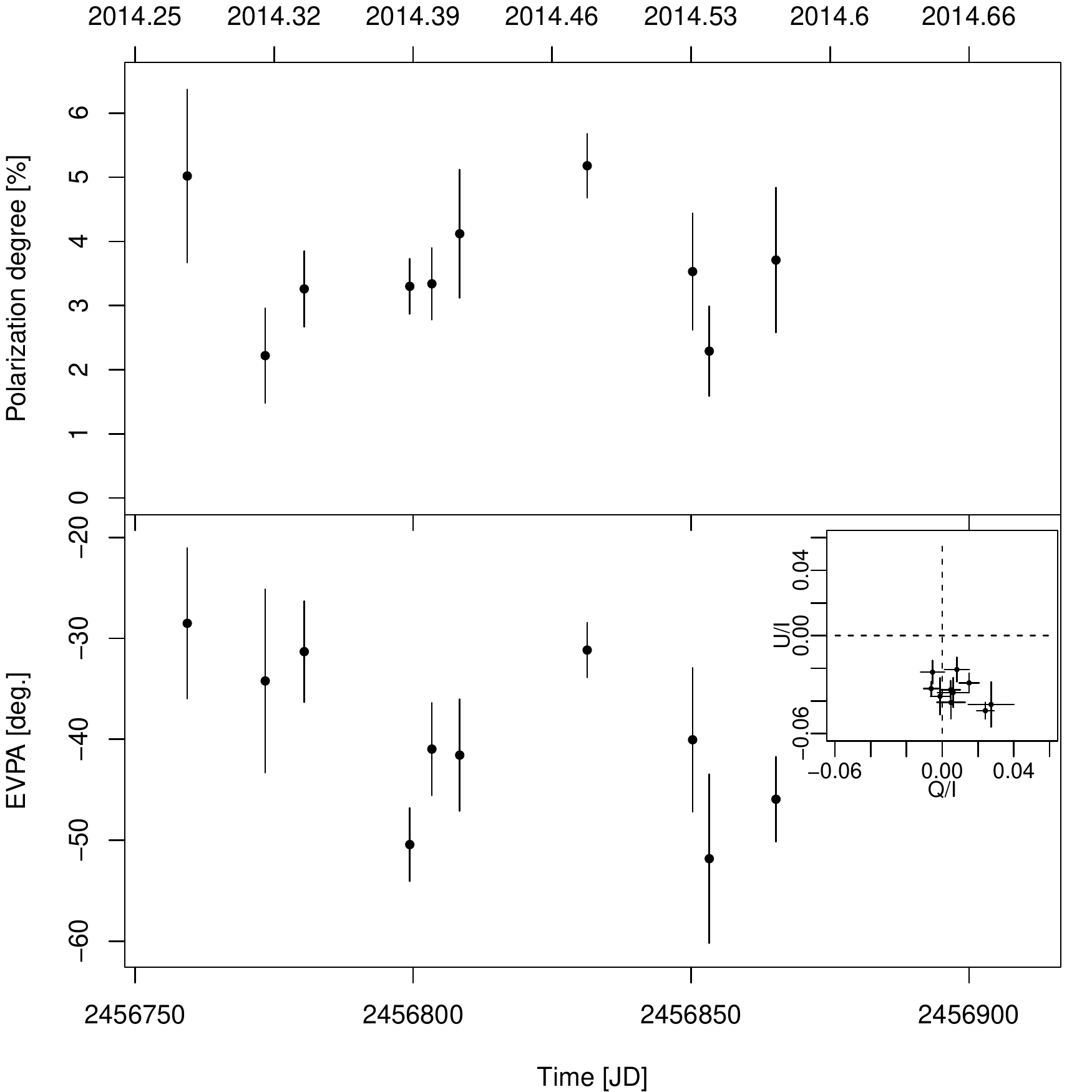}
\caption{Fractional polarization (top) and EVPA (bottom) of the TeV
  source J1136+7009. The inset in the lower panel shows 
  $Q/I$ vs. $U/I$. EVPA data are shown only for
  observations where the signal-to-noise in the polarization fraction
  $\ge 3$.}
\label{Fig:lc13}
\end{figure}

\begin{figure}
\includegraphics[width=0.45\textwidth]{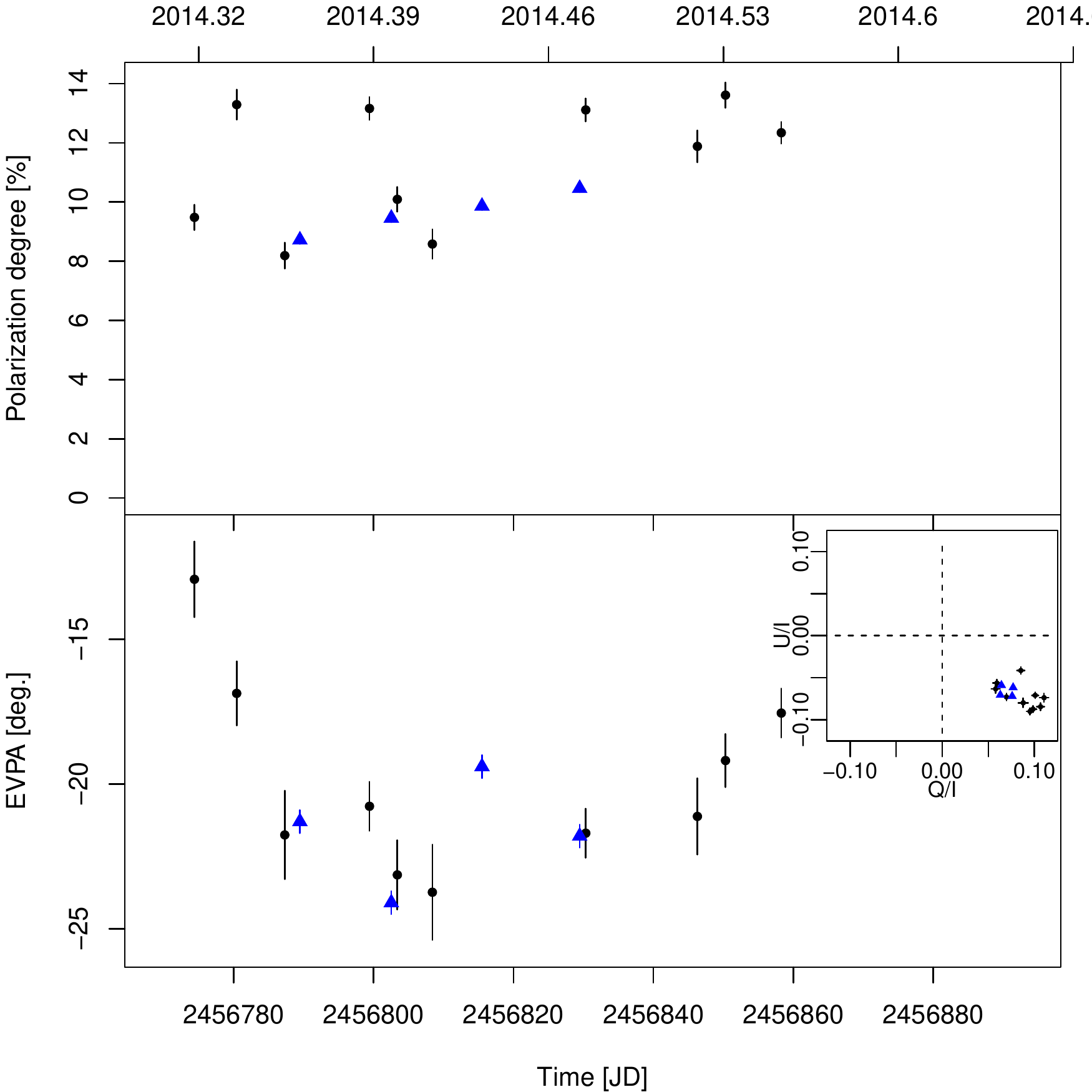}
\caption{Fractional polarization (top) and EVPA (bottom) of the TeV
  source J1217+3007. Black circles are RoboPol data and blue
  triangles NOT data. The inset in the lower panel shows 
  $Q/I$ vs. $U/I$. EVPA data are shown only for
  observations where the signal-to-noise in the polarization fraction
  $\ge 3$.}
\label{Fig:lc14}
\end{figure}

\begin{figure}
\includegraphics[width=0.45\textwidth]{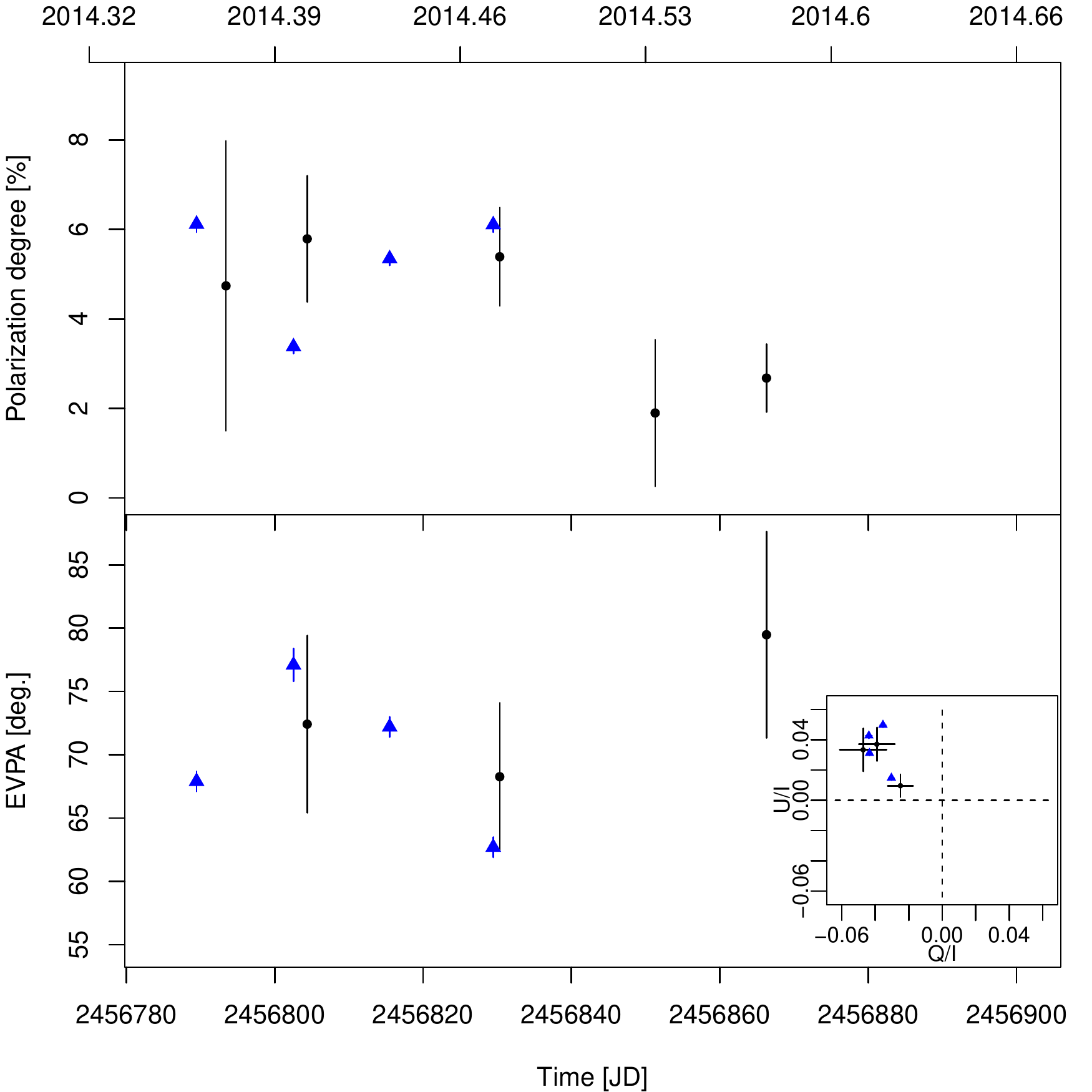}
\caption{Fractional polarization (top) and EVPA (bottom) of the TeV
  source J1221+3010. Black circles are RoboPol data and blue
  triangles NOT data. The inset in the lower panel shows 
  $Q/I$ vs. $U/I$. EVPA data are shown only for
  observations where the signal-to-noise in the polarization fraction
  $\ge 3$.}
\label{Fig:lc15}
\end{figure}

\begin{figure}
\includegraphics[width=0.45\textwidth]{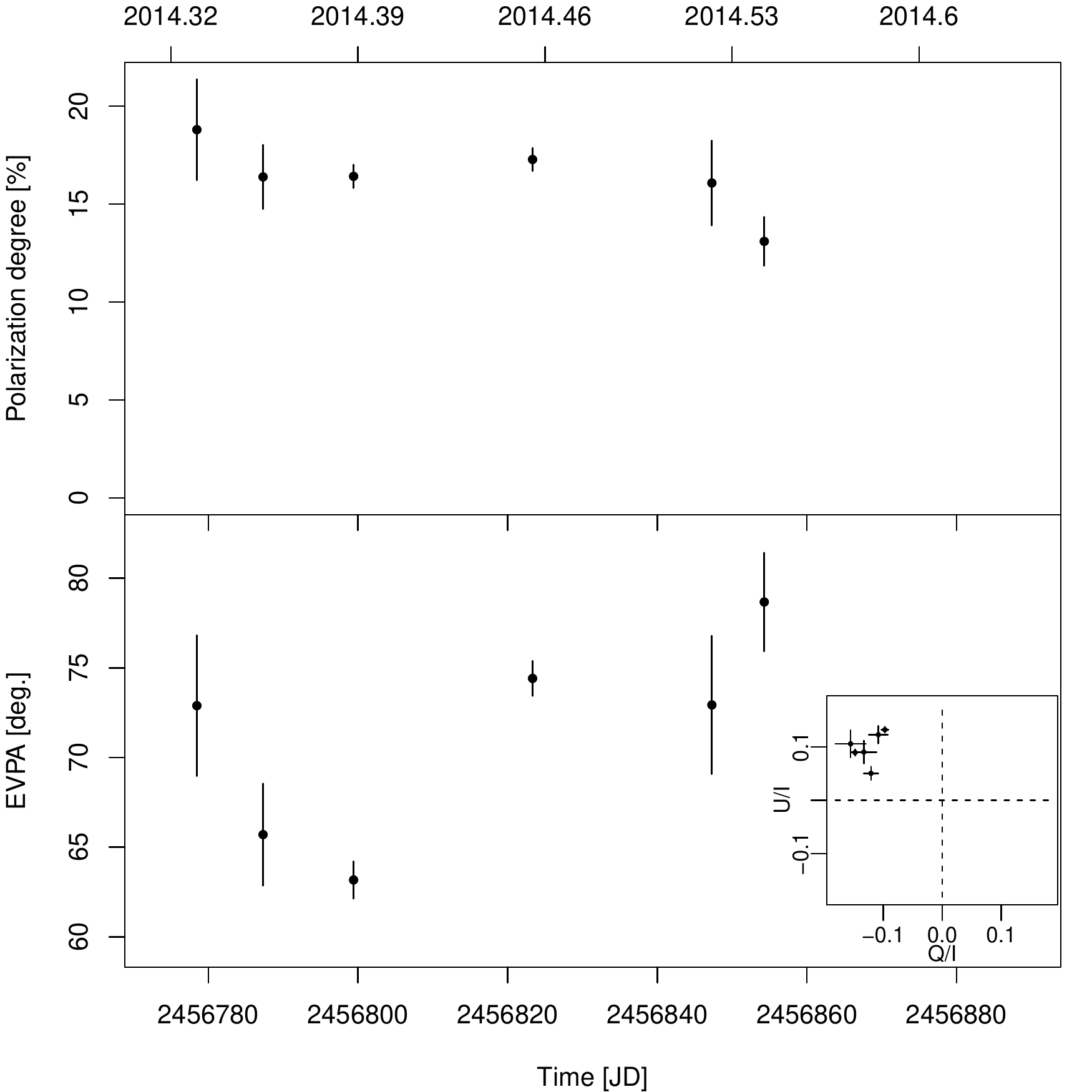}
\caption{Fractional polarization (top) and EVPA (bottom) of the TeV
  source J1221+2813. The inset in the lower panel shows 
  $Q/I$ vs. $U/I$. EVPA data are shown only for
  observations where the signal-to-noise in the polarization fraction
  $\ge 3$.}
\label{Fig:lc16}
\end{figure}

\clearpage

\begin{figure}
\includegraphics[width=0.45\textwidth]{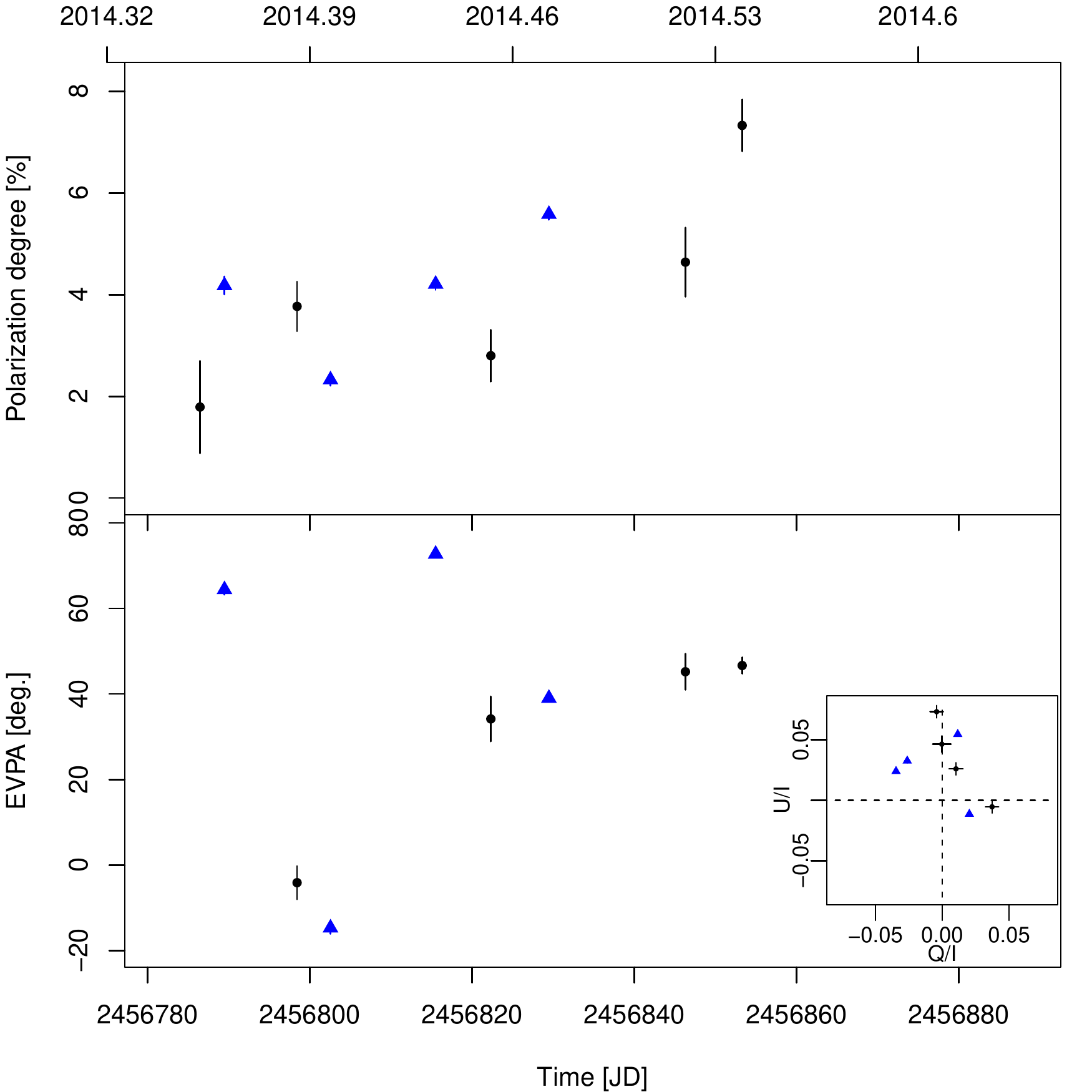}
\caption{Fractional polarization (top) and EVPA (bottom) of the TeV
  source J1224+2436. Black circles are RoboPol data and blue
  triangles NOT data. The inset in the lower panel shows 
  $Q/I$ vs. $U/I$. EVPA data are shown only for
  observations where the signal-to-noise in the polarization fraction
  $\ge 3$.}
\label{Fig:lc17}
\end{figure}

\begin{figure}
\includegraphics[width=0.45\textwidth]{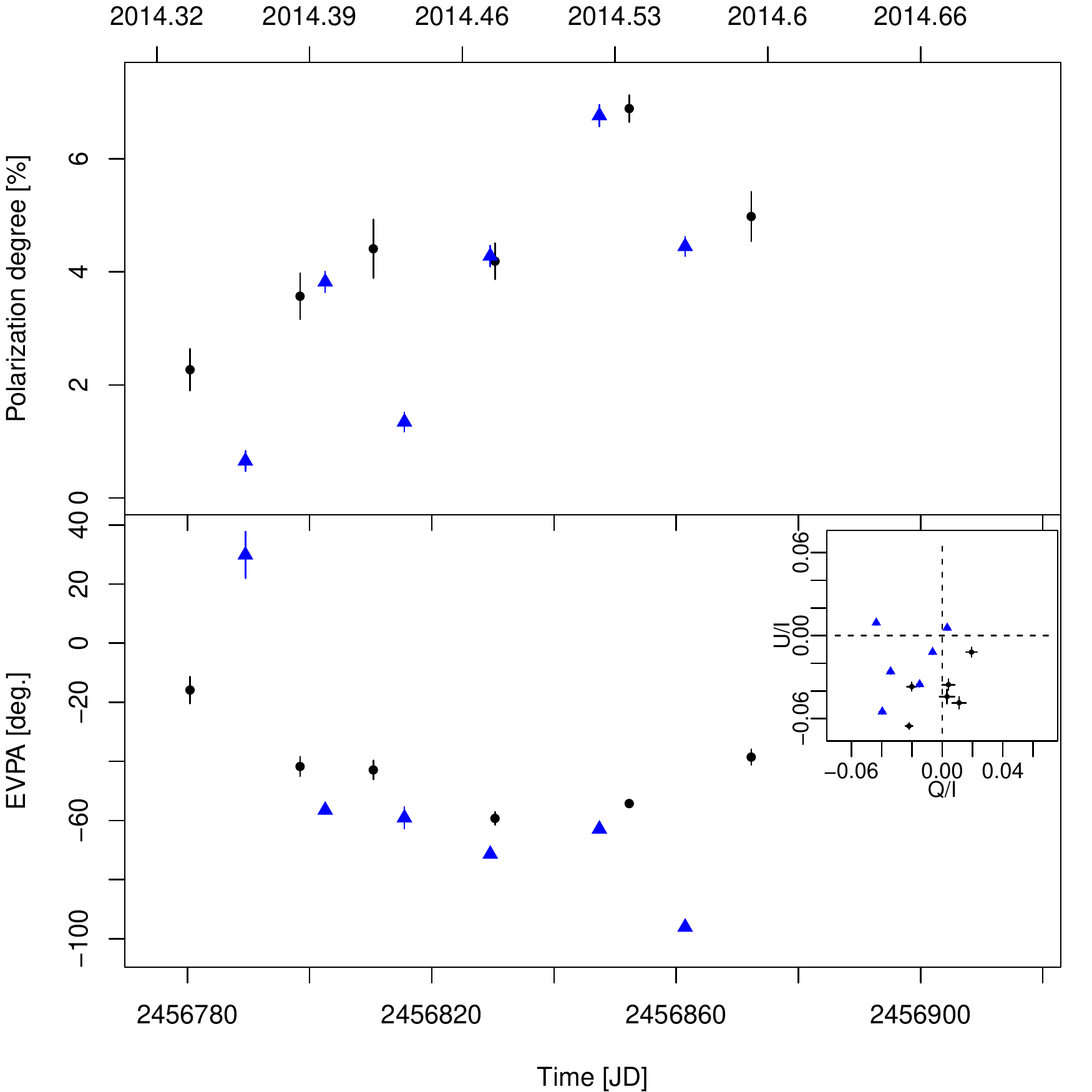}
\caption{Fractional polarization (top) and EVPA (bottom) of the TeV
  source J1427+2348. Black circles are RoboPol data and blue
  triangles NOT data. The inset in the lower panel shows
  $Q/I$ vs. $U/I$. EVPA data are shown only for
  observations where the signal-to-noise in the polarization fraction
  $\ge 3$.}
\label{Fig:lc18}
\end{figure}

\begin{figure}
\includegraphics[width=0.45\textwidth]{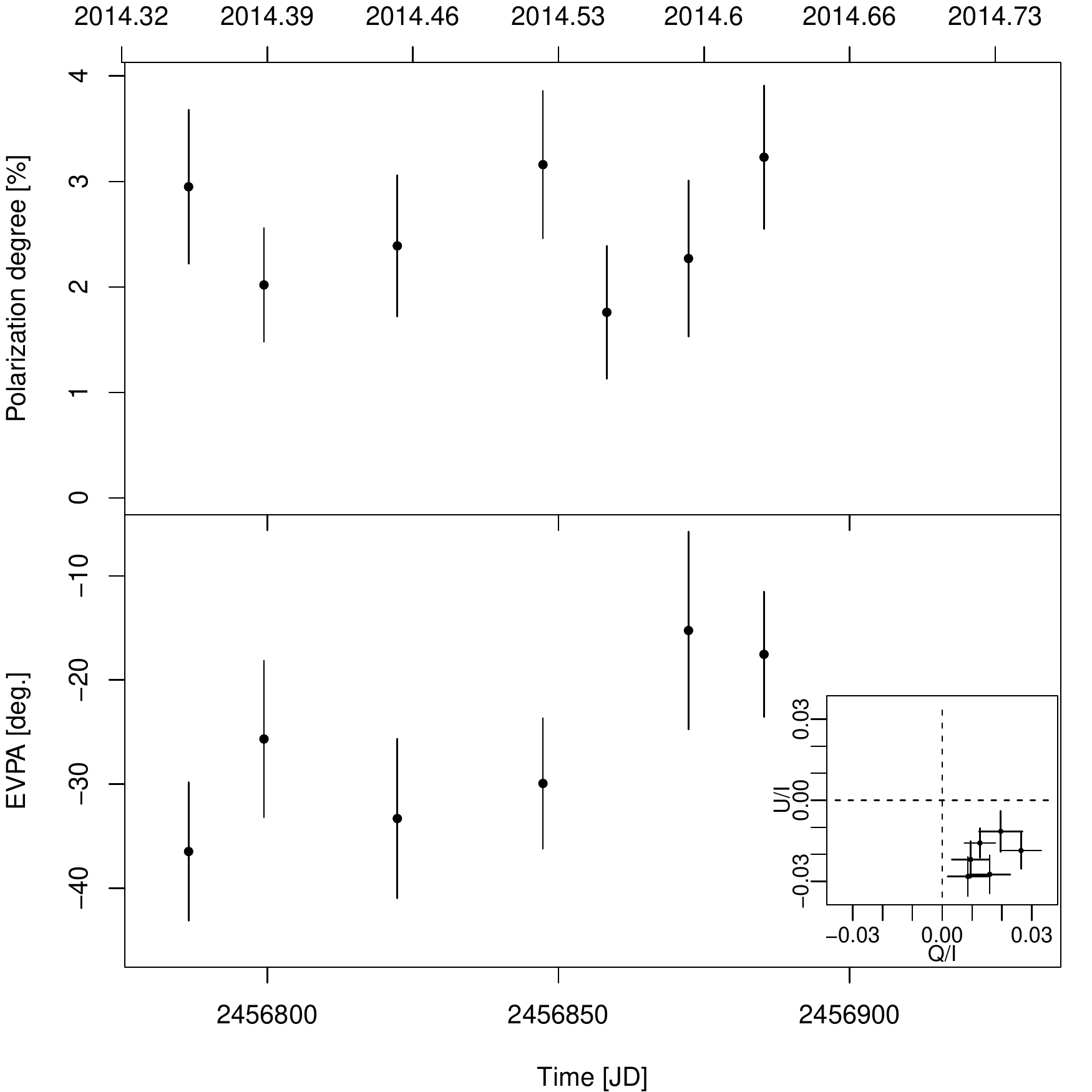}
\caption{Fractional polarization (top) and EVPA (bottom) of the TeV
  source J1428+4240. The inset in the lower panel shows
  $Q/I$ vs. $U/I$. EVPA data are shown only for
  observations where the signal-to-noise in the polarization fraction
  $\ge 3$.}
\label{Fig:lc19}
\end{figure}

\begin{figure}
\includegraphics[width=0.45\textwidth]{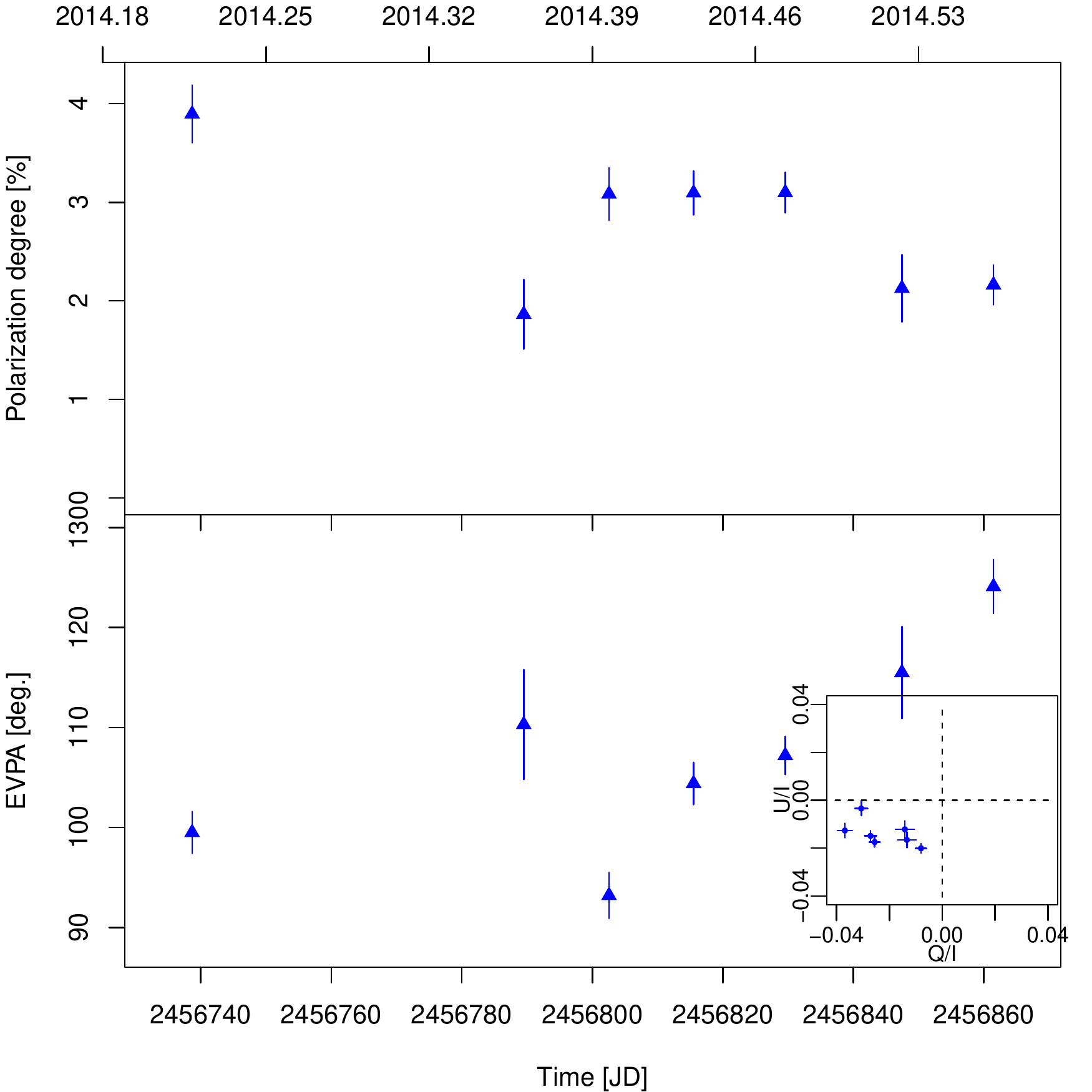}
\caption{Fractional polarization (top) and EVPA (bottom) of the TeV
  source J1442+1200. The inset in the lower panel shows 
  $Q/I$ vs. $U/I$. EVPA data are shown only for
  observations where the signal-to-noise in the polarization fraction
  $\ge 3$. }
\label{Fig:lc20}
\end{figure}

\begin{figure}
\includegraphics[width=0.45\textwidth]{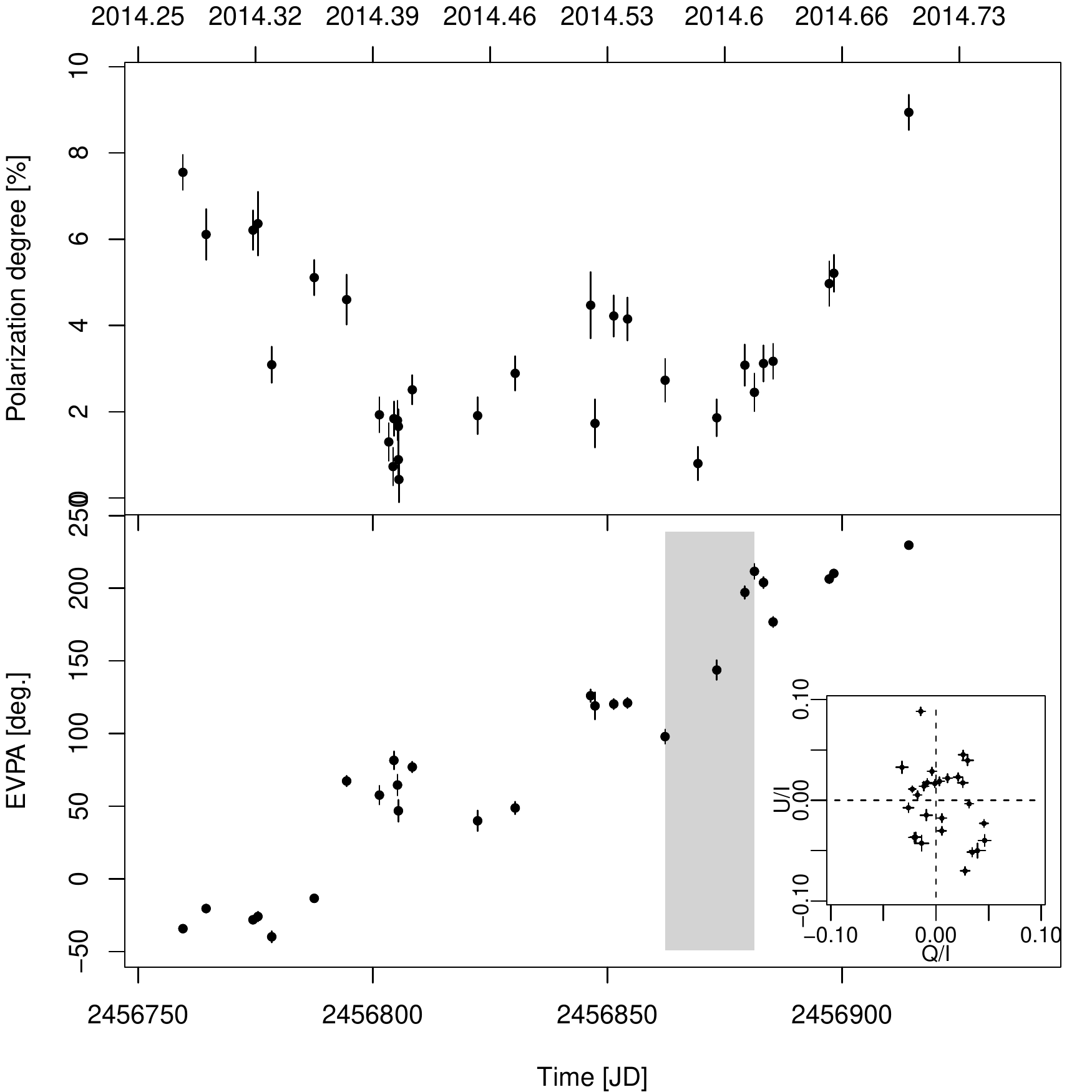}
\caption{Fractional polarization (top) and EVPA (bottom) of the TeV
  source J1555+1111. The shaded region shows the period of a
  significant EVPA rotation. The inset in the lower panel shows 
  $Q/I$ vs. $U/I$. EVPA data are shown only for
  observations where the signal-to-noise in the polarization fraction
  $\ge 3$.}
\label{Fig:lc21}
\end{figure}

\begin{figure}
\includegraphics[width=0.45\textwidth]{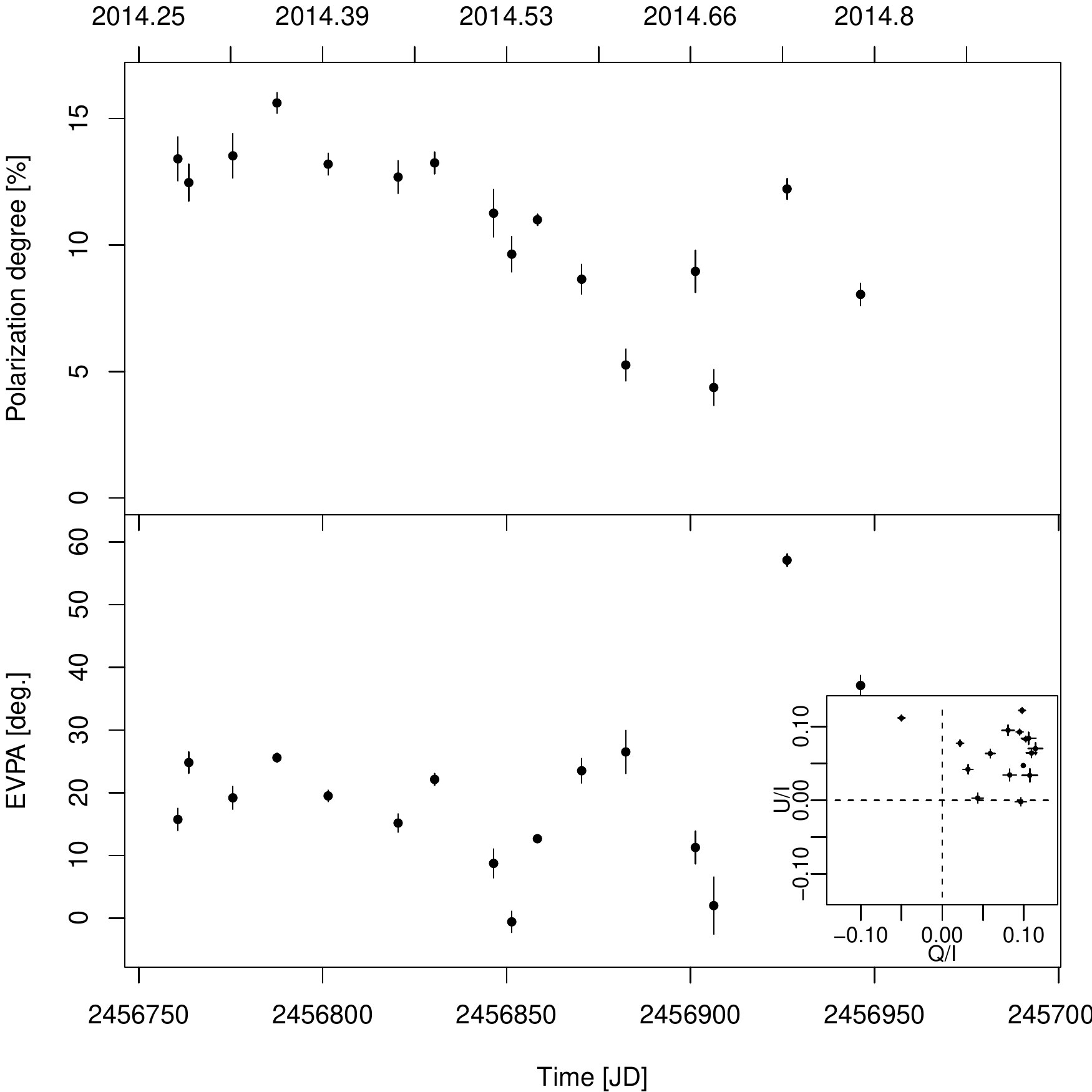}
\caption{Fractional polarization (top) and EVPA (bottom) of the TeV
  source J1725+1152. The inset in the lower panel shows 
  $Q/I$ vs. $U/I$. EVPA data are shown only for
  observations where the signal-to-noise in the polarization fraction
  $\ge 3$.}
\label{Fig:lc22}
\end{figure}

\begin{figure}
\includegraphics[width=0.45\textwidth]{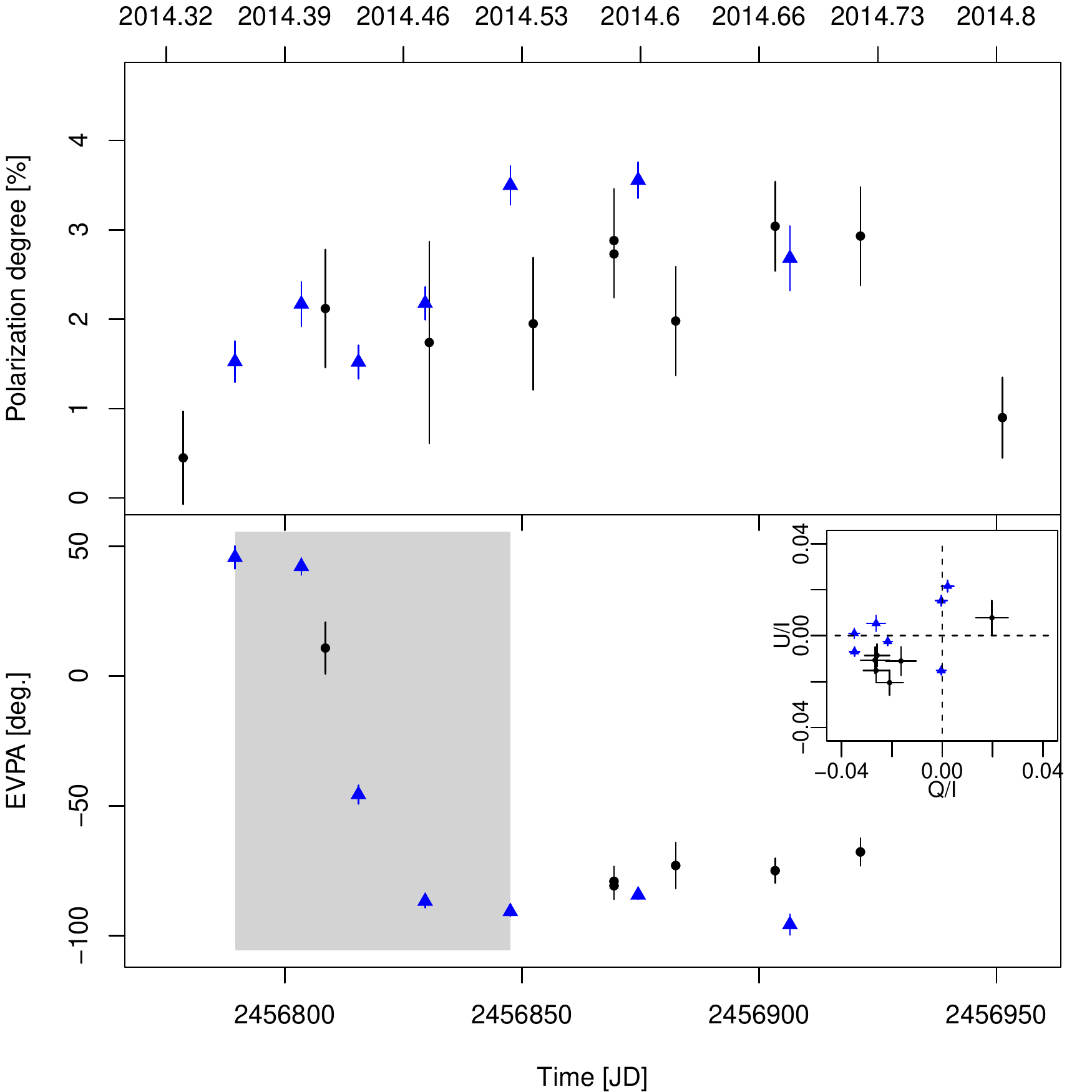}
\caption{Fractional polarization (top) and EVPA (bottom) of the TeV
  source J1728+5013. Black circles are RoboPol data and blue
  triangles NOT data. The shaded region shows the period of a
  significant EVPA rotation. The inset in the lower panel shows 
  $Q/I$ vs. $U/I$. EVPA data are shown only for
  observations where the signal-to-noise in the polarization fraction
  $\ge 3$.}
\label{Fig:lc23}
\end{figure}

\begin{figure}
\includegraphics[width=0.45\textwidth]{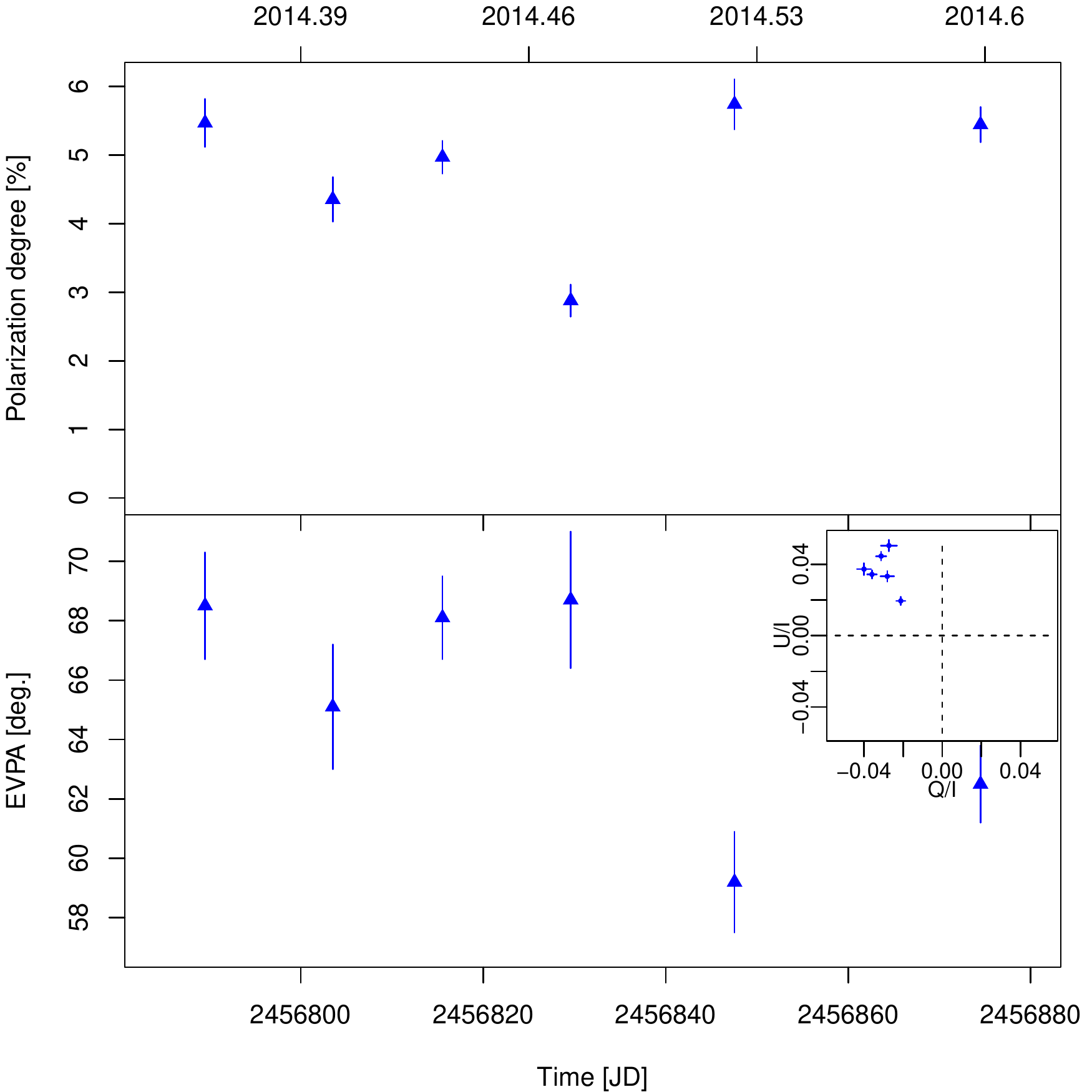}
\caption{Fractional polarization (top) and EVPA (bottom) of the TeV
  source J1743+1935. The inset in the lower panel shows 
  $Q/I$ vs. $U/I$. EVPA data are shown only for
  observations where the signal-to-noise in the polarization fraction
  $\ge 3$. }
\label{Fig:lc24}
\end{figure}

\begin{figure}
\includegraphics[width=0.45\textwidth]{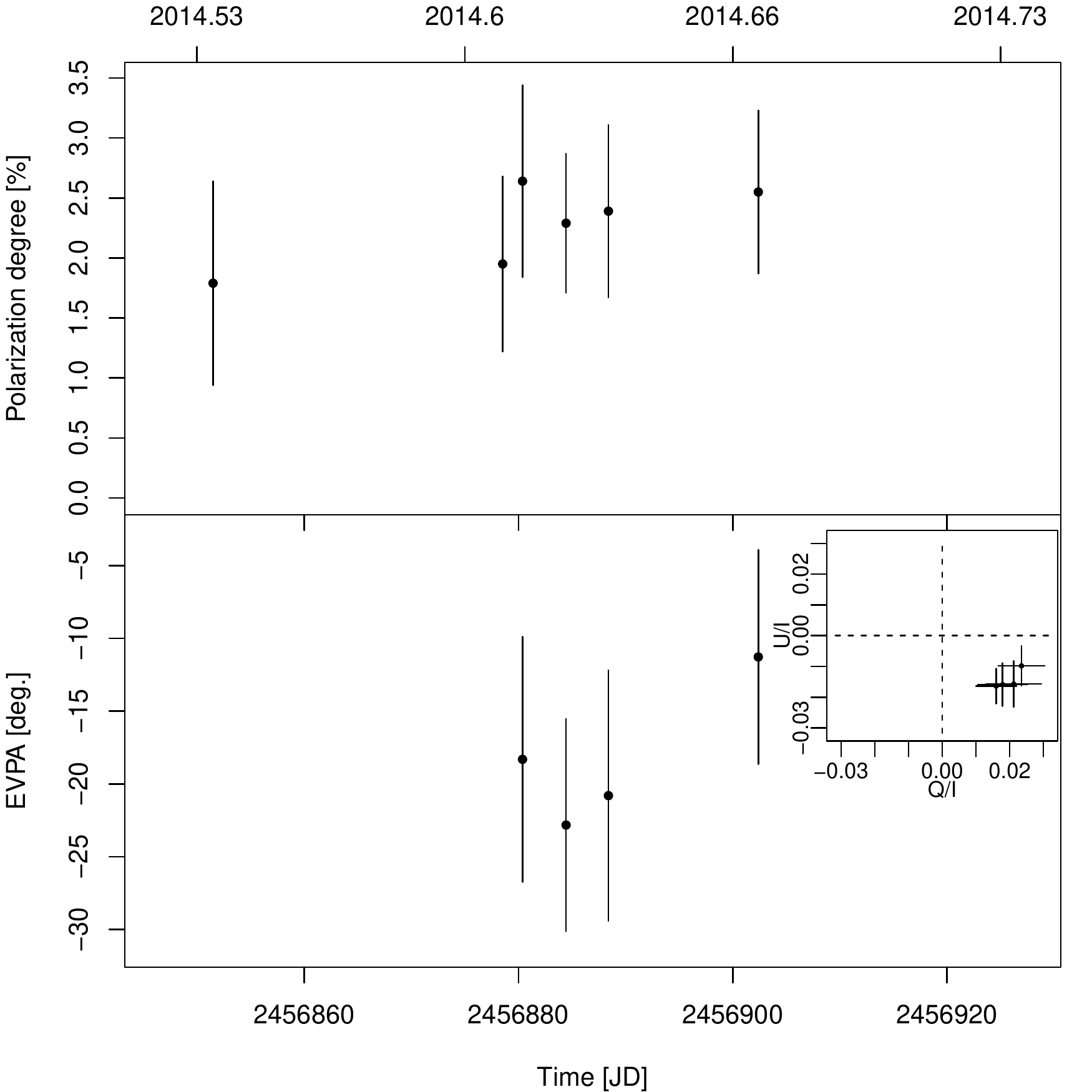}
\caption{Fractional polarization (top) and EVPA (bottom) of the TeV
  source J1943+2118. The inset in the lower panel shows 
  $Q/I$ vs. $U/I$. EVPA data are shown only for
  observations where the signal-to-noise in the polarization fraction
  $\ge 3$.}
\label{Fig:lc25}
\end{figure}

\begin{figure}
\includegraphics[width=0.45\textwidth]{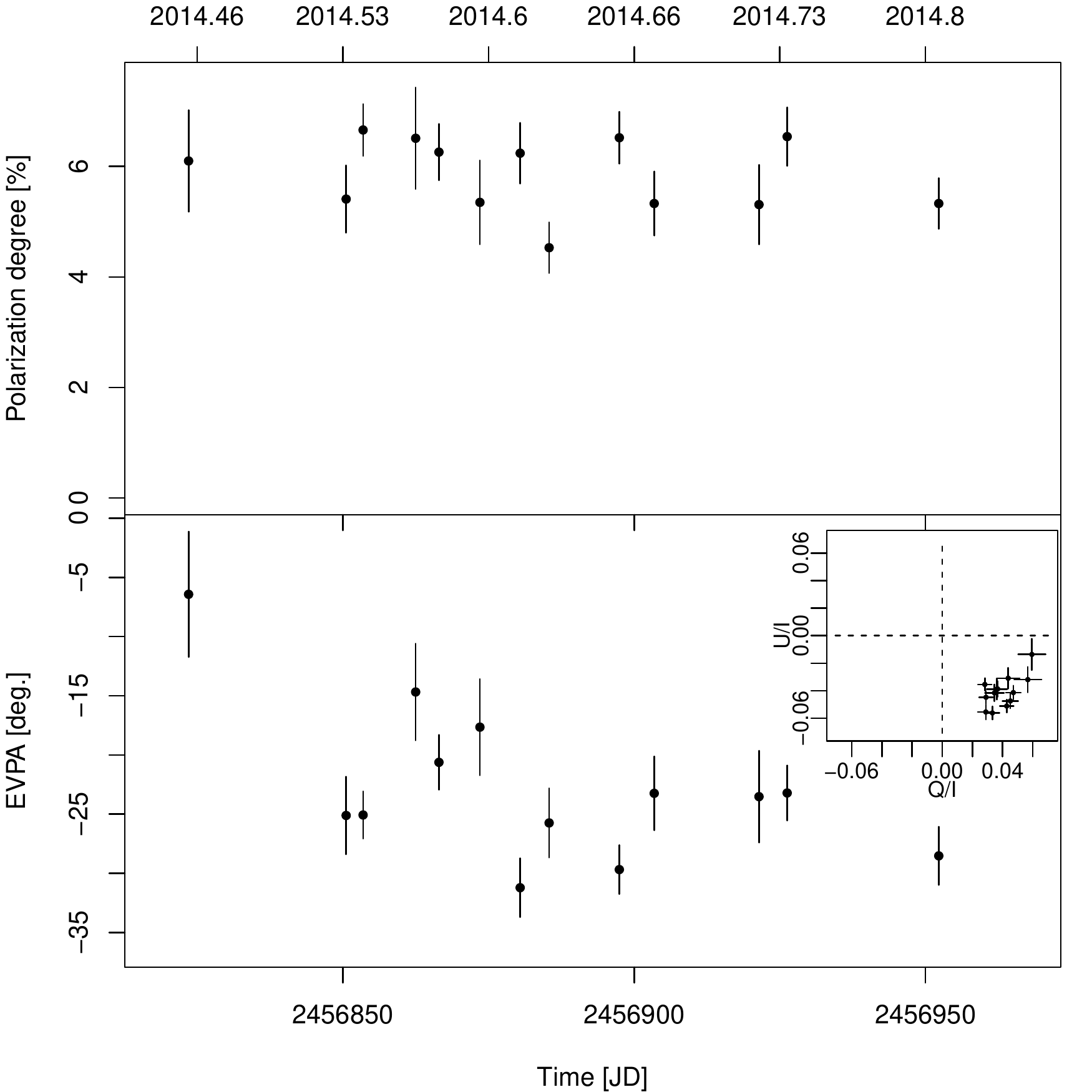}
\caption{Fractional polarization (top) and EVPA (bottom) of the TeV
  source J1959+6508. The inset in the lower panel shows 
  $Q/I$ vs. $U/I$. EVPA data are shown only for
  observations where the signal-to-noise in the polarization fraction
  $\ge 3$.}
\label{Fig:lc26}
\end{figure}

\begin{figure}
\includegraphics[width=0.45\textwidth]{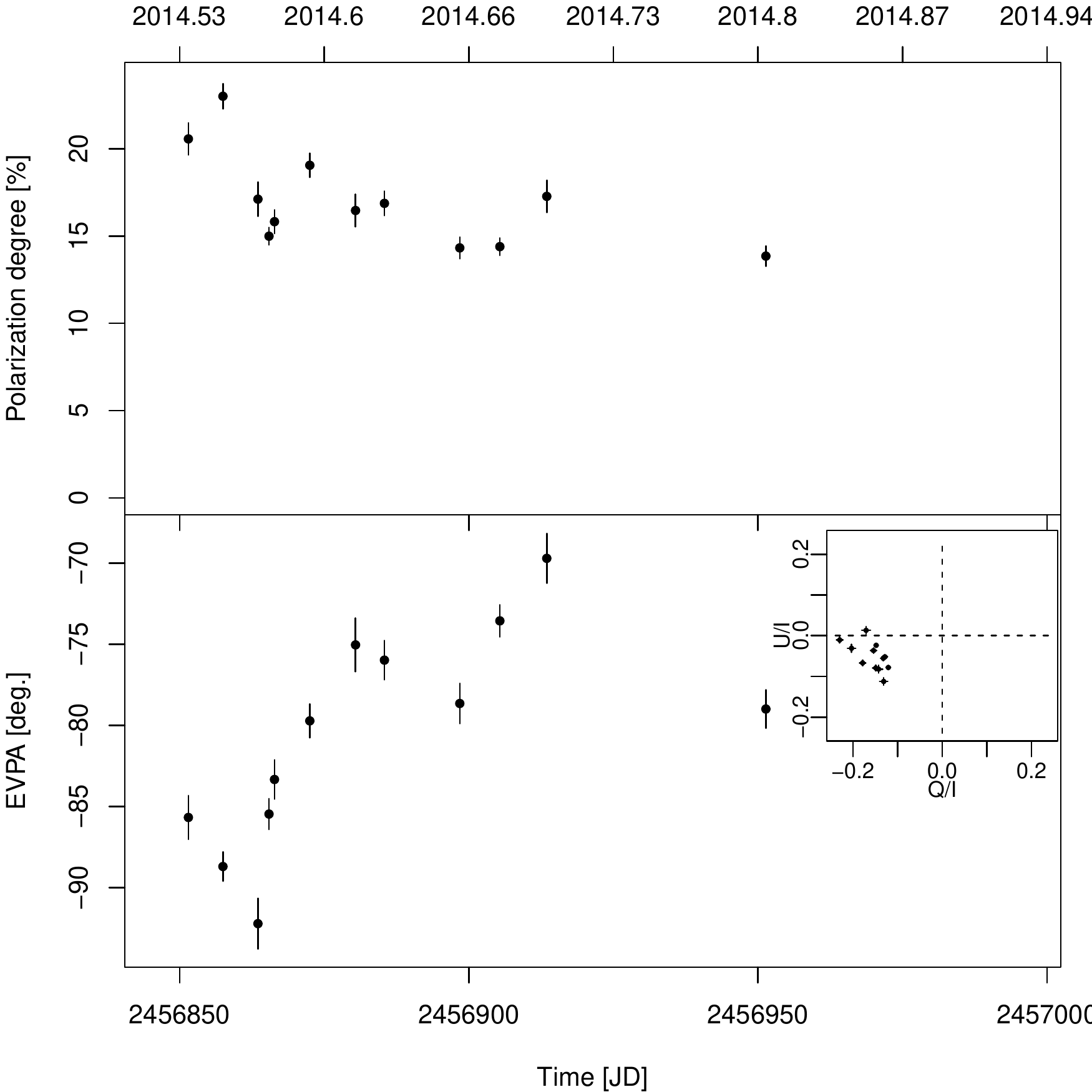}
\caption{Fractional polarization (top) and EVPA (bottom) of the TeV
  source J2001+4352. The inset in the lower panel shows 
  $Q/I$ vs. $U/I$. EVPA data are shown only for
  observations where the signal-to-noise in the polarization fraction
  $\ge 3$.}
\label{Fig:lc27}
\end{figure}

\begin{figure}
\includegraphics[width=0.45\textwidth]{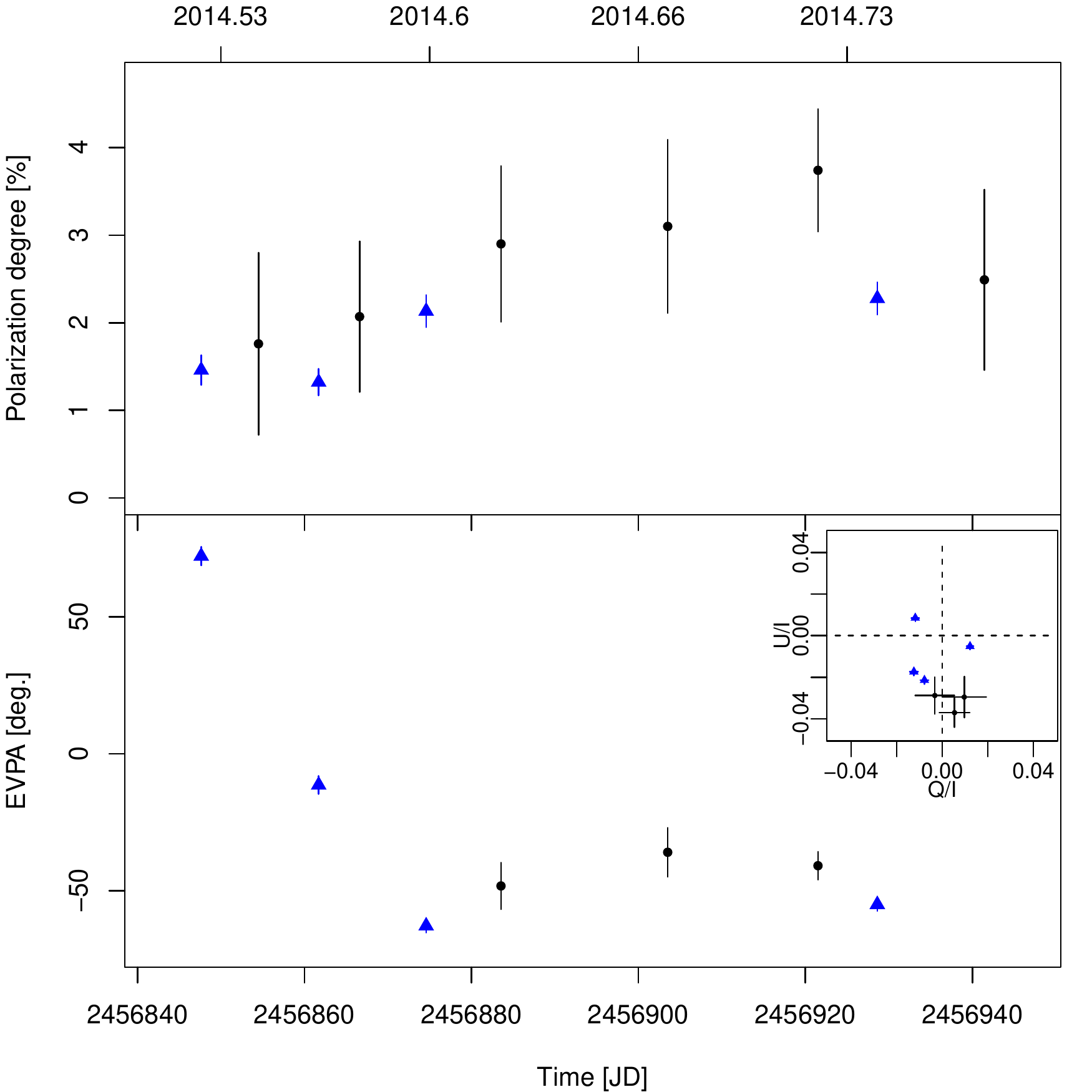}
\caption{Fractional polarization (top) and EVPA (bottom) of the TeV
  source "J2250+3825. Black circles are RoboPol data and blue
  triangles NOT data. The inset in the lower panel shows
  $Q/I$ vs. $U/I$. EVPA data are shown only for
  observations where the signal-to-noise in the polarization fraction
  $\ge 3$.}
\label{Fig:lc28}
\end{figure}

\begin{figure}
\includegraphics[width=0.45\textwidth]{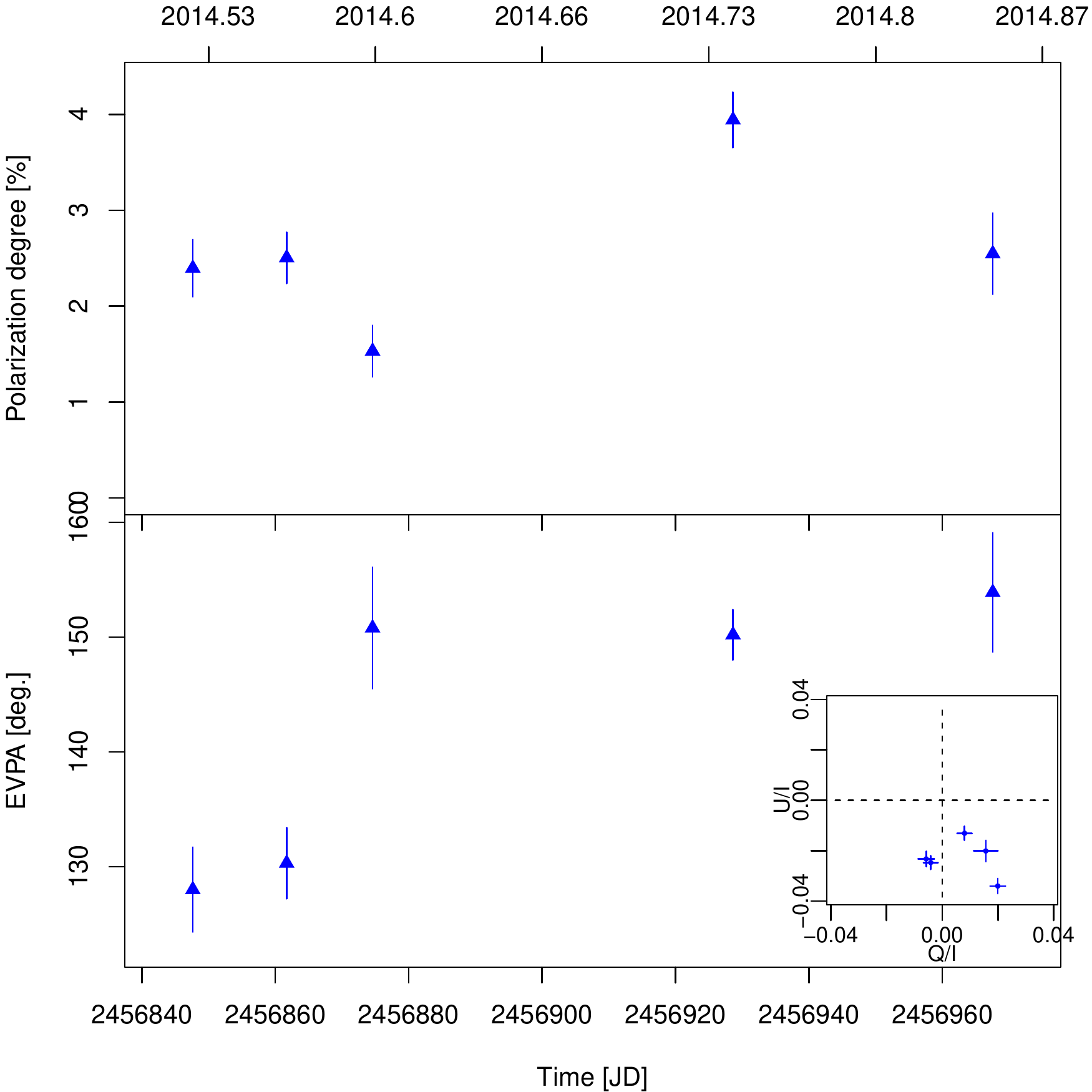}
\caption{Fractional polarization (top) and EVPA (bottom) of the TeV
  source J2347+5142. The inset in the lower panel shows 
  $Q/I$ vs. $U/I$. EVPA data are shown only for
  observations where the signal-to-noise in the polarization fraction
  $\ge 3$. }
\label{Fig:lc29}
\end{figure}

\clearpage
%non-TeV sources

\begin{figure}
\includegraphics[width=0.45\textwidth]{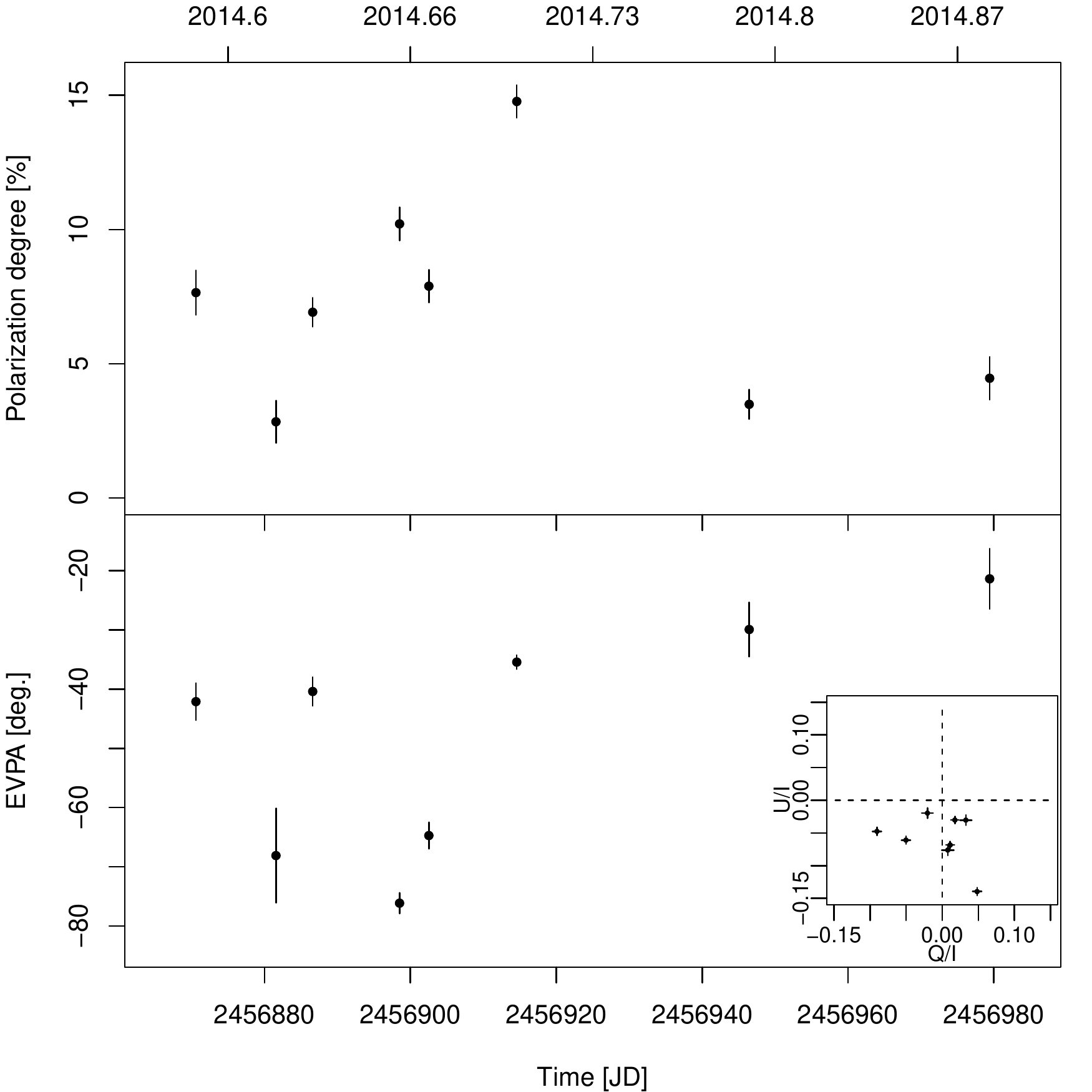}
\caption{Fractional polarization (top) and EVPA (bottom) of the  non-TeV
  source J0114+1325. The inset in the lower panel shows the
  $Q/I$ vs. $U/I$ from the RoboPol data. EVPA data are shown only for
  observations where the signal-to-noise in the polarization fraction
  $\ge 3$.}
\label{Fig:lcc1}
\end{figure}

\begin{figure}
\includegraphics[width=0.45\textwidth]{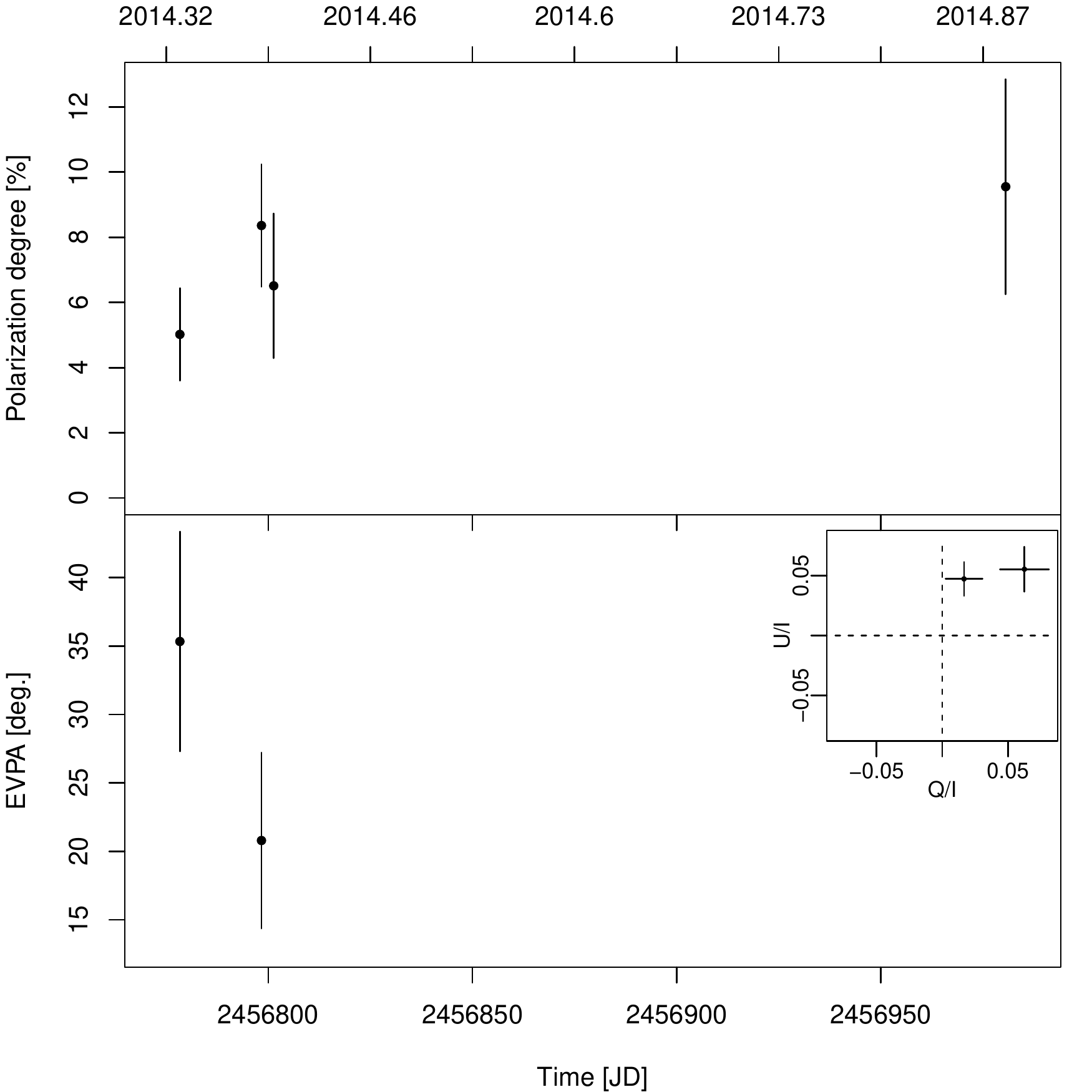}
\caption{Fractional polarization (top) and EVPA (bottom) of the  non-TeV
  source J0848+6606. The inset in the lower panel shows the
  $Q/I$ vs. $U/I$ from the RoboPol data. EVPA data are shown only for
  observations where the signal-to-noise in the polarization fraction
  $\ge 3$.}
\label{Fig:lcc2}
\end{figure}

\begin{figure}
\includegraphics[width=0.45\textwidth]{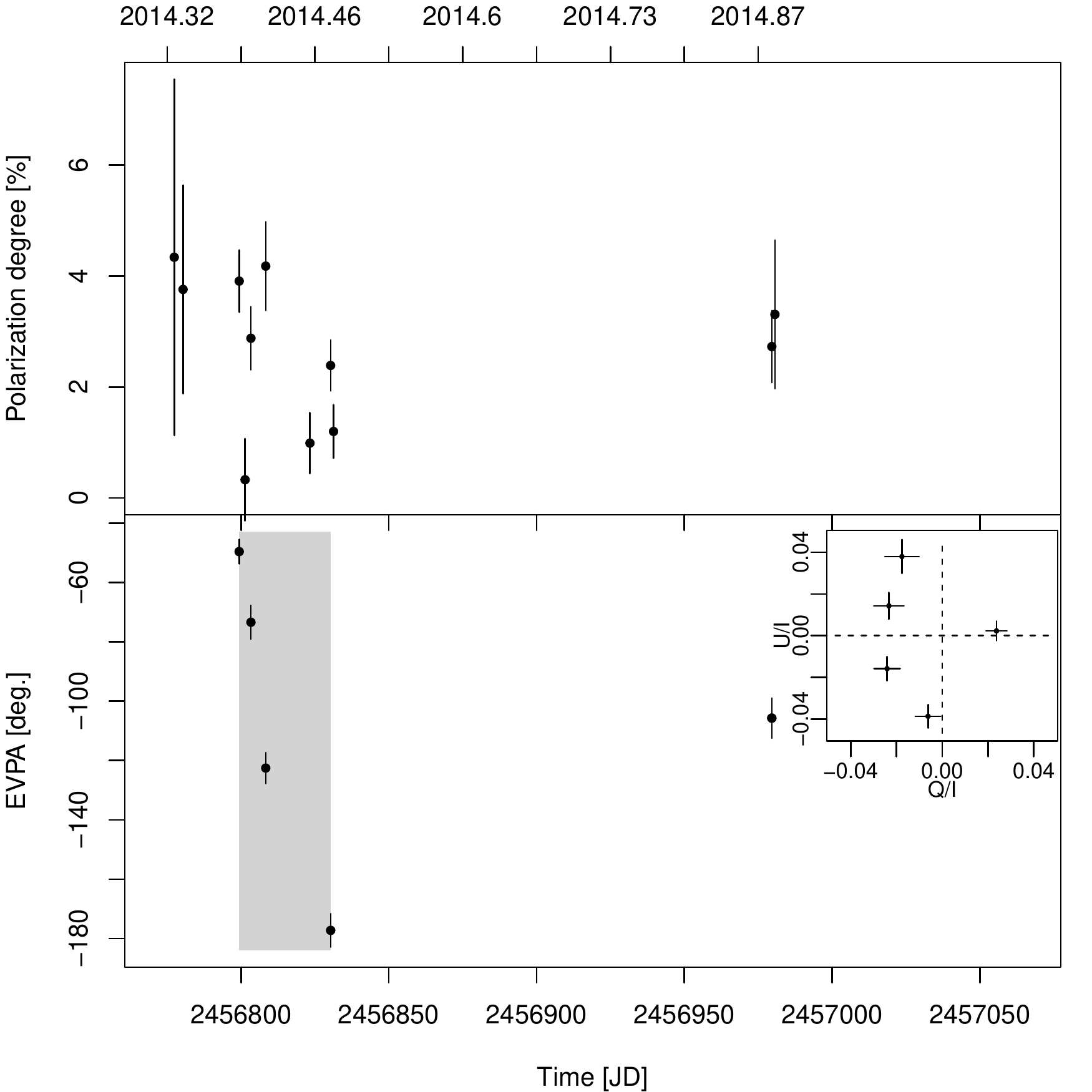}
\caption{Fractional polarization (top) and EVPA (bottom) of the  non-TeV
  source J1037+5711. The shaded region shows the period of a
  significant EVPA rotation. The inset in the lower panel shows the
  $Q/I$ vs. $U/I$ from the RoboPol data. EVPA data are shown only for
  observations where the signal-to-noise in the polarization fraction
  $\ge 3$.}
\label{Fig:lcc3}
\end{figure}

\begin{figure}
\includegraphics[width=0.45\textwidth]{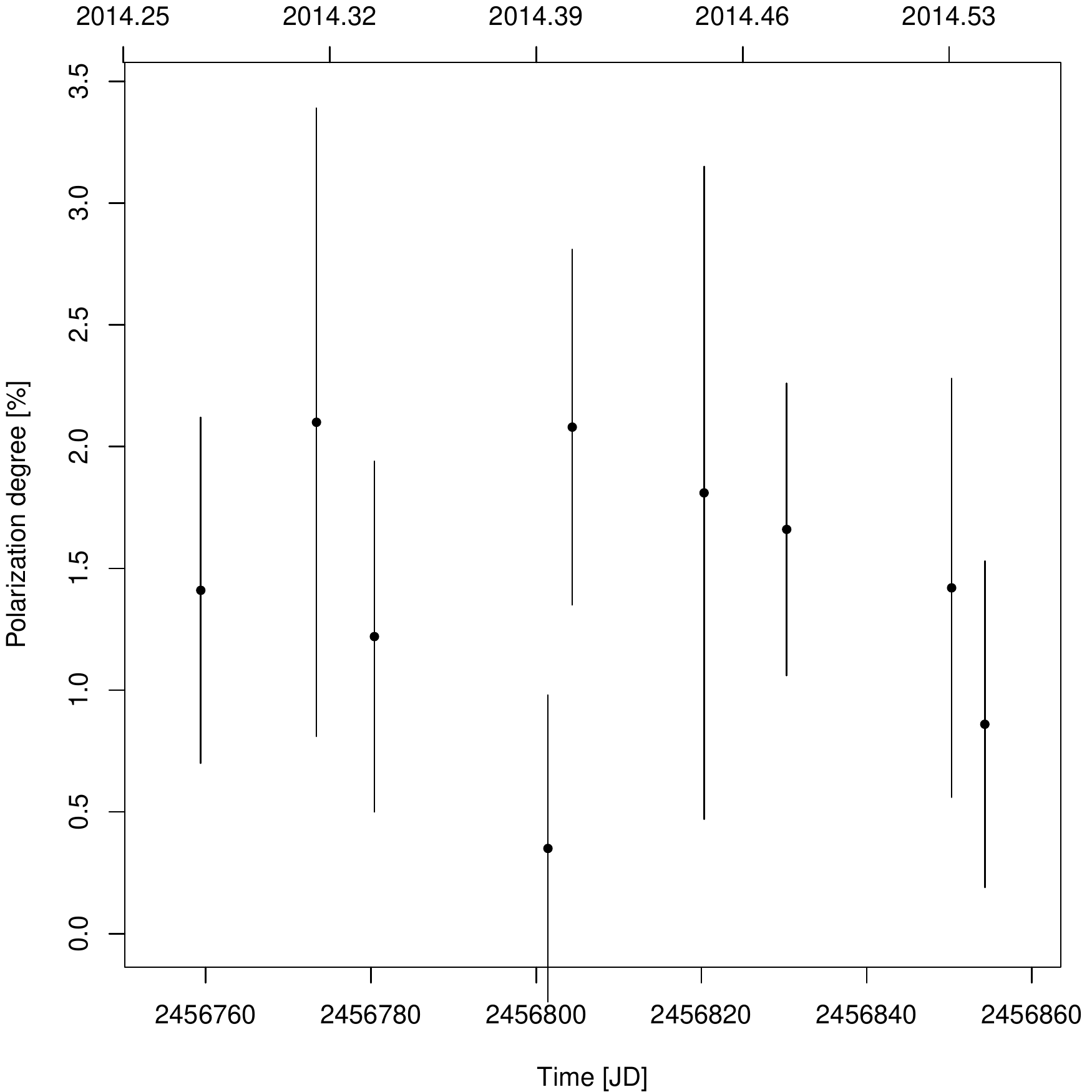}
\caption{Fractional polarization (top) and EVPA (bottom) of the  non-TeV
  source J1203+6031. None of the polarization observations have signal-to-noise 
  $\ge 3$ and no EVPA data are shown.}
\label{Fig:lcc4}
\end{figure}

\begin{figure}
\includegraphics[width=0.45\textwidth]{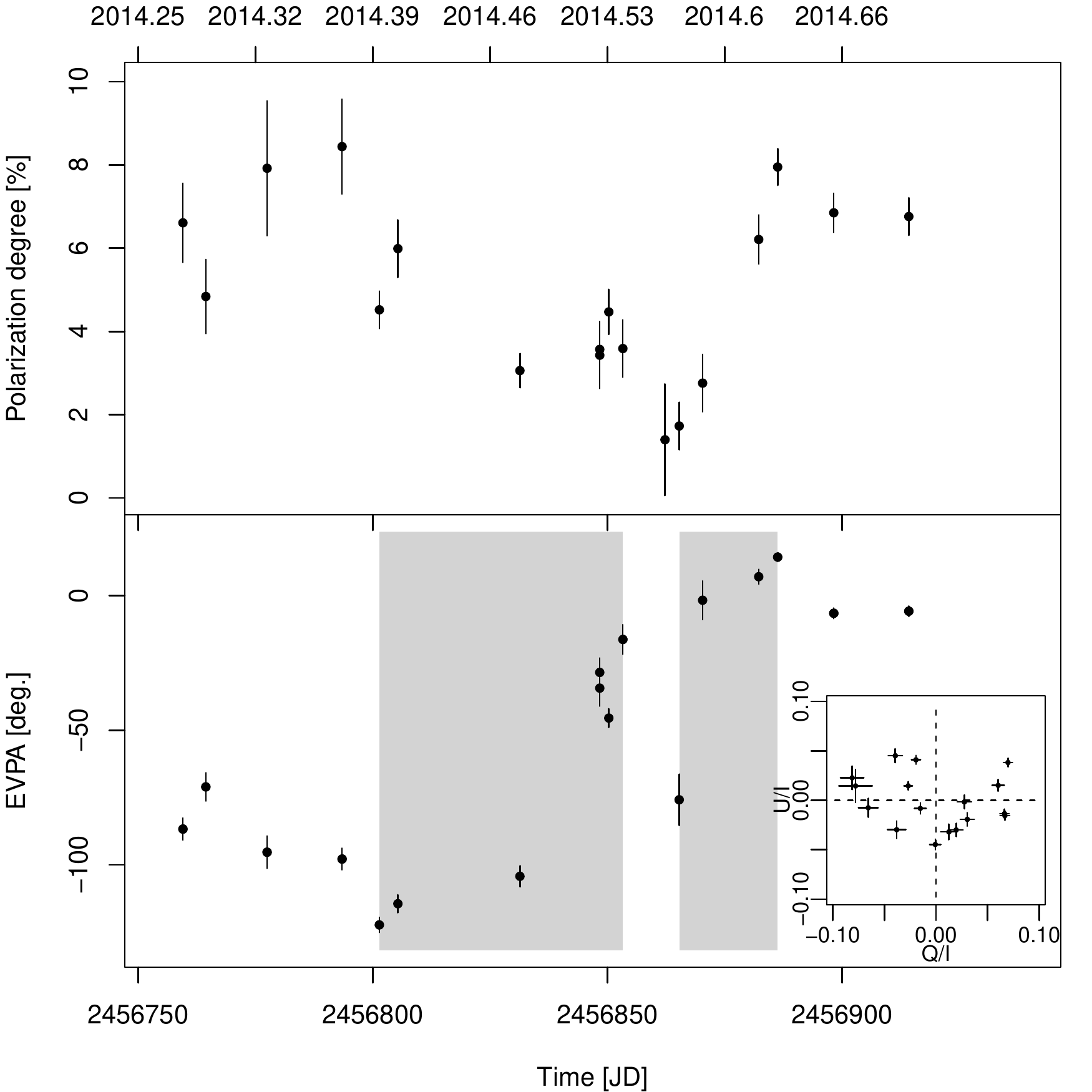}
\caption{Fractional polarization (top) and EVPA (bottom) of the  non-TeV
  source J1542+6129. The shaded region shows the period of a
  significant EVPA rotation. The inset in the lower panel shows the
  $Q/I$ vs. $U/I$ from the RoboPol data. EVPA data are shown only for
  observations where the signal-to-noise in the polarization fraction
  $\ge 3$.}
\label{Fig:lcc5}
\end{figure}

\begin{figure}
\includegraphics[width=0.45\textwidth]{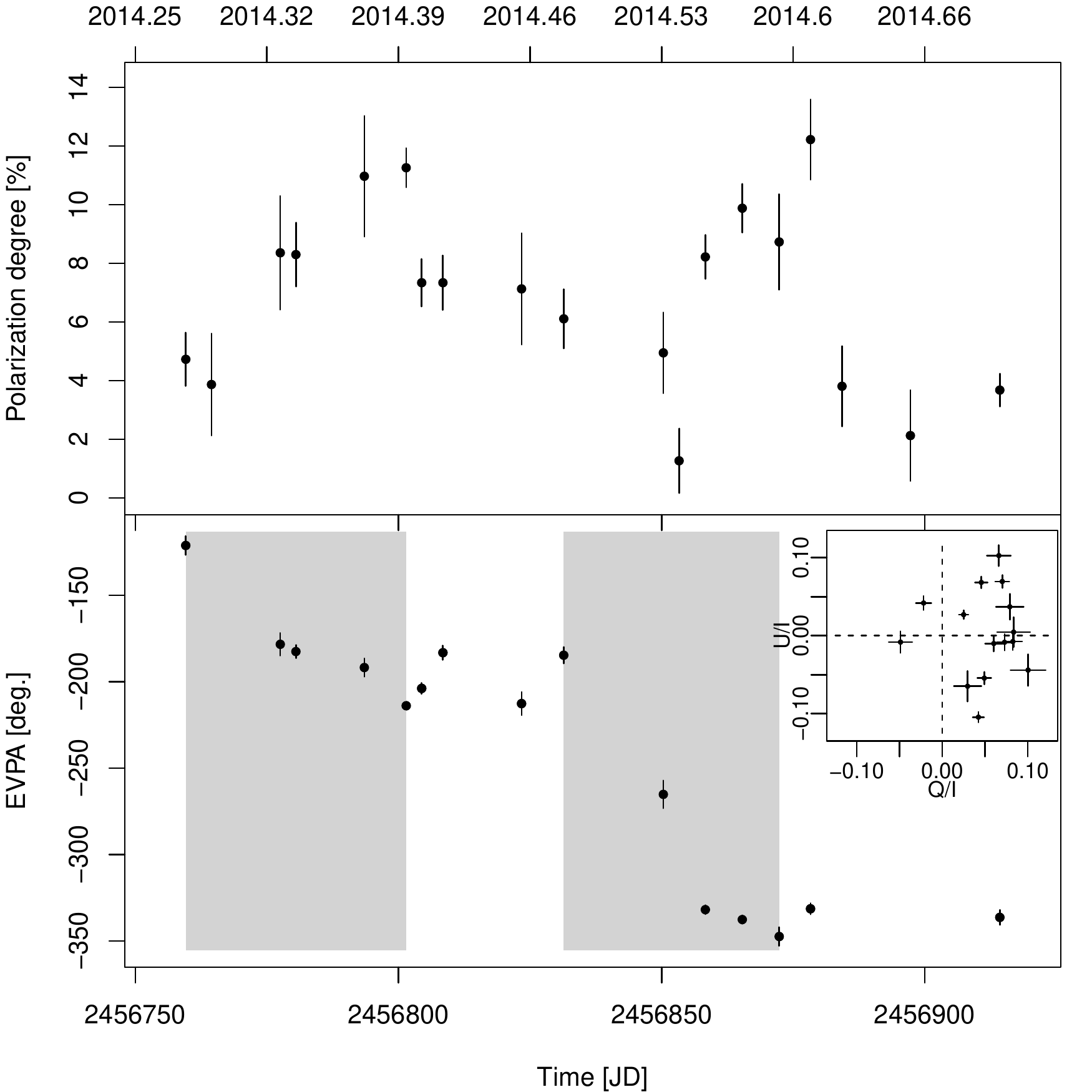}
\caption{Fractional polarization (top) and EVPA (bottom) of the  non-TeV
  source J1558+5625. The shaded region shows the period of a
  significant EVPA rotation. The inset in the lower panel shows the
  $Q/I$ vs. $U/I$ from the RoboPol data. EVPA data are shown only for
  observations where the signal-to-noise in the polarization fraction
  $\ge 3$.}
\label{Fig:lcc6}
\end{figure}

\begin{figure}
\includegraphics[width=0.45\textwidth]{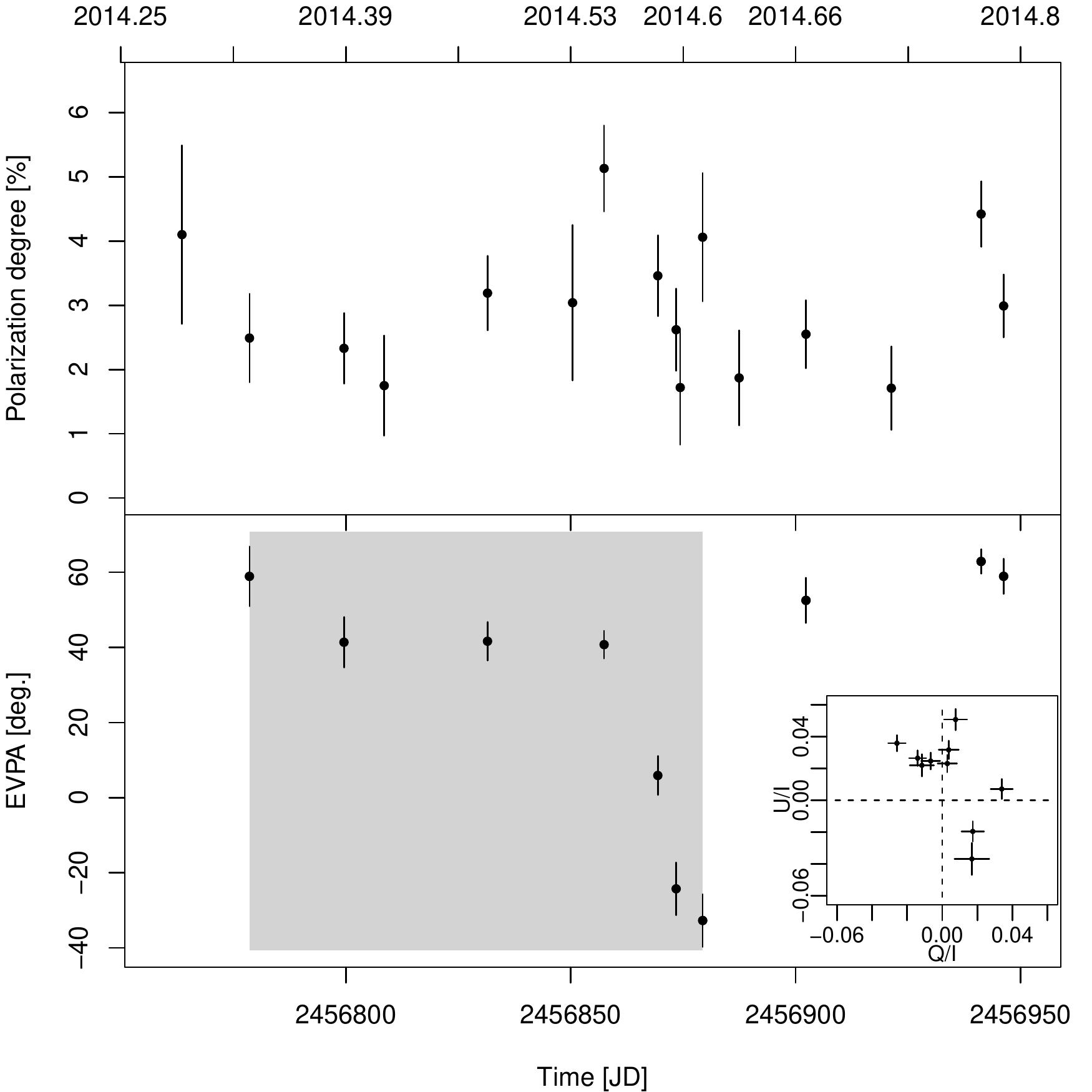}
\caption{Fractional polarization (top) and EVPA (bottom) of the  non-TeV
  source J1649+5235. The shaded region shows the period of a
  significant EVPA rotation. The inset in the lower panel shows the
  $Q/I$ vs. $U/I$ from the RoboPol data. EVPA data are shown only for
  observations where the signal-to-noise in the polarization fraction
  $\ge 3$.}
\label{Fig:lcc7}
\end{figure}

\begin{figure}
\includegraphics[width=0.45\textwidth]{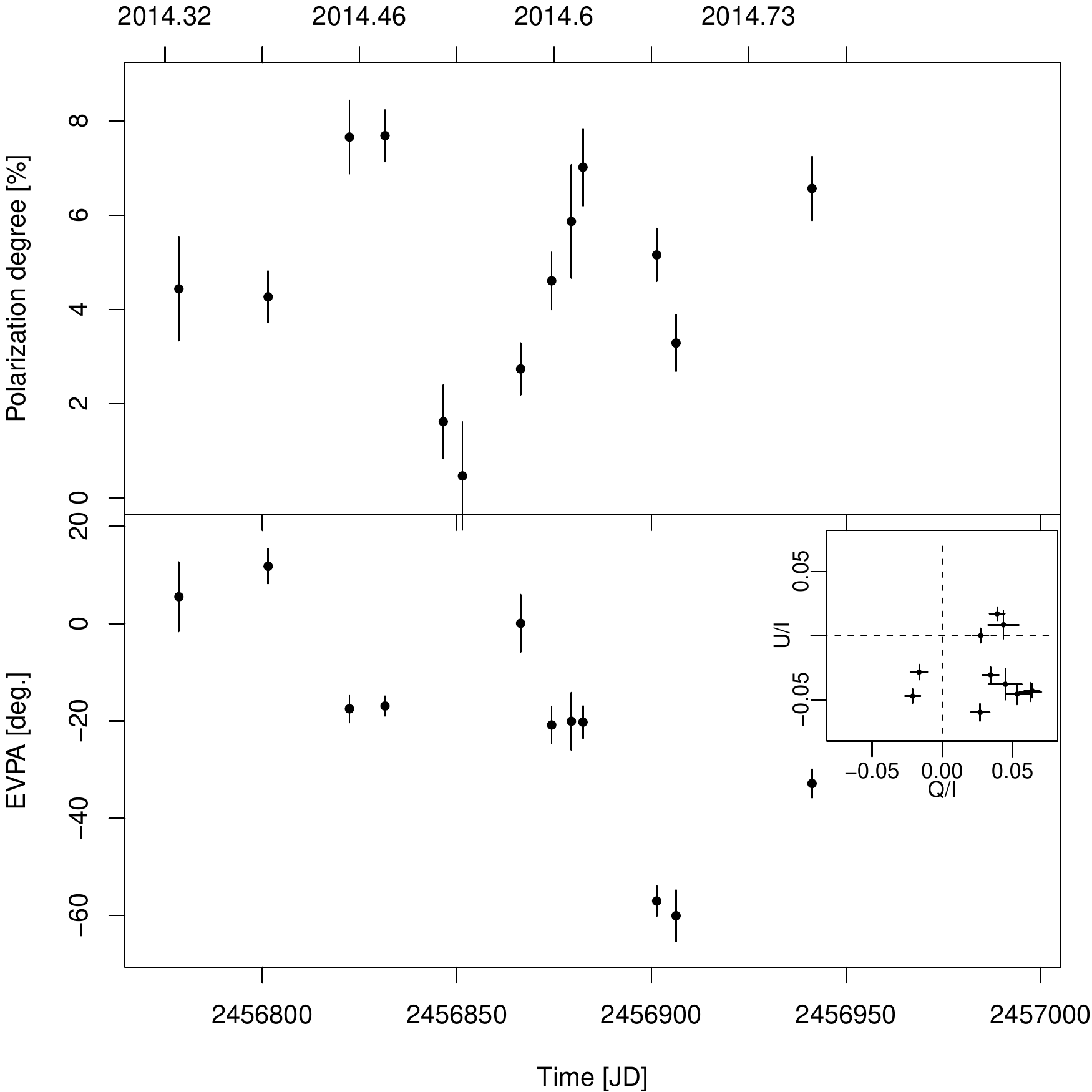}
\caption{Fractional polarization (top) and EVPA (bottom) of the  non-TeV
  source J1754+3212. The inset in the lower panel shows the
  $Q/I$ vs. $U/I$ from the RoboPol data. EVPA data are shown only for
  observations where the signal-to-noise in the polarization fraction
  $\ge 3$.}
\label{Fig:lcc8}
\end{figure}

\clearpage

\begin{figure}
\includegraphics[width=0.45\textwidth]{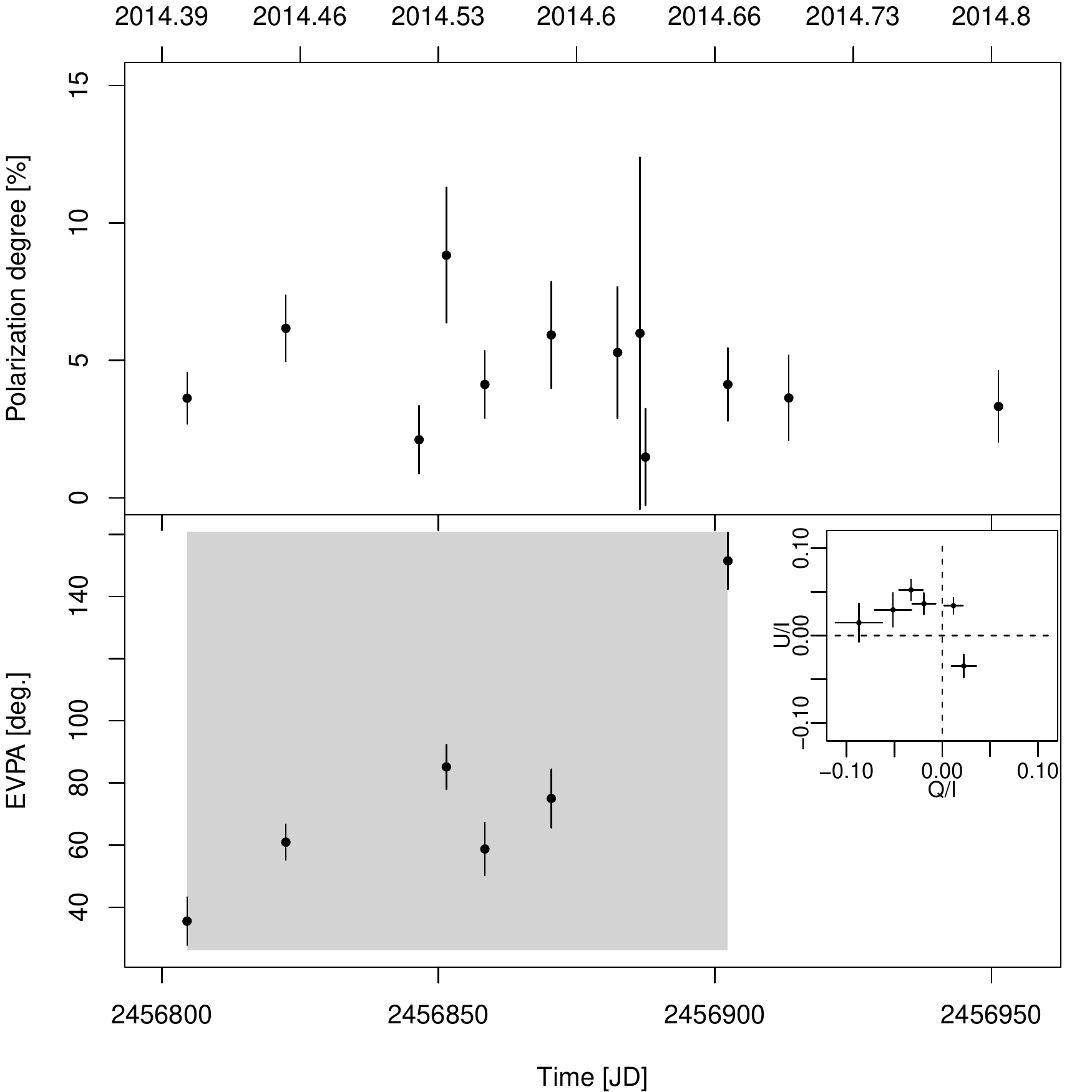}
\caption{Fractional polarization (top) and EVPA (bottom) of the  non-TeV
  source J1809+2041. The shaded region shows the period of a
  significant EVPA rotation. The inset in the lower panel shows the
  $Q/I$ vs. $U/I$ from the RoboPol data. EVPA data are shown only for
  observations where the signal-to-noise in the polarization fraction
  $\ge 3$.}
\label{Fig:lcc9}
\end{figure}

\begin{figure}
\includegraphics[width=0.45\textwidth]{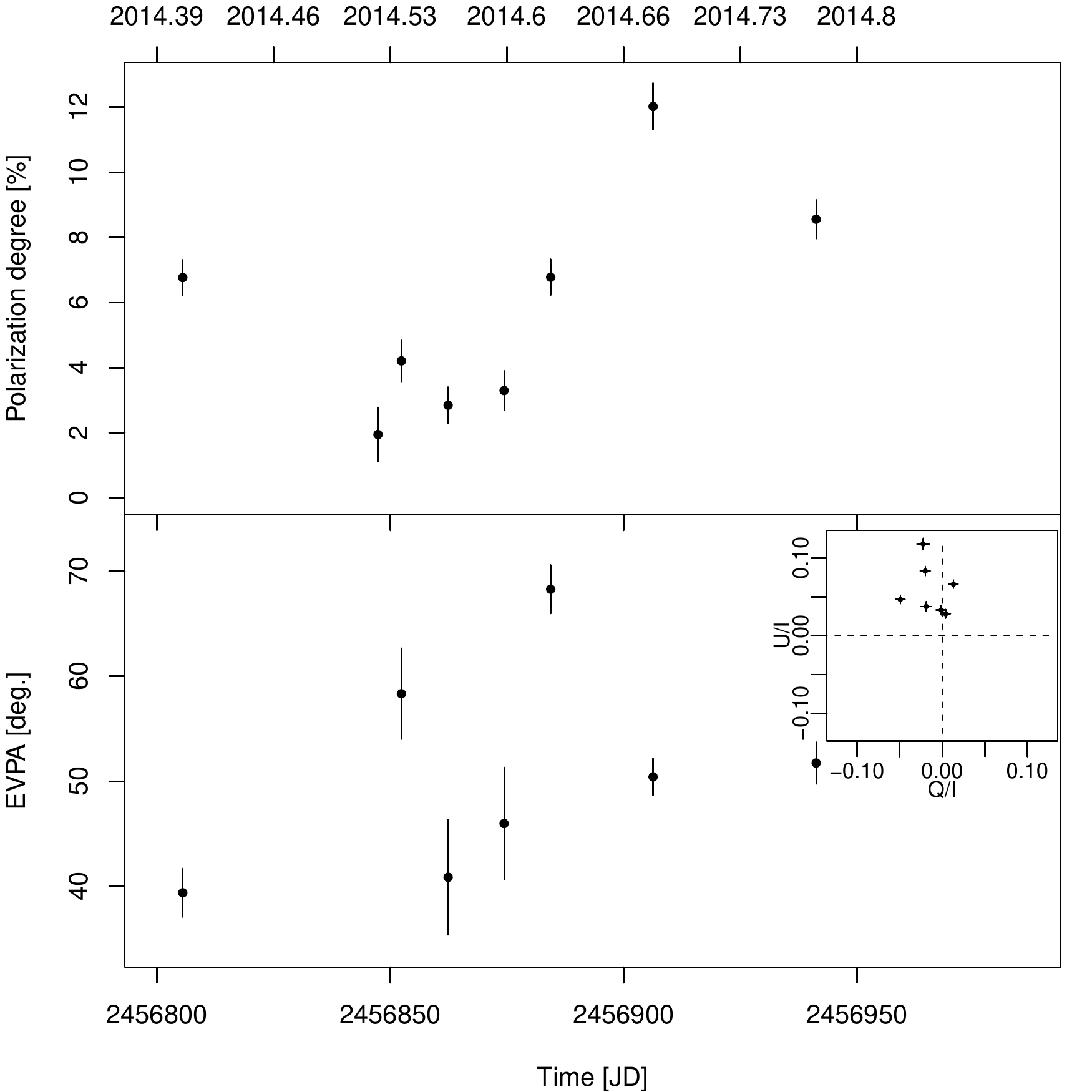}
\caption{Fractional polarization (top) and EVPA (bottom) of the  non-TeV
  source J1813+3144. The inset in the lower panel shows the
  $Q/I$ vs. $U/I$ from the RoboPol data. EVPA data are shown only for
  observations where the signal-to-noise in the polarization fraction
  $\ge 3$.}
\label{Fig:lcc10}
\end{figure}

\begin{figure}
\includegraphics[width=0.45\textwidth]{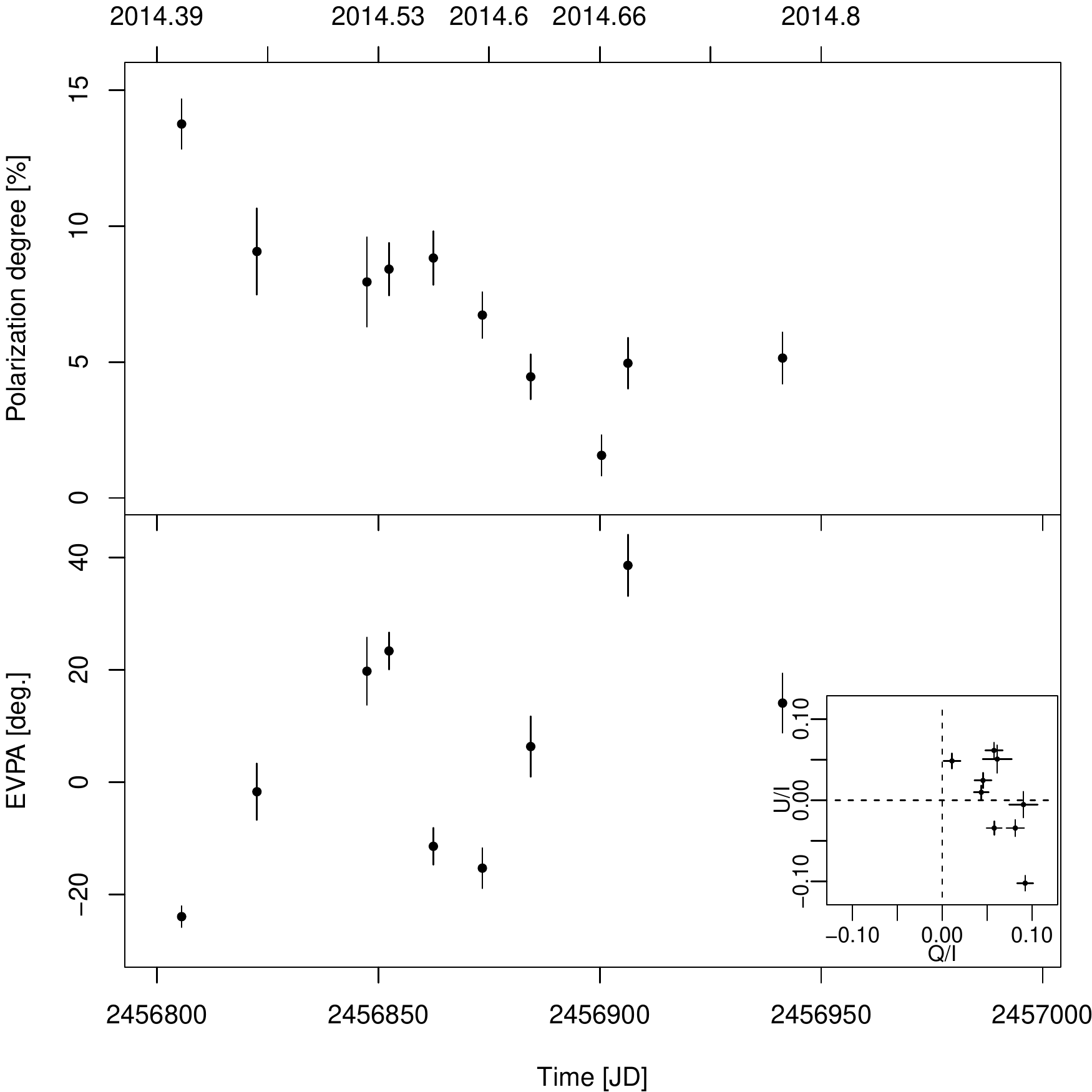}
\caption{Fractional polarization (top) and EVPA (bottom) of the  non-TeV
  source J1836+3136. The inset in the lower panel shows the
  $Q/I$ vs. $U/I$ from the RoboPol data. EVPA data are shown only for
  observations where the signal-to-noise in the polarization fraction
  $\ge 3$.}
\label{Fig:lcc11}
\end{figure}

\begin{figure}
\includegraphics[width=0.45\textwidth]{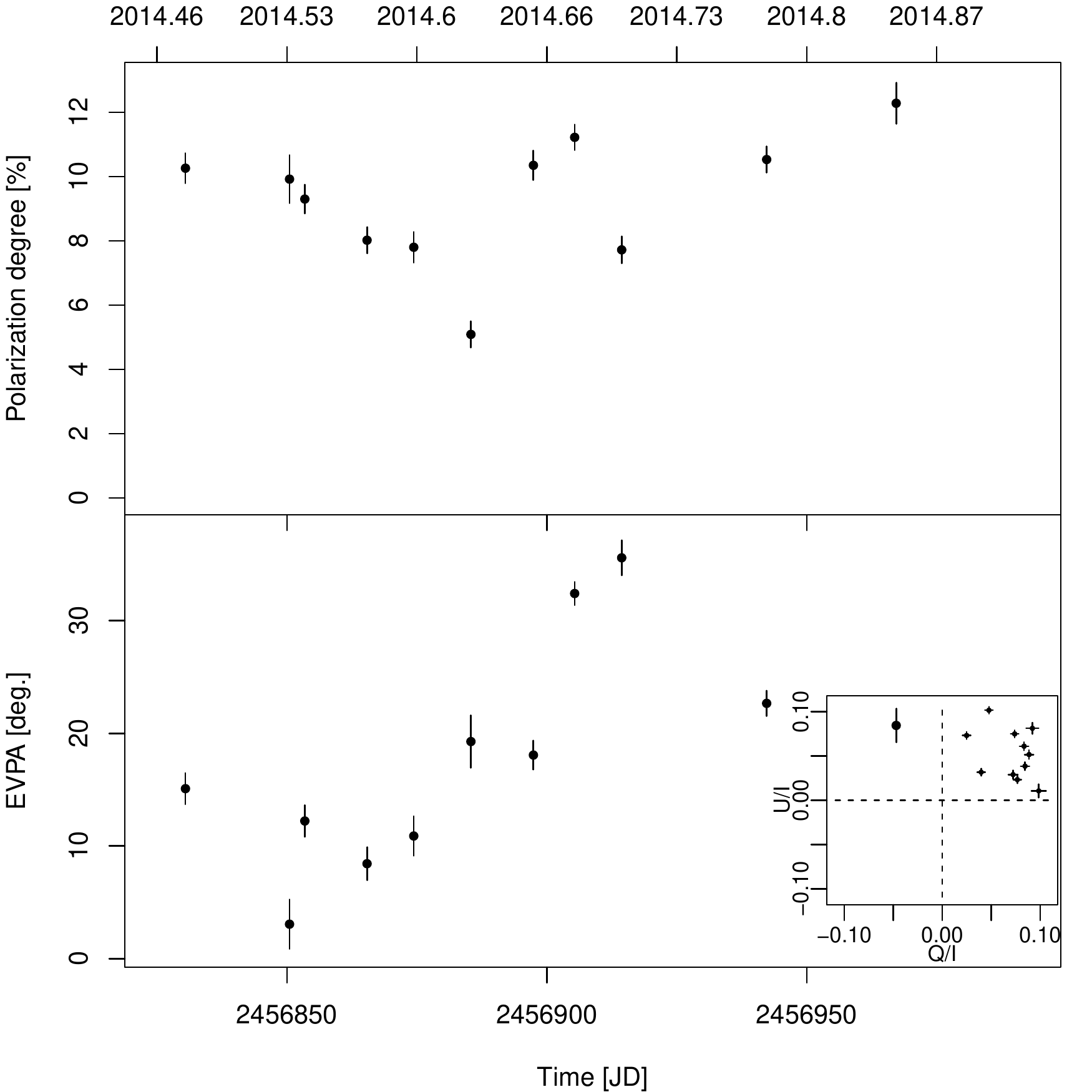}
\caption{Fractional polarization (top) and EVPA (bottom) of the  non-TeV
  source J1838+4802. The inset in the lower panel shows the
  $Q/I$ vs. $U/I$ from the RoboPol data. EVPA data are shown only for
  observations where the signal-to-noise in the polarization fraction
  $\ge 3$.}
\label{Fig:lcc12}
\end{figure}

\begin{figure}
\includegraphics[width=0.45\textwidth]{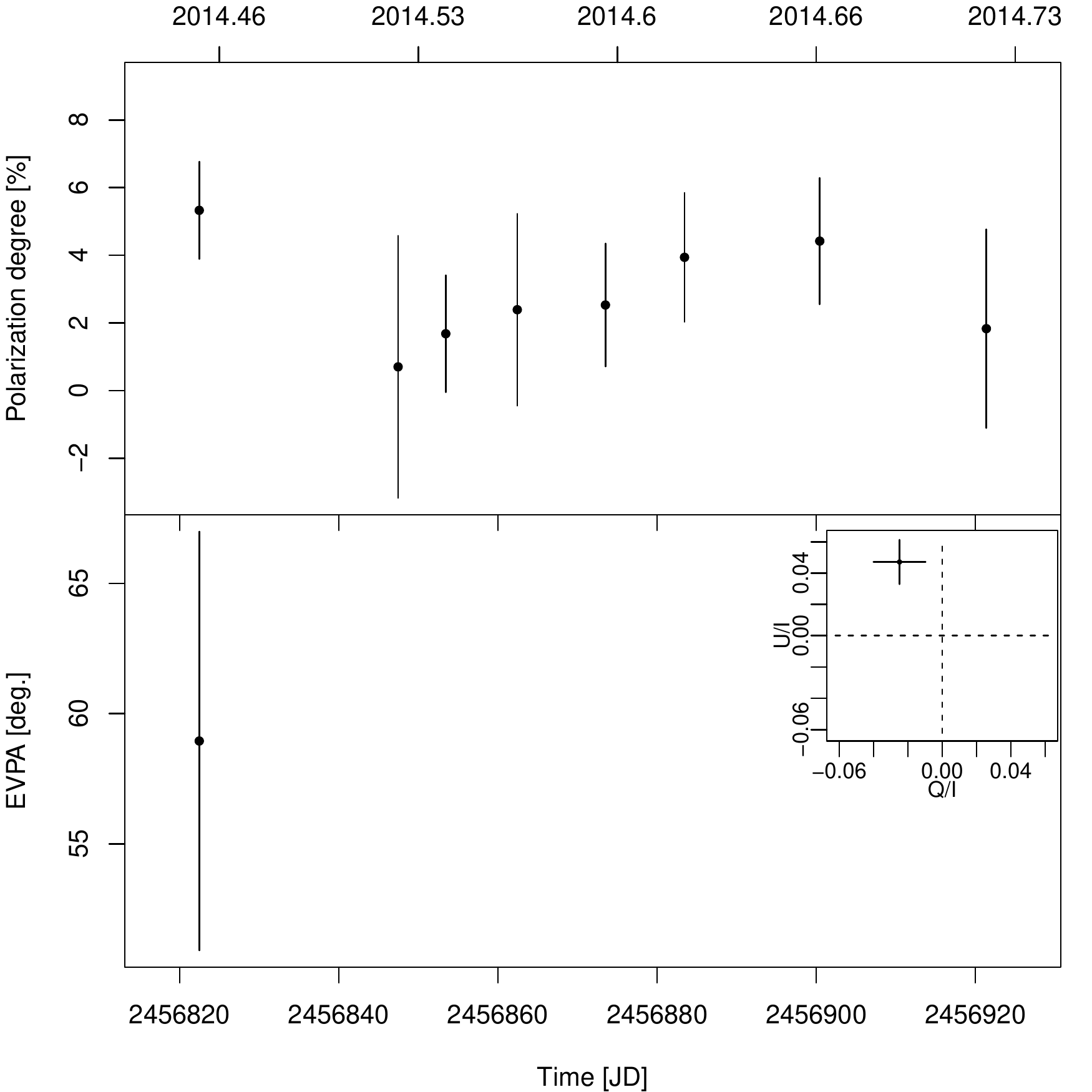}
\caption{Fractional polarization (top) and EVPA (bottom) of the  non-TeV
  source J1841+3218. The inset in the lower panel shows the
  $Q/I$ vs. $U/I$ from the RoboPol data. EVPA data are shown only for
  observations where the signal-to-noise in the polarization fraction
  $\ge 3$.}
\label{Fig:lcc13}
\end{figure}

\begin{figure}
\includegraphics[width=0.45\textwidth]{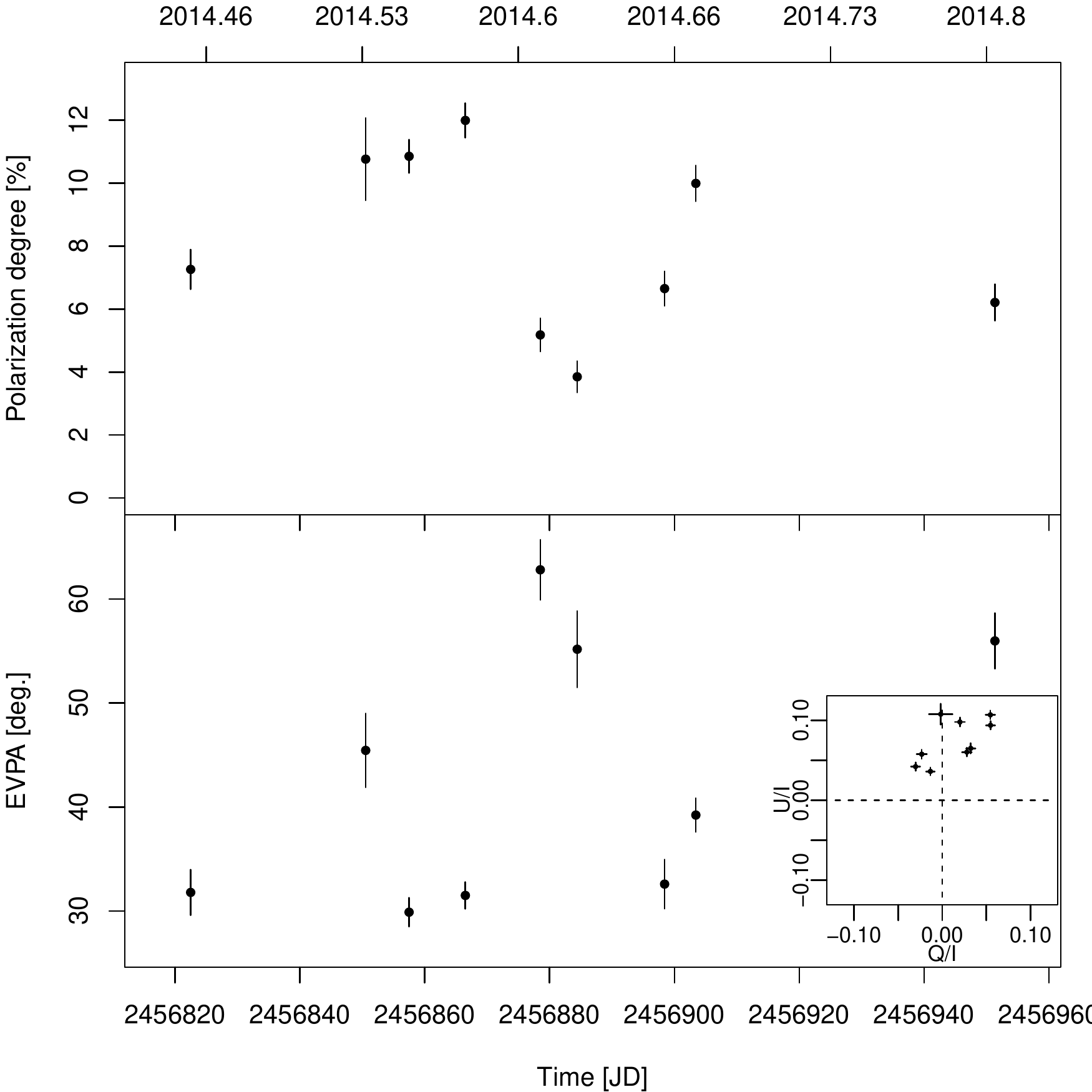}
\caption{Fractional polarization (top) and EVPA (bottom) of the  non-TeV
  source J1903+5540. The inset in the lower panel shows the
  $Q/I$ vs. $U/I$ from the RoboPol data. EVPA data are shown only for
  observations where the signal-to-noise in the polarization fraction
  $\ge 3$.}
\label{Fig:lcc14}
\end{figure}

\begin{figure}
\includegraphics[width=0.45\textwidth]{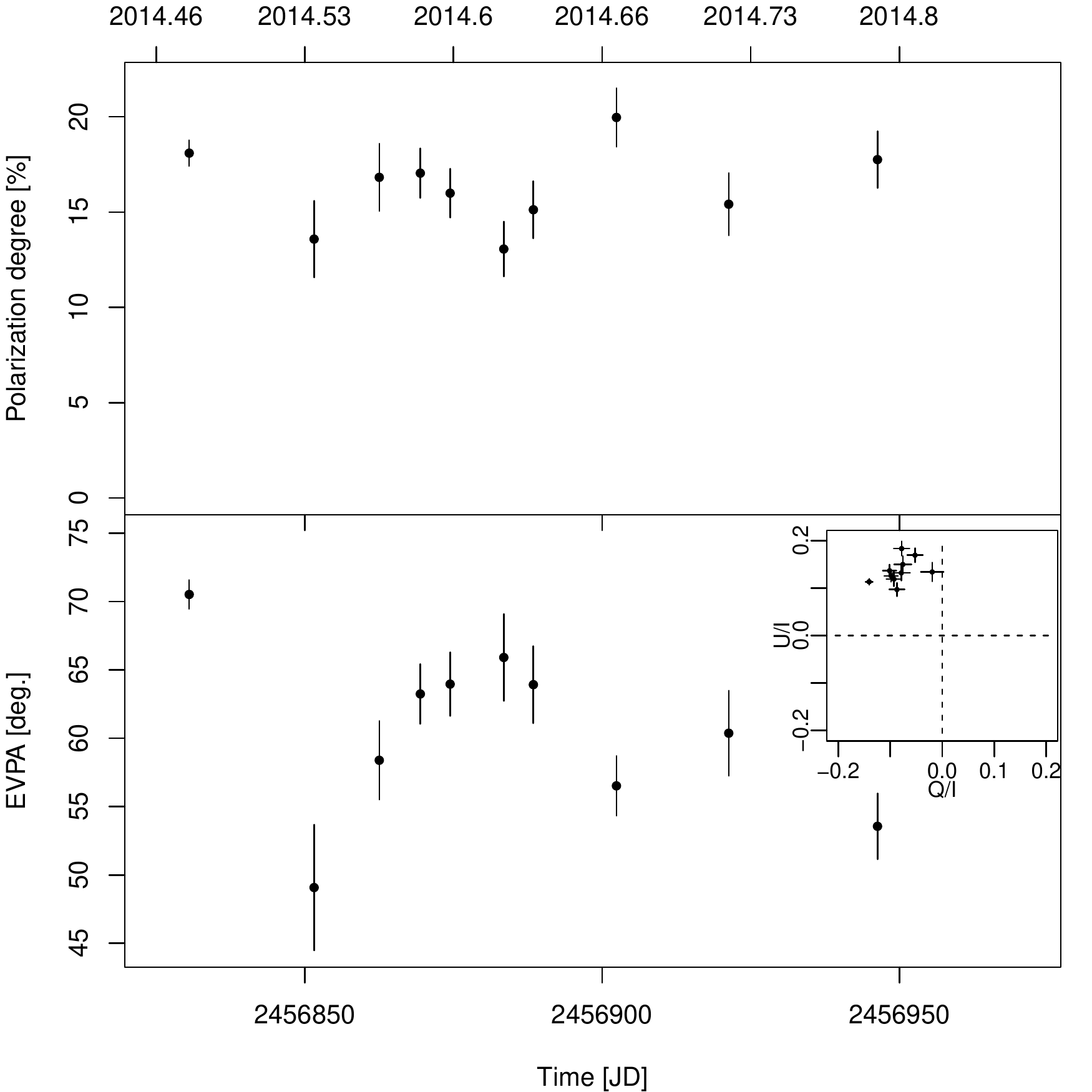}
\caption{Fractional polarization (top) and EVPA (bottom) of the  non-TeV
  source J2015-0137. The inset in the lower panel shows the
  $Q/I$ vs. $U/I$ from the RoboPol data. EVPA data are shown only for
  observations where the signal-to-noise in the polarization fraction
  $\ge 3$.}
\label{Fig:lcc15}
\end{figure}

\begin{figure}
\includegraphics[width=0.45\textwidth]{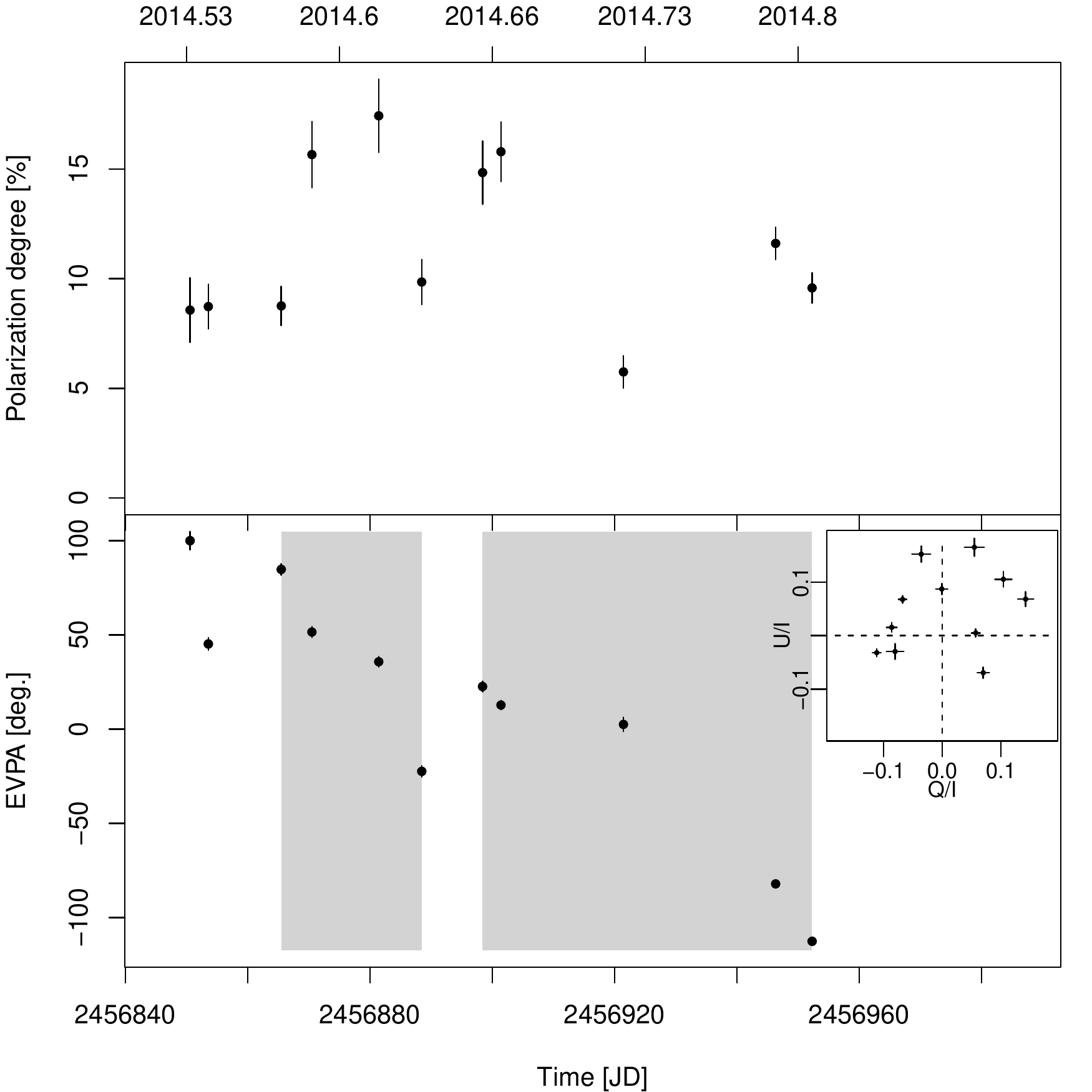}
\caption{Fractional polarization (top) and EVPA (bottom) of the  non-TeV
  source J2022+7611. The shaded region shows the period of a
  significant EVPA rotation. The inset in the lower panel shows the
  $Q/I$ vs. $U/I$ from the RoboPol data. EVPA data are shown only for
  observations where the signal-to-noise in the polarization fraction
  $\ge 3$.}
\label{Fig:lcc16}
\end{figure}

\begin{figure}
\includegraphics[width=0.45\textwidth]{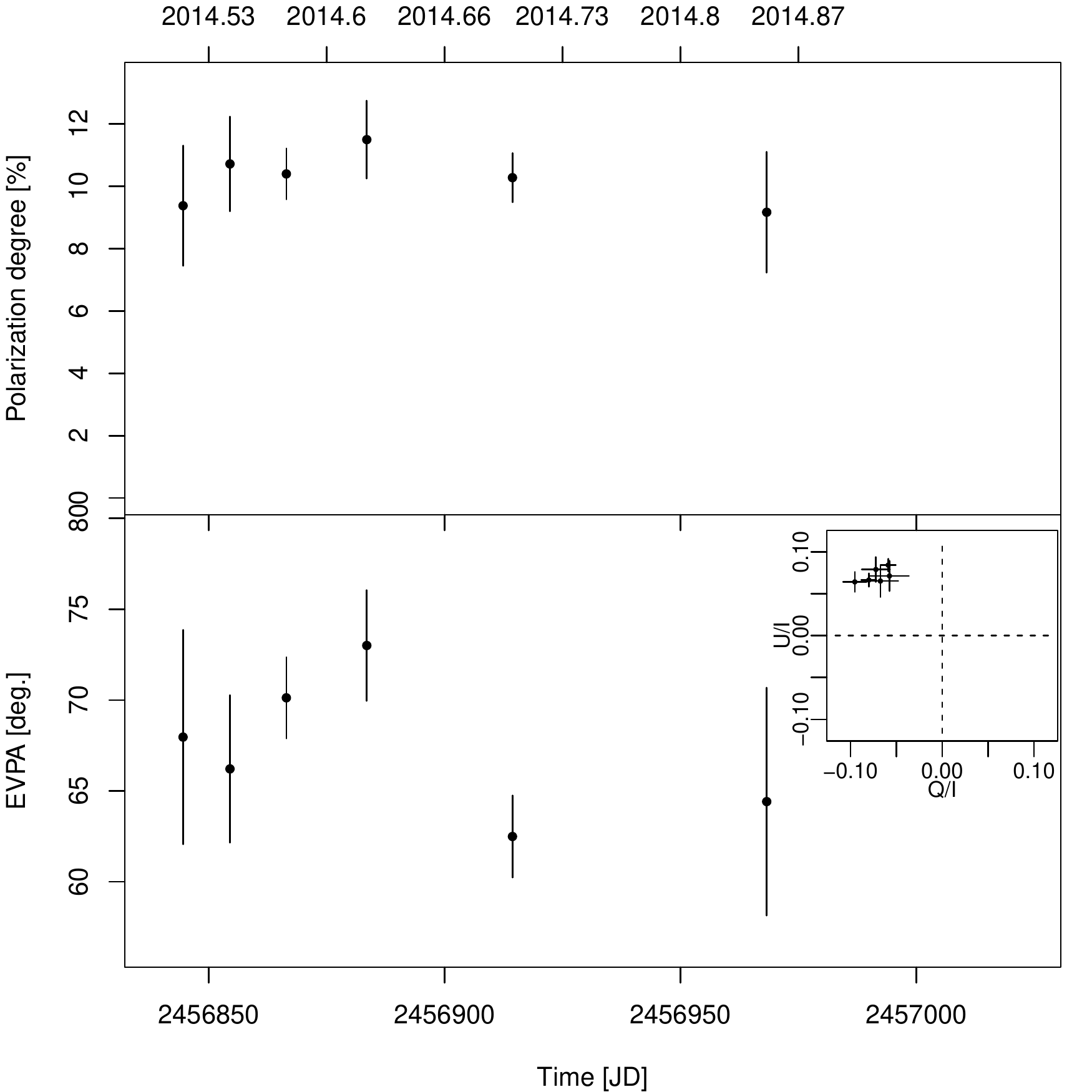}
\caption{Fractional polarization (top) and EVPA (bottom) of the  non-TeV
  source J2131-0915. The inset in the lower panel shows the
  $Q/I$ vs. $U/I$ from the RoboPol data. EVPA data are shown only for
  observations where the signal-to-noise in the polarization fraction
  $\ge 3$.}
\label{Fig:lcc17}
\end{figure}

\begin{figure}
\includegraphics[width=0.45\textwidth]{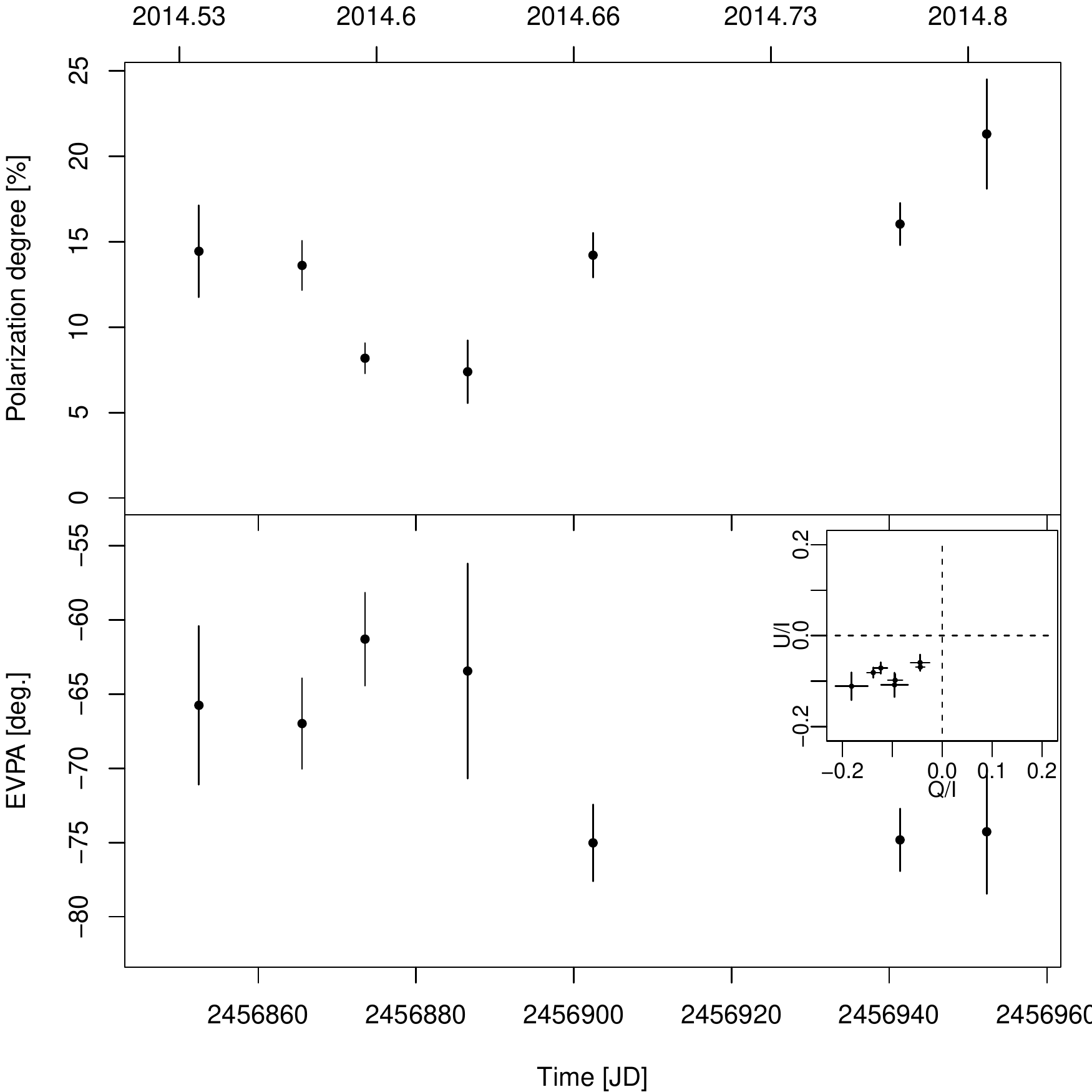}
\caption{Fractional polarization (top) and EVPA (bottom) of the  non-TeV
  source J2149+0322. The inset in the lower panel shows the
  $Q/I$ vs. $U/I$ from the RoboPol data. EVPA data are shown only for
  observations where the signal-to-noise in the polarization fraction
  $\ge 3$.}
\label{Fig:lcc18}
\end{figure}

\begin{figure}
\includegraphics[width=0.45\textwidth]{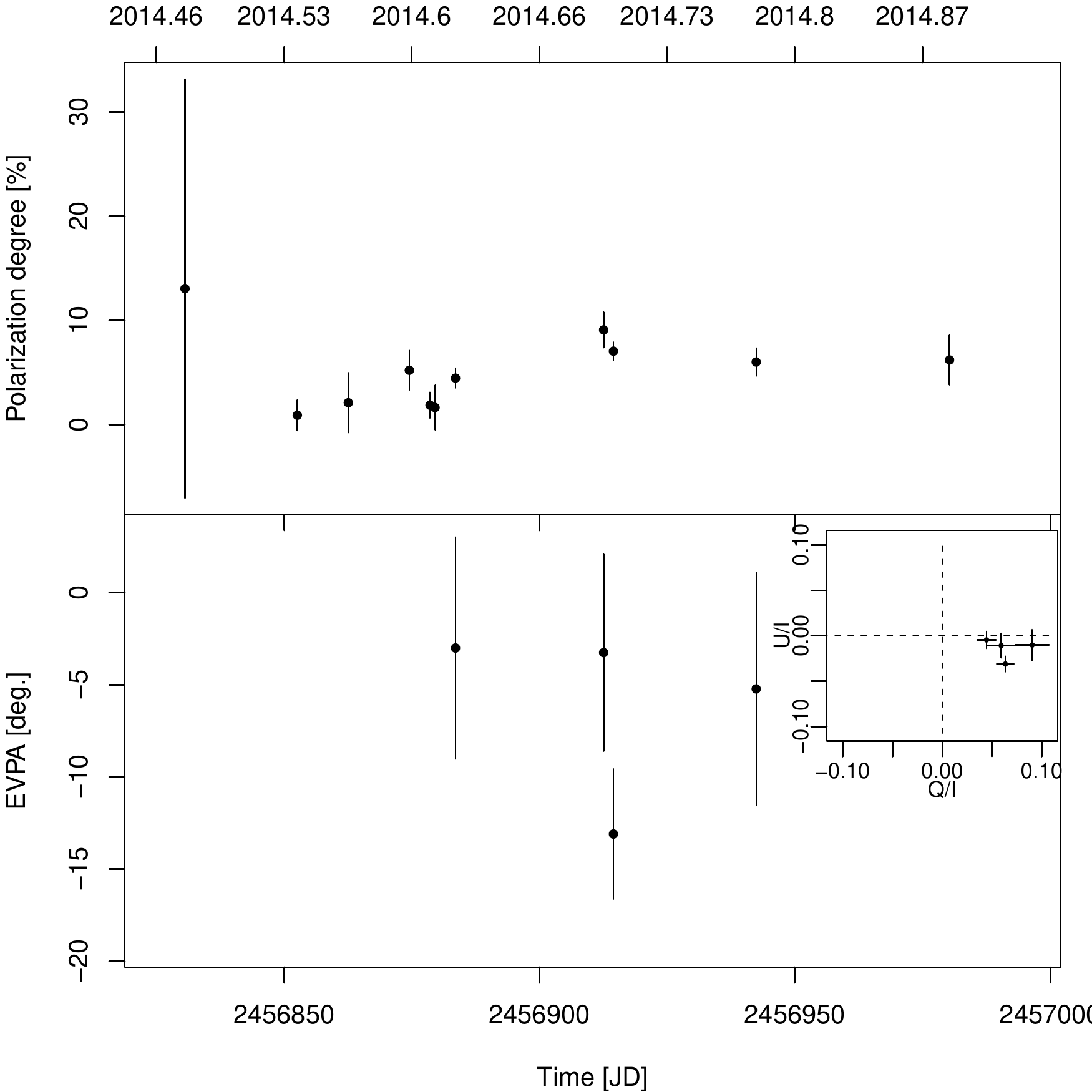}
\caption{Fractional polarization (top) and EVPA (bottom) of the  non-TeV
  source J2340+8015. The inset in the lower panel shows the
  $Q/I$ vs. $U/I$ from the RoboPol data. EVPA data are shown only for
  observations where the signal-to-noise in the polarization fraction
  $\ge 3$.}
\label{Fig:lcc19}
\end{figure}

\end{document}